\documentclass[article,twocolumn,preprintnumbers,bibnotes10pt,superscriptaddress]{revtex4}

\usepackage{graphicx,subfigure}
\usepackage{float}
\usepackage{color}
\usepackage{epstopdf}
\usepackage{amsmath}
\usepackage{amsthm}
\usepackage{amssymb}
\usepackage{amsfonts}
\usepackage{graphicx}
\usepackage{txfonts}
\usepackage{ulem}
\usepackage{mathtools}
\usepackage{bm}

\usepackage{amsmath}
\usepackage{graphicx}
\usepackage{ulem}
\usepackage{dcolumn}
\usepackage{epsfig}
\usepackage{bm}
\usepackage{array}
\usepackage{hyperref}
\hypersetup{
colorlinks=true,
citecolor=black,
linkcolor=red,
urlcolor=black
 , bookmarks=true,
 pdfmenubar=true
}

\newcommand{\bra}[1]{\ensuremath{\langle#1|}}
\newcommand{\ket}[1]{\ensuremath{|#1\rangle}}

\newcommand{\mean}[1]{\ensuremath{\big\langle #1 \big\rangle}}

\newcommand{\makeref}[1]{(\ref{#1})}

\newcommand{\vect}[1]{\bm{#1}}

\newcommand{\be}{\begin{equation}}
\newcommand{\ee}{\end{equation}}

\newcommand{\beq}{\begin{eqnarray}}
\newcommand{\eeq}{\end{eqnarray}}

\newcommand{\FQ}{\mathcal{F}_{\rm Q}}

\allowdisplaybreaks

\begin{document}

\newtheorem{proposition}{Proposition}
\theoremstyle{plain}

\title{Distributed Quantum Sensing with Squeezed-Vacuum Light \\ in a Configurable Network of Mach-Zehnder Interferometers}
\author{Marco Malitesta}
\affiliation{QSTAR, INO-CNR and LENS, Largo Enrico Fermi 2, 50125 Firenze, Italy} 
\affiliation{Universit\`a degli Studi di Napoli ”Federico II”, Via Cinthia 21, 80126 Napoli, Italy}

\author{Augusto Smerzi} 
\affiliation{QSTAR, INO-CNR and LENS, Largo Enrico Fermi 2, 50125 Firenze, Italy} 

\author{Luca Pezz\`e} 
\affiliation{QSTAR, INO-CNR and LENS, Largo Enrico Fermi 2, 50125 Firenze, Italy} 

\begin{abstract}
We study a sensor network of distributed Mach-Zehnder interferometers (MZIs) for the parallel (simultaneous) estimation of an arbitrary number $d \geq 1$ of phase shifts. 
The scheme uses a squeezed-vacuum state that is split between $d$ modes by a quantum circuit (QC).
Each output mode of the QC is the input of one of $d$ MZIs, the other input of each MZI being a coherent state.
In particular, {\it i}) we identify the optimal configuration of the sensor network that allows the estimation of any linear combination of the $d$ phases with maximal sensitivity.
The protocol overcomes the shot-noise limit and reaches Heisenberg scalings with respect to the total average number of particles in the overall probe state, the multiphase estimation only requiring local photocounting.
Furthermore, the parallel multiphase estimation overcomes optimal separable strategies for the estimation of any linear combination of the phases: the sensitivity gain being a factor $d$, at most.
Viceversa, {\it ii}) given a specific QC, we identify the optimal linear combination of the phases that maximizes the sensitivity and show that results are robust against random choices of the QC.
Our scheme paves the ways to a variety of applications in distributed quantum sensing. 
\end{abstract}

\maketitle
\date{\today}

\section{Introduction}

Optical interferometry exploiting squeezed light~\cite{LoudonJMO1987, ScullyBOOK, BreitbachNATURE1997, AndersenPS2016} has been -- since the pioneer 40-years-old proposal by Caves~\cite{CavesPRD1981} -- a cornerstone of  theoretical~\cite{ParisPLA1995, BarnettEPJD2003, PezzePRL2008, LangPRL2013, Ruo-BercheraPRA2015, SparaciariPRA2016, BondurantPRD1984} and experimental~\cite{WuPRL1986, XiaoPRL1987, GrangierPRL1987, PolzikPRL1992, GodaNATPHYS2008} photonic quantum sensing~\cite{SchnabelPR2017, PirandolaNATPHOT2018, LawrieACSP2019, PolinoARXIV}.
A Mach-Zehnder interferometer (MZI) with a high-power coherent state in one input port and a low-intensity squeezed-vacuum light in the other input can reach a phase estimation uncertainty $\Delta^2 \theta = e^{-2r}/\bar{n}_T$~\cite{CavesPRD1981}, where $r\geq 0$ is the squeeze parameter, $\bar{n}_T$ is the total average number of photons and $\theta$ is the relative phase shift between the two arms of the interferometer.
This scheme can overcome the shot-noise (SN) limit $\Delta^2 \theta_{\rm SN} = 1/\bar{n}_T$ by an amount depending on the squeezing strength $r$.
Currently, squeeze factors of more than 10 dB have been observed in several experiments~\cite{VahlbruchPRL2016, SchonbeckOPTLETT2018, SchnabelPR2017, LawrieACSP2019}.
Furthermore, when the coherent and the squeezed-vacuum input states have approximately the same intensity, the MZI  can achieve~\cite{PezzePRL2008} the Heisenberg limit (HL) $\Delta^2 \theta_{\rm HL} = 1/\bar{n}_T^2$.
This prediction has been associated~\cite{HofmanPRA2007, PezzePRL2008} to the onset of NOON states after the first beam splitter of the MZI, as verified experimentally~\cite{AfekSCIENCE2010}. 
Enhancing the sensitivity by replacing the normally-empty input port with squeezed-vacuum light is relevant when there are constraints limiting the total light intensity inside the interferometer. 
Gravitational wave detection is an important application~\cite{SchnabelNATCOMM2010, RafalPRA2013, ChuaCQG2014, AbadieNATPHYS2011, AasiNATPHOT2013, TsePRL2019, AcernesePRL2019}, where the quantum-enhancement offered by squeezing allows to boost substantially the expected rate of detectable events.
Quantum imaging~\cite{MoreauNPR2019, Ruo-BercheraMETROLOGIA2019}, 
microscopy~\cite{CasacioNATURE2021} and the probing of biological samples~\cite{TaylorNATPHOT2013} are other relevant applications that require high resolution but low probe power~\cite{TaylorPHYSREP2016}.
Squeezed-vacuum states can be also generated via spin-changing collisions in a Bose-Einstein condensate~\cite{GrossNATURE2011, HamleyNATPHYS2012, PeiseNATCOMM2015} and used to enhance the sensitivity of atomic MZIs~\cite{KrusePRL2016, PezzeRMP2018}.

\begin{figure*}[t!]
\centering
\includegraphics[width=1\textwidth]{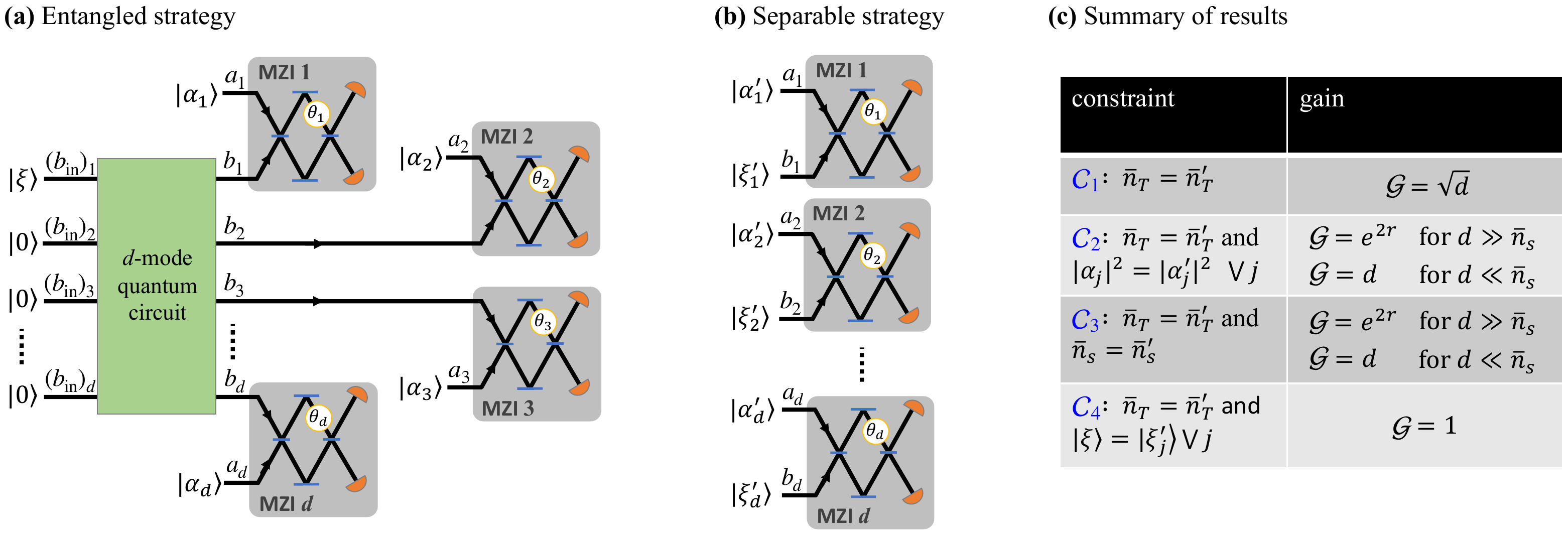}
\caption{ 
A distributed quantum sensor for the estimation of $d$ relative phases $\theta_1, ..., \theta_d$ can follow an entangled (a) or a separable (b) strategy.
Panel (a) shows the Mach-Zehnder sensor network composed by $d$ MZIs.
The input of the $j$th MZI is given by a coherent state $\ket{\alpha_j}$ in the mode $a_j$, while the other input $b_j$ is one of the outputs of a linear $d$-mode quantum circuit. 
The input of the quantum circuit is a single squeezed-vacuum state $\ket{\xi}$, which is mixed with $d-1$ vacuum states $\ket{0}$.
The output state of the $d$-mode quantum circuit is mode entangled: we thus identify the scheme of panel (a) as an entangled multiphase estimation strategy. 
In panel (b) the different MZIs are independent. 
The $j$th MZI has a coherent state $\ket{\alpha_j'}$ in the input mode $a_j$ and a squeezed-vacuum state $\ket{\xi'}$ in mode $b_j$.
This scheme realizes a separable multiphase estimation strategy.
The table in panel (c) summarizes the main results of the manuscript for what concerns the gain $\mathcal{G}$ of the entangled strategy over the separable one, under different constraints $\mathcal{C}$ (see text for details).
Here, $\bar{n}_T$ is the total average number of particles, $\bar{n}_s$ is the total number of particles in the squeezed state(s), and prime symbols refer to the separable strategy.}
\label{fig1}
\end{figure*}

To date, Mach-Zehnder interferometry using squeezed states has focused on the estimation of a single phase shift~\cite{SchnabelPR2017, LawrieACSP2019, RafalPROGOPT2015, PezzeRMP2018}. 
Yet, several applications require the estimation of multiple phases encoded simultaneously in a network of spatially-separated sensors~\cite{AlbarelliPLA2020}.  
Distributed quantum sensing~\cite{HumphreysPRL2013, LiuJPA2016, CiampiniSCIREP2016, ProctorPRL2018, GePRL2018, ElderidgePRA2018, GessnerPRL2018, ZhuangPRA2018, NicholsPRA2018, OhPRR2020, ZhuangNJP2020, GessnerNATCOMM2020, TriggianiARXIV, GebhartARXIV} is thus  attracting increasing interest.
Recently, different schemes have been experimentally realized using squeezed light~\cite{GuoNATPHYS2020, XiaPRL2020}, photonics  Greenberger–Horne–Zeilinger (GHZ)~\cite{LiuNATPHOT2021}, Bell~\cite{ZhaoPRX2021}, multimode NOON~\cite{HongNATCOMM2021} and single-photon Fock~\cite{PolinoOPTICA2019, ValeriNPJ2020} states.  
In Ref.~\cite{GuoNATPHYS2020} a displaced squeezed state is split among four spatial modes, which then undergo a phase shift and are finally measured by homodyne detection.
Reference~\cite{XiaPRL2020} has instead reported sensing of phase-space displacement using a squeezed vacuum state split among three modes~\cite{ZhuangPRA2018}.  
The sensor network of Refs.~\cite{LiuNATPHOT2021, ZhaoPRX2021} is based on the polarization rotation of photonic qubits.
In particular, Ref.~\cite{ZhaoPRX2021} has realized distributed sensing over large spatial distances.
Finally, Refs.~\cite{PolinoOPTICA2019, ValeriNPJ2020} have considered a multimode interferometer on a photonic chip, where single photons are split among many modes by a generalized beam splitter, phase shifted and recombined by a second multimode beam splitter~\cite{CiampiniSCIREP2016}. 

Surprisingly, so far, little effort has been devoted to study multiphase estimation in a network of distributed MZIs.
This system is relevant in quantum optics, as well as in atom interferometry, with possible applications in arrays of quantum clocks~\cite{KomarNATPHYS2014, PolzikPRA2016} and spatial magnetometry~\cite{BaumgratzPRL2016, AltenburgPRA2017, ApellanizPRA2018, HouPRL2020}. 
A key aspect of the MZI is that phase sensitivity bounds are well defined and  quantified in term of the {\it total} average number of particles in the input state~\cite{PezzePRL2007}, without ambiguities related to the resource cost necessary to establish a phase-reference~\cite{JarzynaPRA2012, PezzePRA2015, GoldbergPRA2020}
(e.g. for homodyne detection).
Total resource counting is crucial to  quantify improved performances of quantum devices over classical strategies and to claim sub-SN sensitivities.

In this manuscript, we study a distributed-sensing scheme that generalizes the single MZI with  coherent$\otimes$squeezed-vacuum input light~\cite{CavesPRD1981,ParisPLA1995, BarnettEPJD2003, PezzePRL2008, LangPRL2013} to a network of $d \geq 1$ spatially-distributed MZIs, see Fig.~\ref{fig1}(a).
In our setup, a single squeezed-vacuum state $\ket{\xi}$ of squeeze parameter $r$ is first split by a quantum circuit (QC) consisting of a $d$-mode beam splitter~\cite{GuoNATPHYS2020, XiaPRL2020}.
The QC is identified by a unitary transformation $\vect{U}$ that can be realized, in practice, by a sequence of two-mode linear operations~\cite{ReckPRL1994, NokkalaNJP2018}. 
The sensor network of Fig.~\ref{fig1}(a) can thus be scaled to an arbitrary number, $d$, of interferometers. 
The linear splitting of the squeezed-vacuum state generates entanglement among the $d$ output modes $b_1, ..., b_d$ of the QC. 
We thus identify the scheme of Fig.~\ref{fig1}(a) as an entangled multiphase estimation strategy. 
Each mode $b_j$ is used as the input mode of a MZI, the other input $a_j$ being a coherent state.
It should be notices that the $d$ coherent states $\vert \alpha_j \rangle$ are phase locked with the squeezed-vacuum state $\ket{\xi}$. 
In the $j$th MZI ($j=1, ..., d$), the two input modes mix at a balanced beam splitter, encode a relative phases $\theta_j$ and are detected by photocounting after a final balanced beam splitter.
Here, we estimate {\it arbitrary linear combinations} $\vect{v} \cdot \vect{\theta} = \sum_{j=1}^d v_j \theta_j$ of the $d$ phase shifts $\vect{\theta} = \{\theta_1, ..., \theta_d\}$, where $\vect{v} = \{ v_1, ..., v_d \}$ is a real vector. 
The estimation method is based on a  multimode moment-matrix approach~\cite{GessnerNATCOMM2020} and the corresponding sensitivity is compared to the multiparameter quantum Cram\'er-Rao bound~\cite{HelstromBOOK, HolevoBOOK, ParrisIJQI2009}.

We identify different regimes 
-- depending on the relative intensity of the squeezed-vacuum and the coherent states -- and predict sub-shot noise sensitivities, up to the Heisenberg limit,  with respect to the total number $\bar{n}_T$ of photons used.
In particular, {\it i}) 
given a specific linear combination $\vect{v}\cdot \vect{\theta}$ that one wants to estimate, we identify the optimal configuration of the sensor network of Fig. \ref{fig1}(a) that minimizes the uncertainty $\Delta^2(\vect{v}\cdot \vect{\theta})$.
The analytical optimization of $\vect{U}$, $\vert \alpha_1 \vert^2, ..., \vert \alpha_d \vert^2$ and $r$ is supported by numerical calculations.
In particular, we identify conditions for which the strategy of Fig.~\ref{fig1}(a) is never surpassed (for any $\vect{v} \cdot \vect{\theta}$) by the separable strategy of Fig.~\ref{fig1}(b), which uses independent MZIs for the estimation of each $\theta_j$.
It should be also noticed that the separable strategy uses $d$ squeezed-vacuum states (one for each MZI), while the entangled strategy uses a single squeezed state. 
The table in Fig.~\ref{fig1}(c) summarizes our results concerning the maximum gain $\mathcal{G}$ of the entangled strategy over the separable one, under different constraints $\mathcal{C}$. 
There, $\mathcal{G}$ is understood as optimized over all $\vect{v}$.
When $\bar{n}_s = \sinh^2 r \gg d$, we find a maximum gain given by a factor $d$. 
In the opposite limit, for $\bar{n}_s \ll d$ -- in particular for $d\to \infty$ -- we find a finite gain $e^{2r}$. 
The later result is surprising since, in this case, the squeezed vacuum is mixes, at the QC, with a diverging number of vacuum states.
{\it Viceversa}, {\it ii}) for any given QC transformation $\vect{U}$, we identify optimal and orthogonal linear combinations $\vect{v} \cdot \vect{\theta}$ that can be estimated with the highest possible sensitivity. 
This optimization problem leads us to introduce the useful concepts of Fisher and squeezing spectra. 
In particular, we show that results are robust against random choices of $\vect{U}$. 
Our findings pave the way toward distributed multi-phase estimation in a network of MZIs, using quantum states and detection capabilities that are common to many laboratories.

The paper is structured as it follows.
Secion~\ref{IntroA} introduces basic and general notions of distributed sensing.
In particular, we recall the multimode moment-matrix approach used in this work and the multiparameter Cram\'er-Rao bound. 
Section~\ref{MZSN} illustrates in details the Mach-Zehnder sensor network of Fig.~\ref{fig1}(a). 
Section~\ref{OptSens} presents the optimization for arbitrary $\vect{v}$.
First, we provide, in Sec.~\ref{Sec.bounds}, upper and lower bounds to the sensitivity and present different sensitivity scalings with respect to the total average number of particles. 
We then compare, in Sec.~\ref{Sec.compare}, optimal entangled and separable strategies under different constraints, providing a detailed discussion of the results presented in the table of Fig. \ref{fig1}(c). 
Section~\ref{Optv} studies the optimal linear combination of phases that can be estimated for a given configuration of the Mach-Zehnder sensor network. 
We first introduce the notion of Fisher and squeezing spectra, in Sec. \ref{OptSpect}, and then apply this formalism to calculate the sensitivity achievable for random choices of the QC, in Sec. \ref{OptRand}.
We finally, in Sec.~\ref{conclusion}, compare our finding with the literature and conclude.

\section{Distributed quantum sensing} 
\label{IntroA}

\subsection{Definition and strategies}

In a distributed quantum sensing problem, $d$ unknown parameters $\theta_1, ..., \theta_d$, are encoded in independent (e.g. spatially-separated) modes or interferometers.
Parameter encoding is described by commuting transformations. 
In the ideal noiseless scenario, this is given by  the unitary evolution 
$e^{-i \vect{\theta} \cdot \hat{\vect{H}}} = \otimes_{j=1}^d e^{-i \hat{H}_j \theta_j}$, where $\hat{\vect{H}} = \{ \hat{H}_1, ..., \hat{H}_d \}$ is a set of commuting Hermitian operators, $[\hat{H}_i, \hat{H}_j] =0$ for $i,j=1, ...,d$, and $\vect{\theta} \cdot \hat{\vect{H}} = \sum_{j=1}^d \theta_j \hat{H}_j$.

The sensing scheme can follow an {\it entangled} (also indicated as parallel, or global, in the literature) or a {\it separable} (sequential, or local) strategy~\cite{ProctorPRL2018, GePRL2018, GessnerPRL2018, KnottPRA2016, GuoNATPHYS2020, XiaPRL2020}. 
In an entangled strategy, the overall probe state $\hat{\rho}$ of the sensor network is prepared in a mode-entangled state.
In contrast, a {\it separable} strategy uses the product state $\hat{\rho} = \bigotimes_{j=1}^d \hat{\rho}_j$, where $\hat{\rho}_j$ is the probe state of the $j$th sensor:
the different sensors are thus uncorrelated and the parameters $\theta_1, ..., \theta_d$ are estimated independently.
Notice that classical correlations among the different $\hat{\rho}_j$ are not useful to increase the multiparameter sensitivity, in general~\cite{GessnerPRL2018}.
Finally, an interesting possibility is to consider local measurements at each sensor, without requiring a mode entangled measures (although also distributed sensing scheme based on a final recombination of parameter-sensing modes have been considered~\cite{CiampiniSCIREP2016, GePRL2018, TriggianiARXIV, PolinoOPTICA2019, ValeriNPJ2020, OhPRR2020}): local measurements are advantageous when the sensing modes are spatially delocalized~\cite{ZhaoPRX2021}.

\subsection{Figure of merit}

One of the goals of multiparameter estimation, in general, is to infer linear combinations $\vect{v} \cdot \vect{\theta} = \sum_{j=1}^d v_j \theta_j$ of $d$ parameters
$\vect{\theta}$ encoded in the quantum device \cite{GePRL2018, ProctorPRL2018, GessnerPRL2018, XiaPRL2020, GuoNATPHYS2020, LiuNATPHOT2021, ZhaoPRX2021, RubioJPA2020, GrossJPA2021, QianPRA2019}.
In the following we take $v_j$ real (either positive or negative) and $v_j \neq 0$ for all $j=1, ..., d$ to guarantee an irreducible $d$-parameter problem.
We also consider the normalization $\vert \vect{v} \vert^2  = \sum_{j=1}^d v_j^2 = 1/d$, without loss of generality.
An example of linear combination of parameters is the average $\vect{v} \cdot \vect{\theta} = (\theta_1 + \theta_2 + ... + \theta_d)/d$, corresponding to $v_j=1/d$.

The method of moments is a feasible approach to multi-parameter estimation~\cite{GessnerNATCOMM2020}.
Here, it is based on a set of $d$ Hermitian and commuting measurement operators $\hat{X}_j$ whose mean $\mean{\hat{X}_j}$ is a monotonic function of $\theta_j$ only.
The estimation method consists of repeating the measurement of the local observable $\hat{X}_j$ several times. 
Taking the average value $\bar{X}_j$ and inverting the equation $\mean{\hat{X}_j}= \bar{X}_j$ provides an estimate of $\theta_j$.
For the entangled multiparameter scenario, the method achieves an uncertainty~\cite{GessnerNATCOMM2020}
\be \label{emom}
\Delta^2(\vect{v} \cdot \vect{\theta})_{\rm emom} = 
\vect{v}^T \vect{\mathcal{M}}^{-1} \vect{v},
\ee
where $\vect{\mathcal{M}}=\vect{G}^T\vect{\Gamma}^{-1}\vect{G}$, $\vect{G}_{ij}=\partial \langle\hat{X}_i\rangle/\partial \theta_j$, and $\vect{\Gamma}_{ij}=\langle\hat{X}_i\hat{X}_j\rangle-\langle\hat{X}_i\rangle\langle\hat{X}_j\rangle$
are $d\times d$ matrices ($\vect{G}$ being diagonal in this case), and the  expectation values are calculated with respect to the joint output state of the $d$ sensors, $e^{-i \hat{\vect{H}} \cdot \vect{\theta}}\hat{\rho} e^{i \hat{\vect{H}} \cdot \vect{\theta}}$.
The covariance matrix $\vect{\Gamma}$ expresses correlations between measurement observables. 
These correlations are directly linked to the entanglement in the probe state $\hat{\rho}$ and can be engineered to enhanced the sensitivity in the estimation of certain combinations $\vect{v} \cdot \vect{\theta}$.
In Ref.~\cite{GessnerNATCOMM2020} the moment matrix $\vect{\mathcal{M}}$ has been also used to characterize and detect metrological multimode squeezing. 
In separable strategies, $\vect{\Gamma}$ is diagonal and Eq.~(\ref{emom}) becomes
\be \label{smom}
\Delta^2 (\vect{v} \cdot \vect{\theta})_{\rm smom} = \sum_{j=1}^d \frac{v_j^2 \Delta^2 \hat{X}_j}{(d \mean{\hat{X}_j}/d \theta_j)^2},
\ee
where $\Delta^2 \hat{X}_j = \mean{\hat{X}_j^2} - \mean{\hat{X}_j}^2$, with the expectation values calculated on the output state of the $j$th sensor, $e^{-i \hat{H}_j \theta_j} \hat{\rho}_j e^{i \hat{H}_j \theta_j}$.

The ultimate sensitivity limit in the  estimation of $\vect{v} \cdot \vect{\theta}$ is provided by the quantum Cramer-Rao bound~\cite{HelstromBOOK, HolevoBOOK, ParrisIJQI2009}. 
In the entangled setting, we have 
$\Delta^2 (\vect{v} \cdot \vect{\theta})_{\rm emom} \geq \Delta^2(\vect{v} \cdot \vect{\theta})_{\rm eQCR}$, where 
 \be  \label{Dthetan_par}
\Delta^2(\vect{v} \cdot \vect{\theta})_{\rm eQCR}
= \vect{v}^T \vect{\FQ}^{-1} \vect{v},
\ee 
and $\vect{\FQ}$ is the $d\times d$ quantum Fisher information matrix (QFIM)~\cite{notaQFIM}.
If the overall probe state is pure, $\hat{\rho} = \ket{\psi} \bra{\psi}$, then the QFIM is $(\vect{\FQ})_{ij} = 4 ( \langle  \psi \vert \hat{H}_i \hat{H}_j \ket{ \psi } - 
\langle  \psi \vert \hat{H}_i \ket{ \psi }
\langle  \psi \vert \hat{H}_j \ket{ \psi })$
and Eq.~(\ref{Dthetan_par}) can be saturated by optimal measurements and estimators~\cite{MatsumotoJPA2002, PezzePRL2017}.
In the sequential setting, the quantum Cramer-Rao bound is $\Delta^2 (\vect{v} \cdot \vect{\theta})_{\rm smom} \geq \Delta^2(\vect{v} \cdot \vect{\theta})_{\rm sQCR}$, where 
\be  \label{Dthetan_seq}
\Delta^2(\vect{v} \cdot \vect{\theta})_{\rm sQCR}
= \sum_{j=1}^d \frac{ v_j^2}{\mathcal{F}_j},
\ee 
and $\mathcal{F}_j$ is the (scalar) quantum Fisher information~\cite{HelstromBOOK, HolevoBOOK, BraunsteinPRL1994}.
For pure states, 
Eq.~(\ref{Dthetan_seq}) is obtained from Eq.~(\ref{Dthetan_par}) when taking the product $\ket{\psi} = \bigotimes_{j=1}^d \ket{\psi_j}$ such that $\vect{\FQ}$ becomes diagonal with entries 
$\mathcal{F}_j = (\vect{\FQ})_{jj} = 4( \bra{\psi_j} \hat{H}_j^2 \ket{\psi_j} - \bra{\psi_j} \hat{H}_j \ket{\psi_j}^2)$.

We recall that the different terms on the right-hand side of  Eqs.~(\ref{emom})-(\ref{Dthetan_seq}) are understood as divided by the number of repeated independent measurements $m$ used for the estimation. 
In particular, Eqs.~(\ref{emom})-(\ref{Dthetan_seq}) can be saturated, in general, for $m \gg 1$. To simplify the notation, we neglect the factor $m$ here and in the following
(see Ref.~\cite{GebhartARXIV} for a multi-parameter Bayesian estimation analysis including $m$ as a resource). 

\section{Mach-Zehnder sensor network}
\label{MZSN}

The quantum distributed sensing scheme considered in this manuscript consists of a network of $d$ MZIs, see Fig.~\ref{fig1}.
The entangled strategy is shown in Fig.~\ref{fig1}(a), while the separable one in Fig.~\ref{fig1}(b).
In both cases, the $j$th interferometer is described by the unitary phase encoding transformation $e^{-i \theta_j \hat{H}_j}$, where $\hat{H}_j = (\hat{a}_j^\dag \hat{b}_j - \hat{b}_j^\dag \hat{a}_j)/2i$, $\hat{a}_j$ and $\hat{b}_j$ ($\hat{a}_j^\dag$ and $\hat{b}_j^\dag$) are bosonic annihilation (creation) operators for the two interferometer modes, respectively, and $\theta_j$ is a relative phase shift between the two interferometer arms. 
The $d$ MZIs are independent (namely, $[\hat{a}_i, \hat{a}_j^\dag] = [\hat{a}_i, \hat{b}_j^\dag]  = [\hat{a}_i, \hat{b}_j] = 0$ for $i \neq j$), which guarantees 
$[\hat{H}_i, \hat{H}_j]=0$.
Furthermore, we take local measurement observables $\hat{X}_j = (\hat{a}_j^\dag \hat{a}_j -\hat{b}_j^\dag \hat{b}_j)_{\rm out}/2$, counting the relative number of photons at the output port of the $j$th MZI, for $j=1, ..., d$.

In the entangled strategy of  Fig.~\ref{fig1}(a),
the input mode $a_j$ of the $j$th MZI is fed with a coherent state $\ket{\alpha_j}$, where $\alpha_j = \vert \alpha_j \vert e^{i \phi_j}$. 
The other input mode $b_j$ is fed with the state obtained by the multi-mode splitting of a single squeezed-vacuum state $\ket{\xi}$. 
Here, $\xi= r e^{i \varphi}$, $r$ is the squeeze parameter, $\bar{n}_s = \sinh^2r$ is the mean number of photons, and $\varphi$ is the phase of $\ket{\xi}$.
The multi-mode splitting corresponds to a QC described by a unitary $d\times d$ matrix $\vect{U}$.
Denoting as $(\hat{b}_{\textrm{in}})_j$ the annihilation operators associated to the $j$th input mode, see Fig.~\ref{fig1}(a), we have $\hat{b}_{j}=\sum_k \vect{U}^{\dagger}_{jk}(\hat{b}_{\textrm{in}})_k$. The squeezed vacuum is inserted in one input port of the network, that we indicate as port $D$ (ranging from 1 to $d$), while all the other input modes are in the vacuum state. 
The total average number of photons in the full Mach-Zehnder sensor network of Fig.~\ref{fig1}(a) is given by $\bar{n}_T= \sum_{j=1}^d \vert \alpha_j \vert^2 +\bar{n}_s$.

The inverse moment matrix $\vect{\mathcal{M}}^{-1}$ can be calculated analytically, see Appendix A.
Below, we report the explicit expression by assuming the optimal phase-matching conditions $\Im(e^{i\chi_j}u_j)= 0$ for $j=1, ..., d$. 
Here $\Re(x)$ and $\Im(x)$ indicate the real and imaginary part of $x$, respectively, and $u_j=\vect{U}_{Dj}$ ($\sum_{j=1}^d |u_j|^2=1$ being $\vect{U}$ unitary).
The condition $\Im(e^{i\chi_j}u_j)= 0$ can be fulfilled by adjusting the relative phase $\chi_j=\phi_j-\varphi/2$ between the coherent state in mode $a_j$ and the squeezed-vacuum.
In other words, the optimal sensing condition is obtained by matching the phases of each $\ket{\alpha}_j$ relative to that of $\ket{\xi}$.
At the optimal working point $\theta_j = \pi/2$, we have 
\be \label{MOM}
(\vect{\mathcal{M}}^{-1})_{ij}=\frac{\vert \alpha_i \vert  (e^{-2r}-1) \vert \alpha_j \vert \tilde{u}_i \tilde{u}_j
}{(\vert \alpha_i \vert^2-\tilde{u}_i^2\bar{n}_s)(\vert \alpha_j \vert^2-\tilde{u}_j^2\bar{n}_s)} 
+\frac{\vert \alpha_j \vert^2+\tilde{u}_j^2\bar{n}_s}{(\vert \alpha_j \vert^2-\tilde{u}_j^2\bar{n}_s )^2}\delta_{ij},
\ee
where $\delta_{ij}$ is the Dirac delta function and $\tilde{u}_j = \Re(e^{i\chi_j}u_j) =\pm|u_j|$.
We also calculate analytically the QFIM (see Appendix A): 
\be \label{QFIM}
(\vect{\FQ})_{ij} = 
\vert \alpha_i \vert  (e^{2r}-1) \vert \alpha_j \vert \tilde{u}_i \tilde{u}_j   + ( \vert \alpha_j \vert^2 + \tilde{u}_i^2 \bar{n}_s ) \delta_{ij},
\ee
which is independent from $\vect{\theta}$.
We finally notice that Eqs.~(\ref{MOM}) and~(\ref{QFIM}) do not depend on the QC transformation $\vect{U}$ as a whole, but only on the vector $\vect{\tilde{u}}=\{\tilde{u}_1,\dots,\tilde{u}_d\}$.
Given a specific QC transformation $\vect{U}$ and the actual input port $D$ in which $\ket{\xi}$ is inserted, we can identify $2^d$ different (non-orthogonal) vectors $\{\tilde{u}_i\}_{i=1,\dots,d}$, depending on the choice of sign for each $\tilde{u}_i=\pm|\vect{U}_{Di}|$.
We recall that $\vect{U}_{Di}$ is the element of the matrix $\vect{U}$ at 
row $D$ and column $i$.

In the separable strategy of Fig.~\ref{fig1}(b), the $j$th MZI is fed with a coherent state $\ket{\alpha_j'}$ in mode $a_j$ and a squeezed vacuum state $\ket{\xi'}$ in mode $b_j$, where $\alpha_j'=\vert \alpha_j' \vert e^{i \phi_j'}$ and $\xi_j' = r_j' e^{i \varphi_j'}$.
Under the optimal condition $\chi_j' = \phi_j' - \varphi_j'=0$ (requiring phase locking between $\ket{\alpha_j'}$ and $\ket{\xi'}$), we find~\cite{CavesPRD1981, ParisPLA1995, PezzePRL2008} 
\be \label{sMOM}
\frac{\Delta^2 \hat{X}_j}{(d \mean{\hat{X}_j}/d \theta_j)^2} = \frac{\vert \alpha_j' \vert^2 e^{-2r_j'} + (\bar{n}_s')_j}{[\vert \alpha_j' \vert^2 - (\bar{n}_s')_j]^2},
\ee
and the quantum Fisher information~\cite{PezzePRL2008, LangPRL2013} 
\be \label{sQFIM}
\mathcal{F}_j = \vert \alpha_j' \vert^2 e^{2r_j'} + (\bar{n}_s')_j,
\ee
where $(\bar{n}_s')_j = \sinh^2 r_j'$ is the mean number of photon in the state $\ket{\xi'}$.
Differently from the entangled strategy, the separable strategy uses $d$ squeezed-vacuum states. 
The total average number of particles in the separable sensor is thus $\bar{n}_T' = \sum_{j=1}^d \vert \alpha_j' \vert^2 + \bar{n}_s'$, where $\bar{n}_s' = \sinh^2 r_j'$ is the total average number of photons in the $d$ squeezed states.
When $d=1$, Eqs.~\makeref{MOM} and~\makeref{QFIM} agree with Eqs.~\makeref{sMOM} and~\makeref{sQFIM}, respectively, and recover a single MZI with coherent$\otimes$squeezed-vacuum input state \cite{CavesPRD1981,ParisPLA1995, BarnettEPJD2003, PezzePRL2008, LangPRL2013, Ruo-BercheraPRA2015, SparaciariPRA2016}.

\section{Optimal sensing configuration}
\label{OptSens}

In this section, we study the following problem: given a linear combination of parameters $\vect{v} \cdot \vect{\theta}$, we want to find the optimal configuration of the Mach-Zehnder sensor network of Fig.~\ref{fig1}(a) that minimizes the phase uncertainty $\Delta^2(\vect{v} \cdot \vect{\theta})_{\rm emom}$, for instance when using the method of moment as estimation strategy. 
In other words, for a given $\vect{v} \in \mathbb{R}^d$, we search for
\be \label{minimization}
\min_{\vect{U}, \, \vert \alpha_1 \vert^2, ..., \vert \alpha_d \vert^2, r} \Delta^2(\vect{v} \cdot \vect{\theta})_{\rm emom}. 
\ee
We recall that the minimization over the QC transformation $\vect{U}$ corresponds to a minimization over the vector $\vect{\tilde{u}}$ defined above. 
Here, we approach Eq.~(\ref{minimization}) by a direct calculation of the inverse moment matrix which, in some limits, assumes a convenient form suitable for analytical optimization.
Clearly, Eq.~(\ref{minimization}) can be also generalized to the quantum Cramer-Rao bound, namely 
\be \label{minimizationQCR}
\min_{\vect{U},\,\vert \alpha_1 \vert^2, ..., \vert \alpha_d \vert^2, r} \Delta^2(\vect{v} \cdot \vect{\theta})_{\rm eQCR}. 
\ee

In Sec. \ref{Sec.bounds}, we derive upper and lower bounds to Eqs.~(\ref{minimization}) and~(\ref{minimizationQCR}) that hold for every $\vect{v}$. 
We then compare, in Sec.~\ref{Sec.compare}, the optimized entangled and separable strategies, under different constraints.
In particular, numerical studies for $d=2$ and $d=3$ show that the optimal parallel strategy overcomes a corresponding optimal sequential strategy for every $\vect{v}$.

\subsection{Bounds and scalings}
\label{Sec.bounds}

For fixed $\bar{n}_T$ and $\bar{n}_s$, it is possible to find upper and lower bounds to Eq.~(\ref{minimization}), for every $\vect{v}$: 
\be \label{ubound3LB}
\Delta^2(\vect{v} \cdot \vect{\theta})_{\rm emom} \geq \frac{e^{-2r}}{d\bar{n}_T} + \frac{\bar{n}_s}{d\bar{n}_T^2}, 
\ee
that holds for $\bar{n}_T \gg \bar{n}_s$, and 
\be \label{ubound3UB}
\Delta^2(\vect{v} \cdot \vect{\theta})_{\rm emom} \leq 
\frac{e^{-2r}}{\bar{n}_T} + \frac{\bar{n}_s \mathcal{W}}{\bar{n}_T^2},
\ee
that holds for $\bar{n}_T \gg (d+1)\bar{n}_s$, where $\mathcal{W} = d^3 \sum_{j=1}^d v_j^4$.
The above inequalities are derived in Appendix B. 
As shown below, the upper bound is tight for $\vect{v} = \vect{v}_{\rm ave}=(\pm 1, \pm 1, ..., \pm 1)/d$ such that $\vect{v}_{\rm ave} \cdot \vect{\theta} = \sum_{j=1}^d \pm \theta_j/d$ (that, for brevity, we indicate as generalized average). 
The lower bound is tight in the trivial case when $v_j = 1/\sqrt{d}$ and $v_{i\neq j} =0$ such  that $\vect{v} \cdot \vect{\theta} = \theta_j/\sqrt{d}$ (the factor $\sqrt{d}$ is due to consistency with the normalization $\vert \vect{v} \vert^2 = 1/d$). 
In this case, the optimal scheme consists of a single MZI with a squeezed vacuum in one port and a coherent state of $\bar{n}_T - \bar{n}_s$ particles in the other port.
The bounds (\ref{ubound3LB}) and (\ref{ubound3UB}) are characterized by different regimes and scalings. 

\begin{itemize}

\item {\it Regime $\bar{n}_T \gg \bar{n}_s e^{2r} \mathcal{W}$.}
Noticing that $\mathcal{W} \geq 1$, this regime also implies  $\bar{n}_T \gg \bar{n}_s e^{2r}$.
In this case, the first term in both Eqs.~(\ref{ubound3LB}) and~(\ref{ubound3UB}) dominates over the second one, giving \cite{notaEq13}
\be \label{ubound4}
\frac{e^{-2r}}{d\bar{n}_T} \leq \min_{\vect{U}, \vert \alpha_1 \vert^2, ..., \vert \alpha_d \vert^2} \Delta^2(\vect{v} \cdot \vect{\theta})_{\rm emom} \leq
\frac{e^{-2r}}{\bar{n}_T}.
\ee
Both terms correspond to a sub-SN uncertainty with prefactor related to the squeeze parameter $r$.

\item {\it Optimal squeezing.} 
We minimize Eqs.~(\ref{ubound3LB}) and (\ref{ubound3UB}) with respect to $\bar{n}_s$, for a fixed $\bar{n}_T$. 
Considering $\bar{n}_s\gg 1$ (such that $e^{2r} \approx 4 \bar{n}_s$) and taking the derivative with respect to $\bar{n}_s$, one finds 
\be \label{ubound5}
\frac{1}{d\bar{n}_T^{3/2}} \leq \min_{U, \vert \alpha_1 \vert^2, ..., \vert \alpha_d \vert^2, r} \Delta^2(\vect{v} \cdot \vect{\theta})_{\rm emom} \leq
\frac{\sqrt{\mathcal{W}}}{\bar{n}_T^{3/2}}.
\ee
The optimal value of $\bar{n}_s$ minimizing the upper (lower) bound $\bar{n}_s\approx\sqrt{\bar{n}_T/4\mathcal{W}}$ ($\bar{n}_s\approx\sqrt{\bar{n}_T/4}$).
These values are consistent with the validity conditions of Eq.~(\ref{ubound3LB}), namely $\bar{n}_T \gg \bar{n}_s$.
Furthermore, the upper bound in Eq.~(\ref{ubound5}) holds under the additional condition $\bar{n}_s \gg (d+1)/(4 \mathcal{W})$.
Overall, both bounds in Eq.~(\ref{ubound5}) corresponds to a scaling of phase uncertainty faster than the SN. 

\item {\it Transient Heisenberg scaling for $(d+1)\bar{n}_s \ll \bar{n}_T \ll \bar{n}_s e^{2r}$.}
In this regime, which requires $e^{2r} \gg d+1$, both Eq.~(\ref{ubound3LB}) and Eq.~(\ref{ubound3UB}) show a transient Heisenberg scaling \cite{notaEq15}: 
\be \label{ubound6}
\frac{\bar{n}_s}{d\bar{n}_T^{2}} \leq \min_{U, \vert \alpha_1 \vert^2, ..., \vert \alpha_d \vert^2} \Delta^2(\vect{v} \cdot \vect{\theta})_{\rm emom} \leq
\frac{\bar{n}_s \mathcal{W}}{\bar{n}_T^{2}}.
\ee
Equation (\ref{ubound6})  
is understood as $1/\bar{n}_T^2$ scaling that holds as a function of $\bar{n}_T$ in a restricted regime, and for a fixed value of $\bar{n}_s$.

\end{itemize}

In analogy to Eqs.~(\ref{ubound3LB}) and~(\ref{ubound3UB}), we can find an upper and a lower bound to Eq. (\ref{minimizationQCR}):
\be \label{ubound7LB}
\min_{\vect{U}, \,  \vert \alpha_1 \vert^2, ..., \vert \alpha_d \vert^2} \Delta^2(\vect{v} \cdot \vect{\theta})_{\rm eQCR} \geq
\frac{1}{d[\bar{n}_T e^{2r} - \bar{n}_s (e^{2r}-1)]},
\ee
and 
\be \label{ubound7UB}
\min_{\vect{U}, \,  \vert \alpha_1 \vert^2, ..., \vert \alpha_d \vert^2} \Delta^2(\vect{v} \cdot \vect{\theta})_{\rm eQCR} \leq
\frac{e^{-2r}}{\bar{n}_T - \bar{n}_s}.
\ee
The bounds holds for every $\vect{v}$ and do not require additional conditions on $\bar{n}_T$ and $\bar{n}_s$.
The demonstration of the inequalities~(\ref{ubound7UB}) and ~(\ref{ubound7LB}) is detailed in Appendix C.
Also in this case, the upper bound is tight for the estimation of $\vect{v}_{\rm ave} \cdot \vect{\theta}$, while the lower bound is tight for the estimation of a single $\theta_j$.
We distinguish different regimes:

\begin{itemize}
\item For $\bar{n}_T \gg \bar{n}_s$, 
Eqs.~(\ref{ubound7LB}) and~(\ref{ubound7UB}) simplify and we have 
\be \label{ubound8}
\frac{e^{-2r}}{d \bar{n}_T} \leq \min_{\vect{U}, \, \vert \alpha_1 \vert^2, ..., \vert \alpha_d \vert^2} \Delta^2(\vect{v} \cdot \vect{\theta})_{\rm eQCR} \leq \frac{e^{-2r}}{ \bar{n}_T}.
\ee
Equation~(\ref{ubound8}) holds, in particular, also in the regime $\bar{n}_T \gg \bar{n}_s e^{2r} \mathcal{W}$, where the upper and lower bounds to Eq.~(\ref{ubound8}) coincide with that of Eq.~(\ref{ubound4}).

\item {\it Heisenberg limit for $\bar{n}_T \approx 2 \bar{n}_s$.} 
Taking $\bar{n}_s\gg 1$ (so that $\bar{n}_s\approx e^{2r}/4$), we can immediately optimize Eqs.~(\ref{ubound7LB}) and (\ref{ubound7UB}) with respect to $\bar{n}_s$, and for fixed $\bar{n}_T$:
\be \label{sensitivity_same_intensity}
\frac{1}{d \bar{n}_T^2} \leq \min_{\vect{U}, \,  \vert \alpha_1 \vert^2, ..., \vert \alpha_d \vert^2, r} \Delta^2(\vect{v} \cdot \vect{\theta})_{\rm eQCR} \leq \frac{1}{ \bar{n}_T^2}.
\ee
It should be noticed that the value of the squeeze parameter that minimizes both bounds (\ref{ubound7LB}) and (\ref{ubound7UB}) is $\bar{n}_s = \bar{n}_T/2$.
This is different from the value that minimizes the bounds in Eq.~(\ref{ubound5}).
The different optimizations correspond to different scalings, $O(\bar{n}_T^{-3/2})$ and $O(\bar{n}_T^{-2})$, respectively.

\end{itemize}

\begin{figure}[t!]
    \centering
  \includegraphics[width=1\columnwidth]{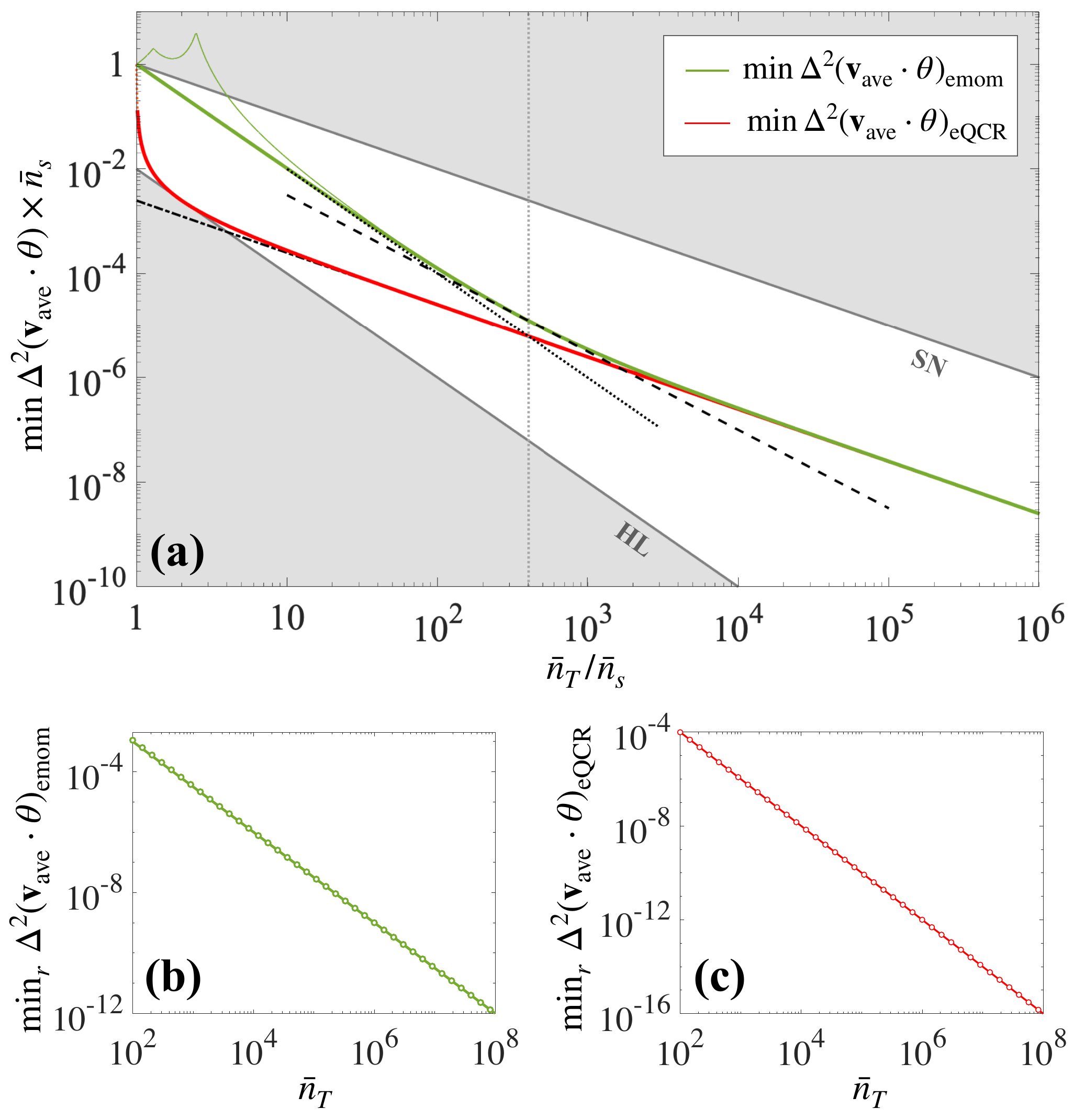}
\caption{(a) Optimized sensitivity for the estimation of the generalized average $ \vect{v}_{\rm ave} \cdot \vect{\theta} = (\pm \theta_1 \pm \theta_2 ... \pm \theta_d)/d$, as a function of the total average number of particles $\bar{n}_T$, for fixed $\bar{n}_s$.
The thin solid green line is $\min \Delta^2 (\vect{\theta}\cdot \vect{v}_{\rm ave})_{\rm emom}$,
while the thin solid red line $\min \Delta^2 (\vect{\theta}\cdot \vect{v}_{\rm ave})_{\rm eQCR}$.
For simplicity here $\min$ indicates the minimum over $\vect{U}$, and $\vert \alpha_1 \vert^2, ..., \vert \alpha_d \vert^2$:
the minimization if performed numerically for $d=2$. 
These numerics are compared with Eq.~(\ref{Dthetaave}) (thick green line) and Eq.~(\ref{DthetaaveQCR}) (thick red line).
The thin red line is barely visible due to the perfect superposition with the thick red line.
The dotted vertical line highlights the point $\bar{n}_T = \bar{n}_s e^{2r}$.
Black lines correspond to different analytical limit: the dot-dashed line is $e^{-2r}/\bar{n}_T$, the dashed line is $1/\bar{n}_T^{3/2}$ and the dotted line is $\bar{n}_s/\bar{n}_T^{2}$. 
We also indicate the SN ($1/\bar{n}_T$) and the HL ($1/\bar{n}_T^2$). 
Here $\bar{n}_s = 100$.
Panels~(b) and~(c) show $\min_r \Delta (\vect{\theta}\cdot \vect{v}_{\rm ave})_{\rm emom}$ and $\min_r \Delta (\vect{\theta}\cdot \vect{v}_{\rm ave})_{\rm eQCR}$ as a function of $\bar{n}_T$.
We use $\min_r$ as shorthand notation to indicate the minimization over $\vect{U}$, $\vert \alpha_1 \vert^2, ..., \vert \alpha_d \vert^2$ and over the squeeze parameter $r$.
Dots are numerical results, while lines are expected analytical behaviours: $\bar{n}_T^{-3/2}$ [in panel (b)] and $\bar{n}_T^{-2}$ [in panel (c)].}
\label{fig2}
\end{figure}

To summarize, through a set of bounds, we have identified different scalings and behaviours that characterize Eqs.~(\ref{minimization}) and~(\ref{minimizationQCR}) in different regimes of parameters. 
Below, we demonstrate the saturation of the upper bounds~(\ref{ubound3UB}) and~(\ref{ubound7UB}). 

For symmetry reasons, the best estimation of the generalized average phase $\vect{v}_{\rm ave} \cdot \vect{\theta} = ( \pm \theta_1   \pm \theta_2 \pm ... \pm \theta_d)/d$ (corresponding to $\vert v_{{\rm ave},j} \vert  = 1/d$) is obtained when all the coherent states have the same intensity, namely  $|\alpha_j|^2=\bar{n}_c, \forall j$. 
Moreover, in the regime $\bar{n}_T\gg (d+1)\bar{n}_s$ it is possible to prove that the QC satisfying the condition $\vect{\tilde{u}}=\sqrt{d}\vect{v}_{\rm ave}$ is optimal~(see Appendix D). 
We have
\be\label{Dthetaave} 
\min_{\vect{U}, \, \vert \alpha_1 \vert^2, ..., \vert \alpha_d \vert^2} \Delta^2(\vect{v}_{\rm ave} \cdot \vect{\theta})_{\rm emom}=
\frac{e^{-2r}}{\bar{n}_T} + \frac{\bar{n}_s}{\bar{n}_T^2},
\ee
that coincides with the upper bound~(\ref{ubound3UB}), when noticing that  $\mathcal{W} = d^3 \sum_{j=1}^2 v_{{\rm ave},j}^4=1$.
Equation~(\ref{Dthetaave}) holds under the same condition of Eq.~(\ref{ubound3UB}), namely $\bar{n}_T\gg (d+1)\bar{n}_s$.
Furthermore, for the uniform QC considered here,
the optimization of the quantum Cramer-Rao bound leads to 
\be\label{DthetaaveQCR} 
\min_{\vect{U}, \, \vert \alpha_1 \vert^2, ..., \vert \alpha_d \vert^2} \Delta^2(\vect{v}_{\rm ave} \cdot \vect{\theta})_{\rm eQCR}=\frac{1}{\bar{n}_Te^{2r}-\bar{n}_s\left(e^{2r}-1\right)},
\ee
that saturates Eq.~(\ref{ubound7UB}) for $r \gg 1$.
Equation~(\ref{DthetaaveQCR}) is proved in Appendix~D for $\bar{n}_T \gg (d+1) \bar{n}_s$, showing that the condition $\vect{\tilde{u}}=\sqrt{d}\vect{v}_{\rm ave}$ is optimal in this regime. 
In Fig.~\ref{fig2}(a) we plot numerical results for $\min_{\vect{U}, \, \vert \alpha_1 \vert^2, ..., \vert \alpha_d \vert^2} \Delta^2(\vect{v}_{\rm ave} \cdot \vect{\theta})_{\rm emom}$ (thin green line) and $\min_{\vect{U}, \, \vert \alpha_1 \vert^2, ..., \vert \alpha_d \vert^2} \Delta^2(\vect{v}_{\rm ave} \cdot \vect{\theta})_{\rm eQCR}$ (thin red line) as a function of $\bar{n}_T$. 
These are compared with Eq.~(\ref{Dthetaave}) and~(\ref{DthetaaveQCR}), shown as thick green and red lines, respectively. 
The perfect superposition between the red lines shows that Eq. (\ref{DthetaaveQCR}) holds in all regimes. 
Equation (\ref{Dthetaave}) instead holds for sufficiently large values of $\bar{n}_T$, as expected.  
In  Fig.~\ref{fig2} we also plot the analytical behaviours   Eqs.~(\ref{ubound4})-(\ref{ubound6}) in the corresponding different regimes.
In particular, Eqs. (\ref{Dthetaave}) and (\ref{DthetaaveQCR}) coincide for $\bar{n}_T\gg\bar{n}_se^{2r}$, indicating that the method of moments is an optimal estimation strategy in that regime.

In Fig.~\ref{fig2}(b) and~(c) we plot, respectively, $\min_{\vect{U}, \, \vert \alpha_1 \vert^2, ..., \vert \alpha_d \vert^2, r} \Delta^2(\vect{v}_{\rm ave} \cdot \vect{\theta})_{\rm emom}$ and $\min_{\vect{U}, \, \vert \alpha_1 \vert^2, ..., \vert \alpha_d \vert^2, r} \Delta^2(\vect{v}_{\rm ave} \cdot \vect{\theta})_{\rm eQCR}$, as a function of $\bar{n}_T$. 
Dots are numerical results.
The solid line in panel (b) [panel (c)] is obtained by minimizing Eq. (\ref{Dthetaave}) [Eq. (\ref{DthetaaveQCR})] with respect to $r$, predicting a sensitivity $1/\bar{n}_T^{3/2}$ [$1/\bar{n}_T^{2}$]. 

\begin{figure*}[ht!]
    \centering
  \includegraphics[width=1\textwidth]{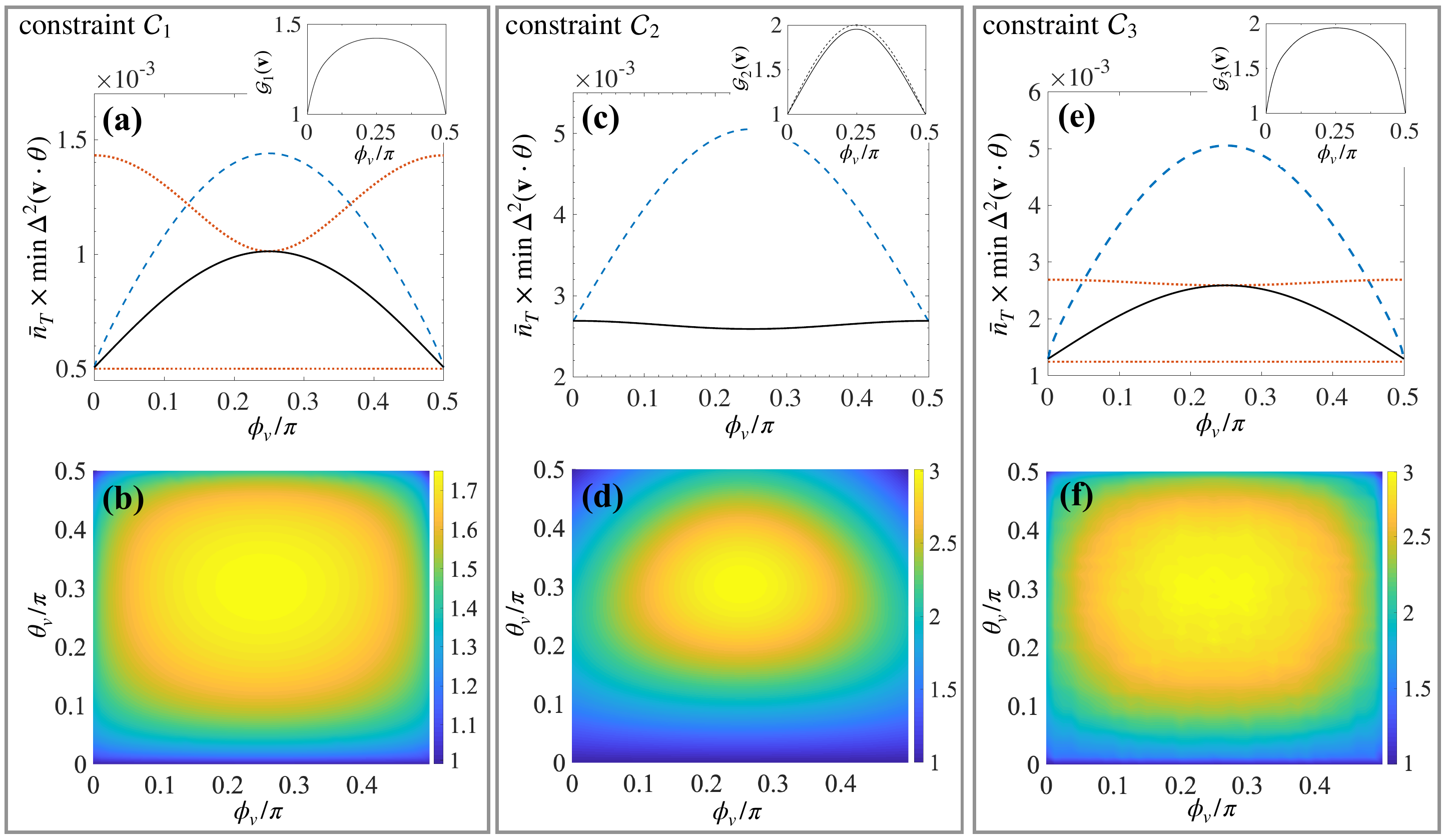} 
\caption{Results of a numerical optimization of the sensor network of Fig.~\ref{fig1}(a) for $d=2$ (upper row panels) and $d=3$ (lower row), under different constraints (as detailed in the text).
Panels (a) and (b) corresponds to the optimization under the constraint $\mathcal{C}_1$, see Sec.~\ref{SubSec.C1}. 
Panel (a) plots $\min_{\vect{U}, \, \vert \alpha_1 \vert^2, ..., \vert \alpha_d \vert^2, \, r} \Delta^2(\vect{v} \cdot \vect{\theta})_{\rm emom}$ (solid black line) and $\min_{(\vert \alpha_1 \vert^2, r_1), ..., (\vert \alpha_d \vert^2, r_d) } \Delta^2(\vect{v} \cdot \vect{\theta})_{\rm smom}$ (dashed blue line).
The dot-dashed orange lines are the upper and lower bounds of Eq.~(\ref{ubound5}).
The gain Eq.~(\ref{gain1}) is shown in the inset of panel (a) for $d=2$ and in panel (b) for $d=3$.
Panels (c) and (d) corresponds to the optimization under the constraint $\mathcal{C}_2$, see Sec.~\ref{SubSec.C2}.
Panel (c) plots $\min_{\vect{U}} \Delta^2(\vect{v} \cdot \vect{\theta})_{\rm emom}$ (solid black line) and $\min_{r_1', ..., r_d'} \Delta^2(\vect{v} \cdot \vect{\theta})_{\rm smom}$ (dashed blue line).
The corresponding gain, Eq.~(\ref{gain2}), is shown as solid line in the inset, together with the analytical prediction Eq.~(\ref{gain2an}) (dashed line).
Panel (d) shows $\mathcal{G}_2(\vect{v})$ for $d=3$.
In panels (c) and (d) we have set $\vert \alpha_j \vert^2 = \vert \alpha_j' \vert^2 = (\bar{n}_T - \bar{n}_s)/d$ for all $j$.
Panels (e) and (f) corresponds to the optimization under the constraint $\mathcal{C}_3$, see Sec.~\ref{SubSec.C3}.
In particular, we consider $\bar{n}_s/\bar{n}_T= 10^{-4}$ for the entangled strategy and $(\bar{n}_s')_j/(\bar{n}_T')_j= 10^{-4}$ ($j=1, ..., d$) for the separable strategy.
Panel (e) plots $\min_{\vect{U}, \, \vert \alpha_1 \vert^2, ..., \vert \alpha_d \vert^2} \Delta^2(\vect{v} \cdot \vect{\theta})_{\rm emom}$ (solid black line) and $\min_{\vert \alpha_1' \vert^2, ..., \vert \alpha_d' \vert^2} \Delta^2(\vect{v} \cdot \vect{\theta})_{\rm smom}$ (dashed blue line). 
The gain Eq.~(\ref{gain3}) is shown in the inset of panel (c) for $d=2$ and in panel (d) for $d=3$.
The dot-dashed orange lines are the lower and upper bounds Eqs.~(\ref{ubound3LB}) and (\ref{ubound3UB}), respectively.
In all panels $\bar{n}_T=\bar{n}_T' = 10^6$.
In panels (c)-(f) $\bar{n}_s=\bar{n}_s'=10^2$.}
\label{fig3}
\end{figure*}

\subsection{Comparison between optimal entangled and separable strategies}
\label{Sec.compare}

In the following, we compare optimal entangled and separable strategies for an arbitrary linear combination of parameters $\vect{v} \cdot \vect{\theta}$. 
Let us define the gain factor 
\be \label{gain}
\mathcal{G}_{\mathcal{C}} =
\frac{\min_{\{\vert \alpha_1' \vert^2, r_1', ..., \vert \alpha_d' \vert^2, r_d' \} \in \mathcal{C}} \Delta^2(\vect{v} \cdot \vect{\theta})_{\rm smom}}{\min_{\{ \vect{U}, \, \vert \alpha_1 \vert^2, ..., \vert \alpha_d \vert^2, r \} \in \mathcal{C}} \Delta^2(\vect{v} \cdot \vect{\theta})_{\rm emom}},
\ee
where $\mathcal{C}$ indicates a common constraint on resources for both cases. 
Different constraints are discussed below.
We recall that the uncertainties $\Delta^2(\vect{v} \cdot \vect{\theta})_{\rm smom}$ and $\Delta^2(\vect{v} \cdot \vect{\theta})_{\rm emom}$ in Eq. (\ref{gain}) are already optimized with respect to the relative phase between the squeezed-vacuum state(s) and the $d$ coherent states. 

The minimizations in Eq. (\ref{gain}) are performed numerically for $d=2$ and $d=3$ by using a variational approach, see Fig. \ref{fig3}.  
Analytical predictions for any $d$ can be derived in interesting cases.   
For $d=2$, the QC consists of a generalized beam-splitter that implements the mode transformation 
$\vect{\hat{b}}={U}^{\dagger}\vect{\hat{b}_{\rm in}}$, where
$U=\begin{pmatrix}\tau&\varrho\\-\varrho^* & \tau^* \end{pmatrix}$.
Without loss of generality, we consider $\tau$ and $\varrho$ being real numbers with  $\tau^2+\varrho^2=1$.
The sensor network is given by two MZIs and 
$\vect{v}^T = ( \cos \phi_v , \sin \phi_v )/\sqrt{2}$ is expressed as a function of $\phi_{v} \in [0, 2\pi]$.
For $d=3$, the sensing scheme consists of three MZIs and the goal is to estimate a linear combinations of three relative phases, $\theta_1$, $\theta_2$ and $\theta_3$. 
The QC is implemented as a general $3\times 3$ orthogonal transformation.
Any such transformation can be realized, in general, by a sequence of three two-mode beam splitters \cite{ReckPRL1994}.
For $d=3$, the vector $\vect{v}$ is parametrized as $\vect{v}^T = ( \sin \theta_v \cos \phi_v , \sin \theta_v \sin \phi_v, \cos \phi_v )/\sqrt{3}$, as a function of $\phi_{v} \in [0, 2\pi]$ and $\theta_v \in [0, \pi]$.

Upon imposing different constraints, we observe
$\mathcal{G}_{\mathcal{C}}(\vect{v}) \geq  1$ for every $\vect{v}$, see Fig. \ref{fig3}.
In particular, we obtain $\mathcal{G}_{\mathcal{C}}(\vect{v})=1$ when the problem reduces to the estimation of a single phase [e.g. for $\phi_v = 0, \pi/2$ and $d=2$, corresponding to $\vect{v} = (1/\sqrt{d},0)$ and $(0,1/\sqrt{d})$, respectively]. 
The maximum gain $\mathcal{G}_{\mathcal{C}}(\vect{v})$ is obtained for the estimation of the generalized
average phase $\vect{v}_{\rm ave} \cdot \vect{\theta} = \sum_{j=1}^d \pm \theta_d/d$, that is $\phi_v = \pm \pi/4$ for $d=2$ and $\phi_v= \pm \pi/4$, $\theta_v=\arccos(1/\sqrt{3})$, for $d=3$ (notice that the figure shows the case $v_j \geq 0$, while $\mathcal{G}_{\mathcal{C}}(\vect{v})$ is symmetric under $v_j\to-v_j$).

\subsubsection{$\mathcal{C}_1$: same total average number of particles}
\label{SubSec.C1}

We impose here the only constraint of having the same total average number of particles for both strategies: $\bar{n}_T = \bar{n}_T'$.
The entangled strategy is thus optimized 
over the QC transformation $\vect{U}$, the average number of particles of each coherent state, $\vert \alpha_j \vert^2$, and the squeeze parameter $r$.
The sequential strategy is instead optimized over the input states of each independent MZI, namely over each $\vert \alpha_j \vert^2$ and squeeze parameter $r_j$, for $j=1, ..., d$.
The gain factor is thus
\be \label{gain1}
\mathcal{G}_1(\vect{v}) = \frac{\min_{\vert \alpha_1' \vert^2, r_1', ..., \vert \alpha_d' \vert^2, r_d' } \Delta^2(\vect{v} \cdot \vect{\theta})_{\rm smom}}{\min_{\vect{U}, \vert \alpha_1 \vert^2, ..., \vert \alpha_d \vert^2, r} \Delta^2(\vect{v} \cdot \vect{\theta})_{\rm emom}}.
\ee
Numerical results for $d=2$ and $d=3$ are shown in Fig.~\ref{fig3}(a) and (b), respectively.

In the case $\vect{v}=\vect{v}_{\rm ave}$, the optimized entangled strategy is $\min \Delta^2(\vect{v_{\rm ave}} \cdot \vect{\theta})_{\rm emom} =\bar{n}_T^{-3/2}$.
The optimized separable strategy is instead $\min \Delta^2(\vect{v_{\rm ave}} \cdot \vect{\theta})_{\rm smom} =
1/d\times(\bar{n}_T/d)^{-3/2}=\sqrt{d}\times(\bar{n}_T)^{-3/2}$, where $\bar{n}_T/d$ is the total average number of photons injected in each MZI. 
The highest gain in this case equals 
\be
\mathcal{G}_1(\vect{v_{\rm ave}})=\sqrt{d}.
\ee
A maximum gain $\sqrt{2}$ and $\sqrt{3}$ can be seen in the inset of panels (a) and (b) of Fig.~\ref{fig3}, respectively.

\subsubsection{$\mathcal{C}_2$: same total average number of particles and same coherent state intensities}
\label{SubSec.C2}

Here, we consider the same total average number of particles for both strategies, $\bar{n}_T = \bar{n}_T'$, and the same coherent state intensities, $\vert \alpha_j \vert^2 = \vert \alpha_j' \vert^2$ for all $j$.
These constraints fix the total average number of photons in the squeezed state(s), namely $\bar{n}_s = \bar{n}_T - \sum_{j=1}^d \vert \alpha_j \vert^2= \bar{n}_s'$. 
The entangled strategy uses a single squeezed-vacuum state and we optimize the multimode splitting in the QC.
The separable strategy is instead optimized with respect to the $d$ squeeze parameters $r_j'$ ($j=1,\dots,d$) with $\sum_{j=1}^d \sinh^2 r_j' = \bar{n}_s' = \bar{n}_s$.
The gain factor writes 
\be \label{gain2}
\mathcal{G}_2(\vect{v}) = \frac{\min_{r_1', ..., r_d'} \Delta^2(\vect{v} \cdot \vect{\theta})_{\rm smom}}{\min_{\vect{U}} \Delta^2(\vect{v} \cdot \vect{\theta})_{\rm emom}}.
\ee   
In the following, for simplicity, we can take $|\alpha_j|^2=\bar{n}_c$ for all $j$.
In Fig.~\ref{fig5}(c) and~(d), we show the results of a numerical optimizations for $d=2$ and $d=3$, respectively. 
They agree well with the analytical prediction 
\be \label{gain2an}
\mathcal{G}_2(\vect{v})=d \,\Bigg( \sum_{i=1}^d|v_i| \Bigg)^2,
\ee
that can be derived under the conditions $r,r_j' \gg 1$ and in the regime $\bar{n}_c \gg \bar{n}_s e^{2r}$, see Appendix~E.
We observe $\mathcal{G}_2(\vect{v}) \geq 1$ for every $\vect{v}$, with maximum gain equal for $\vect{v} = \vect{v}_{\rm ave}$.

\begin{figure}[t!]
    \centering
  \includegraphics[width=1\columnwidth]{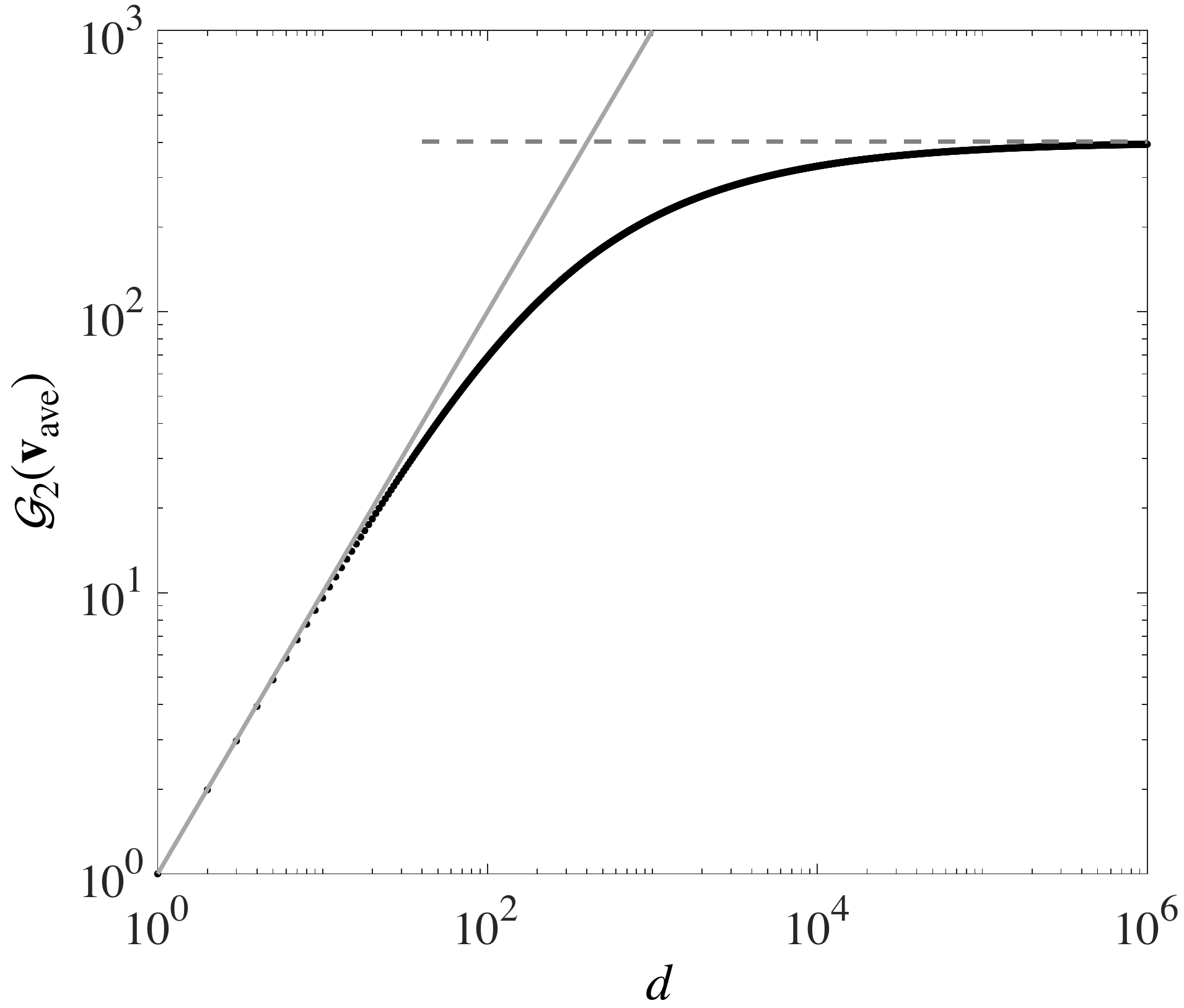} 
\caption{Gain factor $\mathcal{G}_2(\vect{v}_{\rm ave})$, Eq.~(\ref{gaindinf}), as a function of $d$, for $\bar{n}_c = 10^8$ and $\bar{n}_s=10^2$ (dots).
The solid line is $\mathcal{G}_2(\vect{v}_{\rm ave})=d$, which is expected for $d \ll \bar{n}_s$; the dashed line is $\mathcal{G}_2(\vect{v}_{\rm ave})=e^{2r}$, for $d \gg \bar{n}_s$.}
\label{fig4}
\end{figure}

Let us now focus the discussion on the maximum gain point $\vect{v} = \vect{v}_{\rm ave}$ and
$\bar{n}_c \gg \bar{n}_s$.
In this case, the optimal entangled strategy is obtained for $\tilde{u}_j = 1/\sqrt{d}$ and we have 
\be \label{optent2}
\min_{\vect{U}}\Delta^2(\vect{v}_{\rm ave} \cdot \vect{\theta})_{\rm emom} 
 =  \frac{e^{-2r}\bar{n}_c +\bar{n}_s/d}{d\bar{n}_c^2},
\ee
see Eq.~(\ref{Dthetaave}) and demonstration in Appendix D.
The optimal separable strategy is obtained using $d$ squeezed-vacuum states having the same squeeze parameter $r_j' =r' = {\rm arcsinh} \sqrt{\bar{n}_s/d}$ [since $\bar{n}_s$ is fixed, the intensity of the squeezed-vacuum state in each MZI is $(\bar{n}_s')_j = \bar{n}_s/d$]. 
As shown in Appendix~F, for $\bar{n}_c \gg \bar{n}_s$, we obtain 
\be \label{optsep2}
\min_{r'_1, ..., r'_d} \Delta^2 (\vect{v}_{\rm ave} \cdot \vect{\theta})_{\rm smom} = \frac{\bar{n}_c e^{-2 r'} + \bar{n}_s/d}{d\bar{n}_c^2}.
\ee
Finally, the gain factor 
\be \label{gaindinf}
\mathcal{G}_2(\vect{v}_{\rm ave}) = 
\frac{\bar{n}_c e^{-2 r'} + \bar{n}_s/d}{\bar{n}_c e^{-2 r} + \bar{n}_s/d}
\ee
is obtained by taking the ratio between Eqs.~(\ref{optent2}) and~(\ref{optsep2}).
It is interesting to consider different limits of Eq.~(\ref{gaindinf}), see also Fig.~\ref{fig4}.

Let us consider $d$ fixed, $\bar{n}_s/d\gg 1$, such that $r,r' \gg 1$, and $d \bar{n}_c \gg \bar{n}_s e^{2r}$ (which also implies $d \bar{n}_c \gg \bar{n}_s e^{2r'}$).
Notice that these conditions are fulfilled in Fig.~\ref{fig3}(c) and (d).
In this case, Eq.~(\ref{gaindinf}) simplifies to 
\be \label{gaindinf2}
\mathcal{G}_2(\vect{v}_{\rm ave}) = 
\frac{e^{-2 r'}}{e^{-2 r}} = d.
\ee
The gain factor $d$ is obtained taking into account that, for $r,r' \gg 1$, we have $e^{-2 r'}/4 \approx \bar{n}_s'/d$, $e^{-2 r}/4 \approx \bar{n}_s$.
A maximum gain close to $2$ and $3$ can be seen in panels (c) and (d) of Fig.~\ref{fig3}, respectively. 

Let us consider the opposite limit, $\bar{n}_s/d \ll 1$. 
The discussion includes the limit 
$d\to \infty$ that is peculiar of the multiparameter problem. 
First, we notice that, for any fixed value of $\bar{n}_s$, in the limit $d \to \infty$, we have $r' = {\rm arcsinh} \sqrt{\bar{n}_s/d} \to 0$.
The separable sensing scheme thus reduces to $d$ MZIs fed with coherent state in one port and vacuum in the other port.
As shown in Eq.~(\ref{optsep2}), we  recover the shot noise limit, $\min_{r'_1, ..., r'_d} \Delta^2 (\vect{v} \cdot \vect{\theta})_{\rm smom} \to 1/(d \bar{n}_c) = 1/\bar{n}_T$.
It is easy to show that the same sensitivity is also achieved for the optimized quantum Cram\'er-Rao bound $\min_{r'_1, ..., r'_d} \Delta^2 (\vect{v} \cdot \vect{\theta})_{\rm sQCR}$. 
The situation is completely different for the entangled scheme. 
In this case, $r$ remains finite in the limit $d \to \infty$ and, according to Eq.~(\ref{optent2}), we have $\min_{\vect{U}}\Delta^2(\vect{v}_{\rm ave} \cdot \vect{\theta})_{\rm emom} 
 = e^{-2r}/(d \bar{n}_c) =  e^{-2r}/\bar{n}_T$. 
 The entangled strategy still achieves a sub-SN sensitivity, with a gain factor
\be \label{gaindinf3}
\mathcal{G}_2(\vect{v}_{\rm ave}) \to e^{2r}, \qquad {\rm for} \,\, d\to \infty.
\ee
Surprisingly, a finite gain is obtained when a single squeezed-vacuum state is mixed, by the QC, with a diverging number ($d-1$) of vacuum states $\ket{0}$. 
The physical reason for the finite gain Eq.~(\ref{gaindinf3}) is due to quantum correlations in the covariance matrix $\vect{\Gamma}$.  
Under the above conditions $\vect{\Gamma}$ reads
(see Appendix~A for the general expression)
\be \label{Gamma4}
4\vect{\Gamma} = \frac{\bar{n}_c(e^{2r}-1)}{d} 
\begin{pmatrix}
1 & \hdots & 1 \\
\vdots & \ddots & \vdots \\
1 & \hdots & 1
\end{pmatrix}
+ \Big( \bar{n}_c + \frac{\bar{n}_s}{d}\Big)
\begin{pmatrix}
1 & \hdots & 0 \\
\vdots & \ddots & \vdots \\
0 & \hdots & 1
\end{pmatrix}.
\ee
In the limit $d \to \infty$, the prefactor of the all-ones matrix tends to zero while the prefactor of the identity matrix remains finite. 
However, the all-ones matrix is characterized by perfect correlations between all its $d^2$ elements, while the identity matrix does not feature any correlation.
When calculating $ \tfrac{(1,1,...,1)^T}{\sqrt{d}} 4\vect{\Gamma} \tfrac{(1,1,...,1)}{\sqrt{d}} = \bar{n}_c e^{2r} + \bar{n}_s/d \approx \bar{n}_c e^{2r}$ [notice that the vector $(1,1,...,1)/\sqrt{d}$ is normalized to one] the contributions coming from the two terms in Eq. (\ref{Gamma4}) have the same  magnitude with respect to $d$.
The finite mean value of the correlation matrix is responsible for the finite phase sensitivity and gain. 

\subsubsection{$\mathcal{C}_3$: same total average number of particles and same total squeezed-vacuum intensities}
\label{SubSec.C3}

We consider here $\bar{n}_T= \bar{n}_T'$ and the same total average number of particles in the squeezed state(s), namely $\bar{n}_s = \bar{n}_s' = \sum_{j=1}^d (\bar{n}_s')_j$. 
For the entangled strategy, we fix the ratio $\bar{n}_s/\bar{n}_T$ [equal to $10^{-4}$ in Fig.~\ref{fig3}(e) and (f)] and thus optimize $\Delta^2(\vect{v} \cdot \vect{\theta})_{\rm emom}$ over the QC tranformation $\vect{U}$ and the $d$ coherent state intensities $\vert \alpha_1 \vert^2$, ..., $\vert \alpha_d \vert^2$.
For the separable strategy, we fix the ratio $(\bar{n}_s')_j/(\bar{n}_T')_j$ [also equal to $10^{-4}$ for each $j$, in the figure], where $(\bar{n}_T')_j$ and $(\bar{n}_s')_j$ indicate the average number of particles in total and in each squeezed-vacuum state, respectively, as input of the $j$th MZI.  
We then optimize the separable strategy over each $\vert \alpha_j' \vert^2$.
The gain factor is
\be \label{gain3}
\mathcal{G}_3(\vect{v}) = \frac{\min_{\vert \alpha_1' \vert^2, ..., \vert \alpha_d' \vert^2} \Delta^2(\vect{v} \cdot \vect{\theta})_{\rm smom}}{\min_{\vect{U}, \,\vert \alpha_1 \vert^2, ..., \vert \alpha_d \vert^2} \Delta^2(\vect{v} \cdot \vect{\theta})_{\rm emom}}.
\ee
Results $d=2$ and $d=3$ are shown in Fig.~\ref{fig5}(e) and (f), respectively.
It is particularly interesting that the entangled strategy outperforms the separable one in this case, $\mathcal{G}_3(\vect{v}) \geq 1$ for all $\vect{v}$. 
Indeed, it clearly shows that the main responsible for the gain in sensitivity is the use of an optimal entangled multi-mode state in place of a product of squeezed-vacuum states. 

When considering $\vect{v} = \vect{v}_{\rm ave}$, optimal strategies are obtained taking $\vert \alpha_j'\vert^2 = \vert \alpha_j\vert^2 = \bar{n}_c$ for all $j$ and   $\tilde{\vect{u}} = 1/\sqrt{d}$.
In this case, $\mathcal{G}_3(\vect{v}_{\rm ave})$ equals Eq.~(\ref{gaindinf}), for $\bar{n}_c \gg \bar{n}_s$, and the same considerations as above can be obtained. 
In particular, $\mathcal{G}_3(\vect{v}_{\rm ave}) = d$ for $\bar{n}_s \gg d$, while $\mathcal{G}_3(\vect{v}_{\rm ave}) = e^{2r}$ for $\bar{n}_s \ll d$ (in particular, in the limit $d\to \infty$).

\subsubsection{$\mathcal{C}_4$: same total average number of particles and same squeezed-vacuum strength}
 
Here, we compare the entangled strategy using a single squeezed-vacuum state $\ket{\xi}$ with the separable strategy using $d$ copies of the same squeezed-vacuum states $\ket{\xi}$.
In other words, here, $r_j' = r$ for all $j$.
For simplicity, we set $\vert \alpha_j' \vert^2 = \bar{n}_c'$ and $\vert \alpha_j \vert^2 = \bar{n}_c$ for all $j$.
We also consider the same total average number of particles, $\bar{n}_T = \bar{n}_T'$, where $\bar{n}_T = d \bar{n}_c + \bar{n}_s$ and $\bar{n}_T' = d(\bar{n}_c' + \bar{n}_s)$, and recall that $\bar{n}_s = \sinh^2 r$.
In this case, using Eqs.~(\ref{smom}) and~(\ref{sMOM}), the sensitivity achieved with the separable strategy is
\be \label{Dthetasep4}
\Delta^2(\vect{v} \cdot \vect{\theta})_{\rm smom} = \frac{\bar{n}_c' e^{-2r} + \bar{n}_s}{d(\bar{n}_c' - \bar{n}_s)^2}
\ee
for all $\vect{v}$.
One of the characteristic features of Eq. (\ref{Dthetasep4}) is the divergence at $\bar{n}_c' = \bar{n}_s$ (namely $\bar{n}_T = 2 d \bar{n}_s$) and the saturation of the Cramer-Rao bound, $\Delta^2(\vect{v} \cdot \vect{\theta})_{\rm sQCR} = 1/[d(\bar{n}_c' e^{2r} + \bar{n}_s)]$, for $\bar{n}_T \gg e^{2r} \bar{n}_s$ \cite{PezzePRL2008}.
Instead, the entangled strategy, should be optimized over the QC transformation, thus giving 
\be \label{gain4}
\mathcal{G}_4(\vect{v}) = 
\frac{ \Delta^2(\vect{v} \cdot \vect{\theta})_{\rm smom}}{\min_{\vect{U}} \Delta^2(\vect{v} \cdot \vect{\theta})_{\rm emom}}.
\ee

Let us focus on the case $\vect{v} = \vect{v}_{\rm ave}$ and take the limit $\bar{n}_c, \bar{n}_c' \gg \bar{n}_s$ [that, in particular imply $\bar{n}_T \gg (d+1) \bar{n}_s$, $\bar{n}_T \approx d \bar{n}_c$ and $\bar{n}_T' \approx d \bar{n}_c'$].
In this regime, $\min_{\vect{U}} \Delta^2(\vect{v} \cdot \vect{\theta})_{\rm emom}$ is given by Eq.~(\ref{Dthetaave}) which, taking into account Eq.~(\ref{Dthetasep4}), provides
\be \label{gain4_1}
\mathcal{G}_4(\vect{v_{\rm ave}}) = \frac{\bar{n}_T e^{-2r} + d\bar{n}_s}{\bar{n}_T e^{-2r} + \bar{n}_s}.
\ee
If $\bar{n}_T \gg d e^{2r} \bar{n}_s$, Eq. (\ref{gain4_1}) simplifies to 
\be
\mathcal{G}_4(\vect{v_{\rm ave}})=1.
\ee
Although there is no gain in this case, it is still interesting that the parallel strategy using a {\it single} squeezed-vacuum state achieves the same performance as the sequential strategy using $d$ squeezed states with the same squeeze parameter.

The opposite regime, $\bar{n}_T \ll \bar{n}_s e^{2r}$, should be considered with care.
In particular, for $(d+1) \bar{n}_s \ll \bar{n}_T \ll \bar{n}_s e^{2r}$, the second term in both the numerator and denominator of Eq.~(\ref{gain4_1}) dominates, giving
$\mathcal{G}_4(\vect{v_{\rm ave}}) = d$.
Achieving $\mathcal{G}_4(\vect{v_{\rm ave}}) >1$ in this regime is due to the divergence of Eq.~(\ref{Dthetasep4}), as further discussed in Appendix~G.
In particular, a calculation of the gain factor following a numerical optimization of 
$\min_{\vect{U}} \Delta^2(\vect{v} \cdot \vect{\theta})_{\rm emom}$ shows that $\mathcal{G}_4(\vect{v_{\rm ave}})$ diverges when $\bar{n}_T= 2 d \bar{n}_s$.
This is an artifact due to the use of Eq.~(\ref{Dthetasep4}) outside the regime where it saturates the Cramer-Rao bound. 

\section{Optimal linear combination of phases}
\label{Optv}

In the previous section, we have discussed the optimal configuration of the sensor network of Fig. \ref{fig1}(a) that maximizes the sensitivity for the estimation of a fixed a linear combination of phases $\vect{v}\cdot \vect{\theta}$. 
Here we consider the opposite problem. 
Given a specific configuration of the sensor network, namely a specific QC transformation $U$, coherent state intensities and squeezed parameter, we discuss the optimal linear combination of phases $\vect{v}\cdot \vect{\theta}$ that can estimated with the smallest possible uncertainty. 

In the following, we fist provide a general framework for the minimization problem considered here, namely we introduce the notion of Fisher and squeezing spectra. 
We then apply this formalism to the Mach-Zehnder sensor network of Fig. \ref{fig1}(a) for random choices of the QC and discuss different sensitivity limits and regimes.

\subsection{Fisher spectrum and squeezing spectrum}
\label{OptSpect}

Finding the optimal vector $\vect{v} \in \mathbb{R}^d$ that minimizes Eq.~(\ref{emom}) and/or Eq.~(\ref{Dthetan_par}) is solved by calculating the spectrum of the matrices $\vect{\FQ}$ and $\vect{\mathcal{M}}$, that we indicate as Fisher and squeezing spectrum, respectively. 
These spectra contain useful information regarding the multiparameter problem, in general. 
We have 
\be \label{ineq0}
\min_{\vect{v} \in \mathbb{R}^d} \Delta^2(\vect{v} \cdot \vect{\theta})_{\rm emom} = \frac{1}{\mu_{\rm max} d},
\ee
where $\mu_{\rm max}$ is the largest
eigenvalue of $\vect{\mathcal{M}}$.
The corresponding optimal eigenvector $\vect{v}_{\mu_{\rm max}}$ 
gives the linear combination of parameters $\vect{v}_{\mu_{\rm max}} \cdot \vect{\theta}$ that can be 
estimated with the smallest possible uncertainty when using the specific method of moments considered (namely, based on the chosen measurement observables $\hat{X}_j$, probe state and phase encoding transformation).
Following the inequality 
$\Delta^2(\vect{v} \cdot \vect{\theta})_{\rm eQCR} \leq \Delta^2(\vect{v} \cdot \vect{\theta})_{\rm emom}$, 
we have $f_{\rm max} \geq \mu_{\rm max}$, where 
$f_{\rm max}$ is the largest 
eigenvalue of the QFIM and satisfies
\be \label{ineq1}
\min_{\vect{v}\in \mathbb{R}^d} \Delta^2(\vect{v} \cdot \vect{\theta})_{\rm eQCR} = \frac{1}{f_{\rm max} d}.
\ee
The corresponding optimal eigenvector $\vect{v}_{f_{\rm max}}$ (in general, $\vect{v}_{f_{\rm max}}\neq \vect{v}_{\mu_{\rm max}}$) gives the linear combinations of parameters, $\vect{v}_{f_{\rm max}} \cdot \vect{\theta}$, that can be estimated with the highest possible sensitivity (when optimized over all generalized output measurements and all possible estimation strategies) for the given probe state and phase encoding transformation.  
The demonstration of Eqs.~(\ref{ineq0}) and~(\ref{ineq1}) is reported in Appendix~H.
Furthermore, a degeneracy (e.g. in the squeezing spectrum) reveals independent linear combinations of parameters that can be estimated with the same sensitivity.
Specifically, if $d_\mu$ is the degeneracy of the eigenvalue $\mu$ of $\vect{\mathcal{M}}$, then
the sensitivity $\Delta^2 (\vect{v} \cdot \vect{\theta})_{\rm emom} = 1/(\mu d)$
is the same for any $\vect{v}$ given by a linear combination of the $d_\mu$ orthonormal eigenvectors $\vect{v}_\mu^{(1)}, ..., \vect{v}_\mu^{(d_\mu)}$.
In particular, $\vect{\mathcal{M}}^{-1}$ is defined on the subspace of $\mathbb{R}^d$ generated by a basis of eigenvectors of $\vect{\mathcal{M}}$ corresponding to finite eigenvalues (and similarly for $\vect{\FQ}^{-1}$).  
 
\subsection{Random choice of quantum circuit} 
\label{OptRand}

Here, we consider random choices of the QC and find the corresponding optimal $\Delta^2 (\vect{v}\cdot \vect{\theta})_{\rm emom}$. 
To be more explicit, we generate random unitary QC matrices $\vect{U}$ (with uniform deHaar measure) and calculate the largest eigenvalue $\mu_{\rm max}$ of $\vect{\mathcal{M}}$, Eq. (\ref{MOM}). 
Furthermore, without loss of generality, we take the same number of photons in each coherent state, namely $\vert \alpha_j \vert^2= \bar{n}_c$ for all $j$.

\begin{figure}
  \includegraphics[width=\columnwidth]{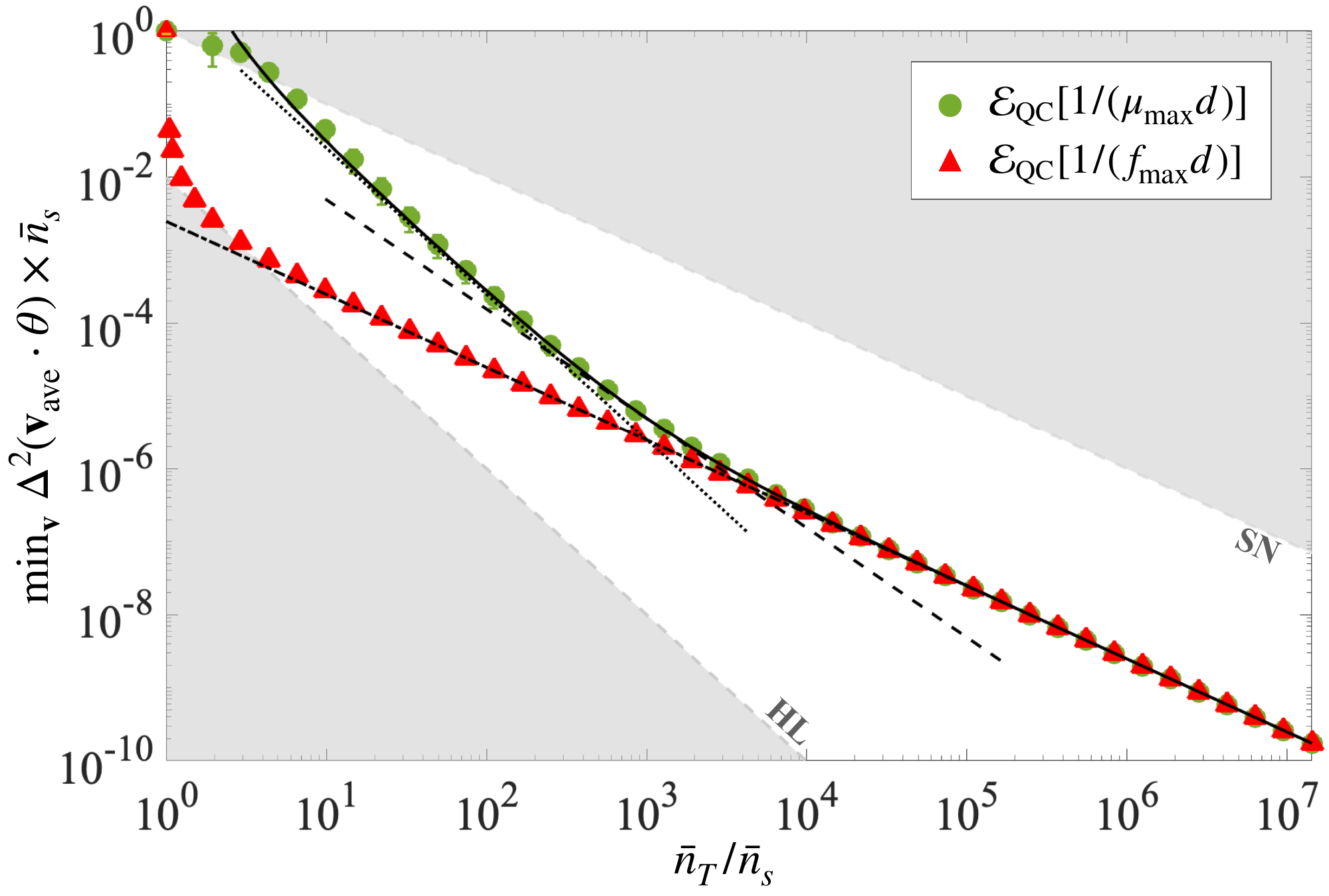} 
  \caption{Optimized phase uncertainties as a function of $\bar{n}_T$.
  Symbols show $\mathcal{E}_{\rm QC}[1/(\mu_{\rm max}d)]$ (green dots) and $\mathcal{E}_{\rm QC}[1/(f_{\rm max}d)]$ (red triangles), where $\mathcal{E}_{\rm QC}[...]$ indicates statistical averaging over random choices of the QC. 
  Error bars are root mean square fluctuations.
  The solid line is Eq.~(\ref{ubound}).
  The dot-dashed line is $e^{-2r}/\bar{n}_T$,  Eq.~(\ref{sensitivity_homodyne_limit}), the dashed line is
  $\sqrt{\mathcal{S}}/\bar{n}_T^{3/2}$, Eq.~(\ref{ubound1}), while the dotted line is $\bar{n}_s \mathcal{S}/\bar{n}_T^2$, Eq.~(\ref{ubound2}).
  The grey regions are defined by $1/\bar{n}_T$ (SN) and $1/\bar{n}_T^2$ (HL). 
  Here, $d=10$, $\bar{n}_s=100$ and statistical averaging is obtained over  $10^4$ random choices of unitary transformation $\vect{U}$.}
  \label{fig5}
\end{figure}

Figure~\ref{fig3} summarizes our findings, while different analytical limits are discussed below.
The figure shows $\mathcal{E}_{\rm QC}[1/(\mu_{\rm max}d)]$ (green dots), where $\mathcal{E}_{\rm QC}[...]$ indicates statistical averaging. 
For comparison, we also consider $\mathcal{E}_{\rm QC}[1/(f_{\rm max}d)]$ (red triangles).
An analytical upper bound to Eq.~(\ref{ineq0}) can be derived by taking $\vect{v} = \vect{\tilde{u}}/\sqrt{d}$, giving
$\min_{\vect{v} \in \mathbb{R}^d} \Delta^2(\vect{v} \cdot \vect{\theta})_{\rm emom} \leq  \vect{\tilde{u}}^T \mathcal{M}(\vect{\tilde{u}})^{-1} \vect{\tilde{u}}/d$.
The inequality is valid for every QC~\cite{nota3} and numerical calculations reveal that it is tight in a wide regimes of parameters. 
In particular, for $\bar{n}_T \gg \bar{n}_s$, and taking the statistical average, we find the simplified expression~\cite{nota3}
\be \label{ubound}
\frac{\mathcal{E}_{\rm QC}\big[\tilde{\vect{u}}^T \mathcal{M}(\tilde{\vect{u}})^{-1} \tilde{\vect{u}}\big]}{d} = \frac{e^{-2r}}{\bar{n}_T} + \frac{\bar{n}_s \mathcal{S}}{\bar{n}_T^2},
\ee
where $\mathcal{S} \equiv \mathcal{E}_{\rm QC}[d \sum_{j=1}^d \tilde{u}_j^4]$. 
Equation (\ref{ubound}) is plot as solid black line in Fig. \ref{fig5}.

{\it Regime $\bar{n}_T \gg \bar{n}_s e^{2r}$.} 
In this regime, Eq.~(\ref{MOM}) becomes \cite{nota2}
\be \label{moment_matrix_homodyne_limit}
\vect{\mathcal{M}}^{-1} = \frac{e^{-2r}-1}{\bar{n}_c} \vect{\tilde{u}}\vect{\tilde{u}}^T + \frac{1}{\bar{n}_c} \vect{I}_d,
\ee
where $\vect{I}_d$ is the $d\times d$ identity matrix. 
Equation~(\ref{moment_matrix_homodyne_limit}) can be diagonalized straightforwardly: we find $\mu_{\max} = \bar{n}_c e^{2r}$, the corresponding eigenvector being $\vect{v}_{\mu_{\rm max}} = \vect{\tilde{u}}/\sqrt{d}$. 
In this case, the upper bound
$\vect{\tilde{u}}^T \mathcal{M}(\tilde{u})^{-1} \vect{\tilde{u}})/\sqrt{d}$ is tight, with the first term in Eq.~(\ref{ubound}) dominating over the second one.
The optimal sensitivity is 
\be \label{sensitivity_homodyne_limit}
\min_{\vect{\vect{v}} \in \mathbb{R}^d} \Delta^2(\vect{v} \cdot \vect{\theta})_{\rm emom}
=\frac{e^{-2r}}{\bar{n}_T},
\ee
shown as dot-dashed line in Fig.~\ref{fig5}.
Equation.~(\ref{sensitivity_homodyne_limit}) holds for any QC.
It is worth noticing that such sensitivity is independent of $\vect{\tilde{u}}$, while $\vect{v}_{\mu_{\rm max}}$ is independent of the numbers of particles $\bar{n}_c$ and $\bar{n}_s$ used.
Equation~(\ref{sensitivity_homodyne_limit}) agrees with the numerical calculations shown in Fig.~\ref{fig5}.
Below we show that $f_{\max} = \mu_{\max} = \bar{n}_c e^{2r}$ in this regime, with corresponding eigenvectors 
$\vect{v}_{f_{\rm max}} =\vect{v}_{\mu_{\rm max}} = \vect{\tilde{u}}/\sqrt{d}$.
The optimal sensitivity predicted by the QFIM is thus saturated by the practical estimation method given by the method of moments: in the present limit, $\mathcal{E}_{\rm QC}[1/(\mu_{\rm max}d)] = \mathcal{E}_{\rm QC}[1/(f_{\rm max}d)] = e^{-r}/\bar{n}_T$ with negligible fluctuations due to random choices of the QC.

\begin{figure}[t!]
  \includegraphics[width=\columnwidth]{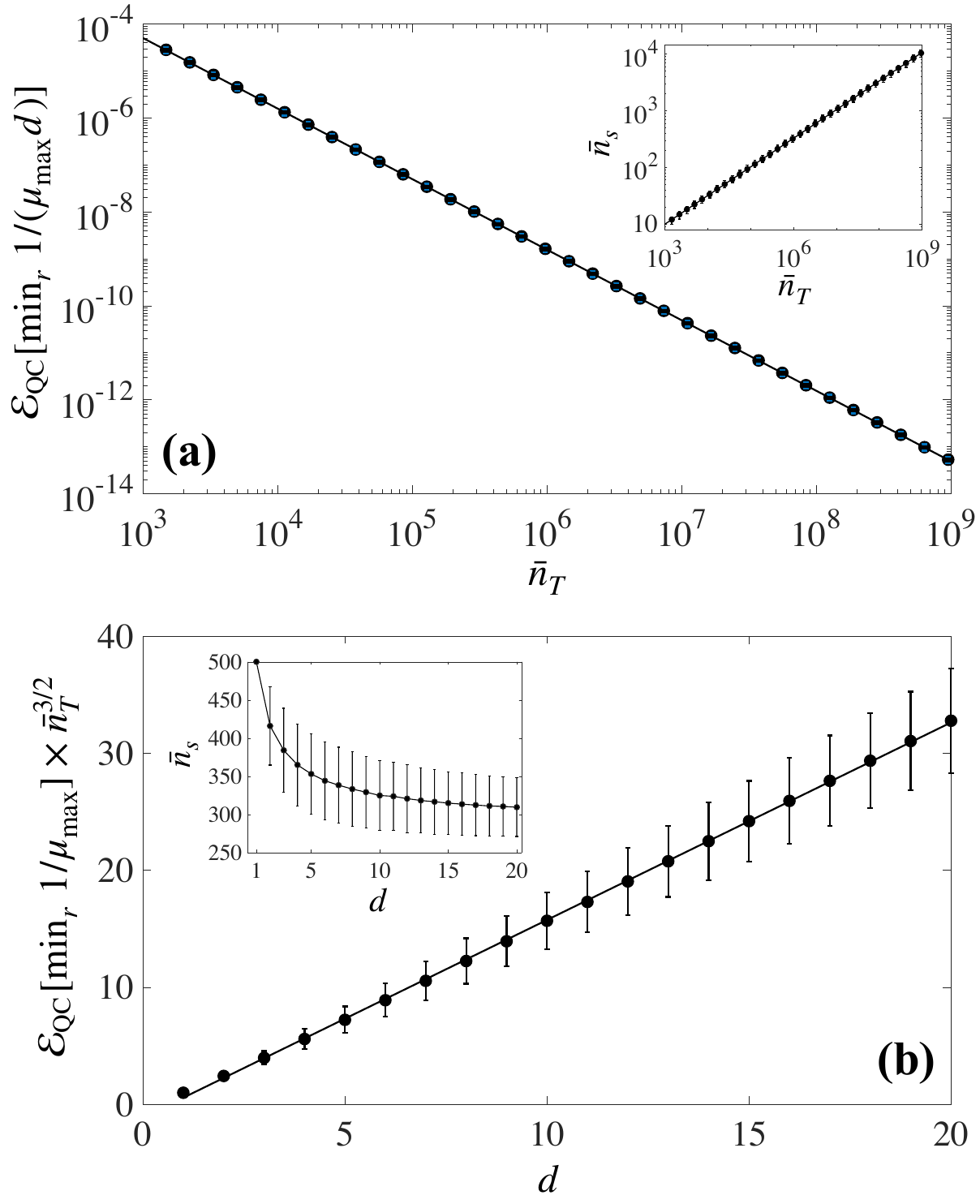} 
\caption{(a) $\mathcal{E}_{\rm QC}[\min_{r} 1/(\mu_{\rm max}d)]$ as a function of $\bar{n}_T$ and for $d=10$~(dots). 
The  solid line is Eq.~(\ref{ubound1}).
The corresponding optimal values of $\bar{n}_s$ are shown in the inset~(dots). 
There, the solid line is $\bar{n}_s = \sqrt{\bar{n}_T}/\sqrt{4 \mathcal{S}}$.
Panel (b) shows $\mathcal{E}_{\rm QC}[\min_{\bar{n}_s} 1/(\mu_{\rm max}d)]\times \bar{n}_T^{3/2}$ as a function of $d$ and for $\bar{n}_T = 10^6$~(dots). 
The solid line is Eq.~(\ref{ubound1}). The corresponding optimal values of $\bar{n}_s$ are shown in the inset, where the solid line corresponds to the theoretical prediction.
$\mathcal{E}_{\rm QC}[]$ indicates statistical average, here over $10^4$ random QC transformations, where error bars are root mean square fluctuations.}
\label{fig6}
\end{figure}

{\it Optimal squeezing for $\bar{n}_T \approx \bar{n}_s e^{2r}$.} 
We now optimize the average number of particles in the squeezed-vacuum state in order to maximize $\mu_{\rm max}$, for a given total average number of particles $\bar{n}_T$ and QC transformation $\hat{U}$.
Such optimization cannot be performed analytically and we rely on a numerical diagonalization of Eq.~\makeref{MOM}.
For each QC, we evaluate numerically the maximum eigenvalue $\mu_{\rm max}$ of the corresponding $\vect{\mathcal{M}}$ and optimize it with respect to $\bar{n}_s$.
Numerical results are compared to the analytical optimization of Eq.~(\ref{ubound}). 
For $\bar{n}_s \gg 1$
(such that $e^{2r} \approx 4 \bar{n}_s$) this predicts
\be \label{ubound1}
\min_{r} \frac{\mathcal{E}_{\rm QC}\big[\tilde{u}^T \mathcal{M}(\tilde{u})^{-1} \tilde{u}\big]}{d} \approx \frac{\sqrt{\mathcal{S}}}{\bar{n}_T^{3/2}}
\ee
for $\bar{n}_s \approx \sqrt{\bar{n}_T/(4 \mathcal{S})}$.
In Fig.~\ref{fig6}(a) we plot $\mathcal{E}_{\rm QC}[\min_{\bar{n}_s} 1/(\mu_{\rm max}d)]$ as a function of $\bar{n}_T$ and for fixed $d$~(dots).
The solid line is Eq.~(\ref{ubound1}).
The inset shows the corresponding optimal values of $\bar{n}_s$~(dots), the solid line being $\bar{n}_s = \sqrt{\bar{n}_T}/\sqrt{4\mathcal{S}}$.
In Fig.~\ref{fig6}(b) we plot $\mathcal{E}_{\rm QC}[\min_{\bar{n}_s} 1/\mu_{\rm max}] \times \bar{n}_T^{3/2}$ as a function of $d$, where the corresponding optimal values of $\bar{n}_s$ are shown in the inset.
The numerical results (dots) are in excellent agreement with Eq.~(\ref{ubound1}) (solid line).
Equation~(\ref{ubound1}) is further shown as dashed line in Fig.~\ref{fig5}.  

{\it Transient Heisenberg scaling for $\bar{n}_s \ll \bar{n}_T \ll \bar{n}_s e^{2r}$.}
In this regime, the first term in Eq.~(\ref{ubound}) can be neglected and we obtain 
\be \label{ubound2}
\frac{\mathcal{E}_{\rm QC}\big[ \tilde{u}^T \mathcal{M}(\tilde{u})^{-1} \tilde{u} \big]}{d} \approx
\frac{\bar{n}_s \mathcal{S}}{\bar{n}_T^2}.
\ee
This predicts a transient Heisenberg scaling, for fixed $\bar{n}_s$, with prefactor approximately given by $\bar{n}_s$.
This prediction is confirmed in Fig.~\ref{fig5} where Eq.~(\ref{ubound2}) is shown as the dotted line. 

\begin{figure}[t!]
  \includegraphics[width=\columnwidth]{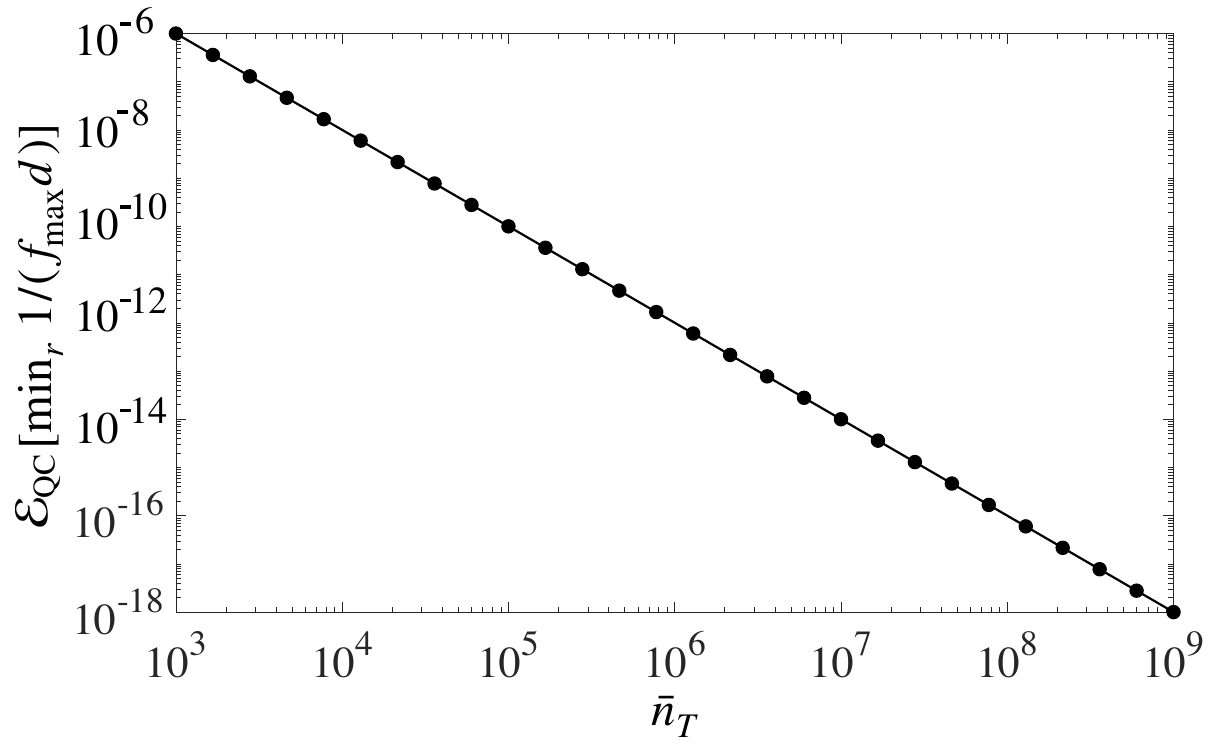} 
\caption{Plot of $\min_{\bar{n}_s} 1/(f_{\rm max} d)$, averaged over $10^4$ random choices of the QC (dots). 
The solid line is the analytical prediction $1/\bar{n}_T^2$.}
\label{fig7}
\end{figure}

{\it Quantum Cramer-Rao bound.}
We now study the QFIM, Eq.~(\ref{QFIM}), for $\vert \alpha_j \vert^2 = \bar{n}_c$ for all $j$. 
In the limit $\bar{n}_c e^{2r} \gg \bar{n}_s$ (which is $\bar{n}_c \gg 1$ for sufficiently large values of $\bar{n}_s$), Eq.~(\ref{QFIM}) assumes the simple form
\be \label{QFIM_homodyne_limit}
\vect{\FQ} = \bar{n}_c (e^{2r}-1) \vect{\tilde{u}}\vect{\tilde{u}}^T + \bar{n}_c \vect{I}_d.
\ee
The maximum eigenvalue is $f_{\rm max} = e^{2r} \bar{n}_c$ and the corresponding eigenvector is $\vect{v}_{f_{\rm max}} = \vect{\tilde{u}}/\sqrt{d}$. 
For $d\bar{n}_c \gg \bar{n}_s$, we have $\bar{n}_T \approx d\bar{n}_c$ and we find $\min_{\vect{\vect{v}}\in \mathbb{R}^d} \Delta^2(\vect{v} \cdot \vect{\theta})_{\rm eQCR} = e^{-2r}/\bar{n}_T$. 
This behaviour holds also for $\bar{n}_T \gg  \bar{n}_s e^{2r}$, where $\min_{\vect{\vect{v}}\in \mathbb{R}^d} \Delta^2(\vect{v} \cdot \vect{\theta})_{\rm eQCR} =
\min_{\vect{\vect{v}} \in \mathbb{R}^d} \Delta^2(\vect{v} \cdot \vect{\theta})_{\rm emom}$ as given in Eq.~(\ref{sensitivity_homodyne_limit}), see Fig.~\ref{fig5}.
Furthermore, taking $\bar{n}_s\gg 1$ (so that $\bar{n}_s\approx e^{2r}/4$), we can optimize $f_{\rm max} = 4\bar{n}_c \bar{n}_s$ with respect to $\bar{n}_s$, for a fixed $\bar{n}_T$:
replacing $\bar{n}_s = \bar{n}_T - \bar{n}_c d$ and taking the derivative with respect to $\bar{n}_c$, we find the optimal condition $d\bar{n}_c = \bar{n}_s = \bar{n}_T/2$. 
This predicts the saturation of the Heisenberg limit 
\be \label{sensitivity_same_intensity_1}
\min_{\vect{\vect{v}}\in \mathbb{R}^d} \Delta^2(\vect{v} \cdot \vect{\theta})_{\rm eQCR} =
\Delta^2(\vect{\tilde{u}} \cdot \vect{\theta})_{\rm eQCR} =\frac{1}{\bar{n}_T^2}
\ee
with respect to the total number of particles $\bar{n}_T$. 
In Fig. \ref{fig7} we show the the statistical average of $\min_{\bar{n}_s} 1/(f_{\rm max} d)$ (dots) as a function of $\bar{n}_T$.
The solid line is $1/\bar{n}_T^2$.
In Fig.~\ref{fig5} we plot $1/(f_{\rm max} d)$, averaged on random choices of the QC (triangles). 
Numerical simulations agree well with analytical predictions in the different limits. 
In particular, we see that the QFIM tends to a sub-shot noise scaling faster than the moment matrix, i.e. for smaller values of $\bar{n}_T$.

\section{Conclusions and Discussion}
\label{conclusion}

This work generalizes one of the most important archetype of quantum interferometry -- namely, the single MZI with coherent$\otimes$squeezed-vacuum light~\cite{CavesPRD1981,ParisPLA1995, BarnettEPJD2003, PezzePRL2008, LangPRL2013, Ruo-BercheraPRA2015, SparaciariPRA2016} -- to a distributed sensor network composed by $d$ MZIs, see Fig. \ref{fig1}(a).
The multiphase estimation analysis is based on a method of moments requiring local and independent photocounting at the output of each MZI. 
This avoids to recombine the phase-shifted modes using a second multimode beam splitter~\cite{TriggianiARXIV, OhPRR2020, GePRL2018}. 
The scheme is thus optimal to realize a highly spatially-separated sensor using a multimode entangled state of a large number of particles. 
In particular, our MZI sensor network is characterized by different regimes reaching $\Delta^2 (\vect{v} \cdot \vect{\theta})_{\rm emom} = O(\bar{n}_T^{-3/2})$ and the Heisenberg scaling $\Delta^2 (\vect{v} \cdot \vect{\theta})_{\rm emom} = O(\bar{n}_T^{-2})$.

The manuscript is focused on two different problems:

i) We have optimized the full Mach-Zehnder sensor network in order to minimize the uncertainty, $\Delta^2 (\vect{v} \cdot \vect{\theta})$, for the estimation of an {\it arbitrary} fixed linear combination of phase shifts $\vect{v} \cdot \vect{\theta}$.
Studying the cases $d=2$ and $d=3$, we have shown that the optimized parallel strategy, exploiting the mode entanglement generated by the linear multimode splitting is never surpassed by an optimized sequential strategy using mode-separable squeezed-vacuum states, when considering different constraints. 
For any number of parameters, the maximum gain of the parallel over the sequential strategy is a factor $d$, for the estimation of the generalized average phase $\vect{v}_{\rm ave} \cdot \vect{\theta} = \sum_{j=1}^d \pm \theta_j/d$. While the literature on distributed quantum sensing has mainly focused on the estimation of specific linear combinations of different parameters, the possibility to optimize the sensor network for the estimation of any desired $\vect{v}\cdot \vect{\theta}$, as shown here, is generally highly desirable. 
This possibility has only been shown in the configurable multimode displacement sensor of Ref.~\cite{XiaPRL2020} and for multipass phase sensing~\cite{GebhartARXIV} using a photonic Bell state~\cite{ZhaoPRX2021}.
Further configurable approaches include the splitting and multimode recombination of squeezed-vacuum light~\cite{TriggianiARXIV} and twin-Fock states~\cite{GePRL2018}.
In our case, the optimized sensing scheme uses local measurements and avoids the recombination of the states in a second quantum circuit.

ii) For arbitrary unitary splitting the squeezed vacuum state, we have identified optimal linear combinations of the $d$ phases that minimize the uncertainty $\Delta^2(\vect{v} \cdot \vect{\theta})$.
Typical results and scalings with the total average number of particles $\bar{n}_T$ hold regardless the random choice of unitary QC transformation.

A further interesting problems raised in the context of multiparameter estimation is whether a single sensor network allows the estimation of multiple linear combination of phases at the same time \cite{RubioJPA2020}. 
This problem is solved here by changing the mode $D$ of the QC where the squeezed vacuum state is injected, see Fig.~\ref{fig1}(a).
Indeed, a single $d$-mode QC can be optimized to estimate $d$ different (e.g. orthogonal) linear combinations $\vect{v}_D \cdot \vect{\theta}$ (with $D=1, ..., d$ and $\vect{v}_i \cdot \vect{v}_j = \delta_{ij}$) with the same sensitivity: each input mode $D$ of the QC corresponds to a specific optimal $\vect{v}_D \cdot \vect{\theta}$.

It is also worth comparing here directly with the results of Ref.~\cite{XiaPRL2020} that considered a sensor network based on the linear splitting of a squeezed-vacuum state, displacement operations and homodyne measurements on each output mode. 
It should be noticed that displacement sensing and phase sensing using a network of MZIs (as considered here) corresponds, in general, different parameter-encoding transformations. 
A precise mapping is obtained within a (mean-field) Holstein-Primakof approximation, where the mode operator $\hat{a}_j$, see Fig.~\ref{fig1}, is replaced by the classical number $\vert \alpha_j \vert$.
In this case, the $j$th Mach-Zehnder transformation $e^{-\theta_j (\hat{a}^\dag_j \hat{b}_j - \hat{b}_j^\dag \hat{a}_j)/2}$ reduces to the single-mode displacement operator $\hat{D}(q_j)$ with $q_j = \vert \alpha_j \vert /2$.
Our results predict $\Delta^2 (\vect{v} \cdot \vect{q})_{\rm eQCR} = \Delta^2 (\vect{v} \cdot \vect{\theta})_{\rm eQCR} \times \bar{n}_c/4 =  e^{-2r}/4$
for $\vert \alpha_j \vert^2 = \bar{n}_c$, in agreement with Refs.~\cite{XiaPRL2020, ZhuangPRA2018}. 
Yet, while the phase uncertainty is characterized by a convenient scaling with $\bar{n}_T$, 
$\Delta^2 (\vect{v} \cdot \vect{\theta})_{\rm eQCR} = e^{-2r}/\sqrt{\bar{n}_T}$, the displacement uncertainty does not scale with $\bar{n}_T$.
In this regime, the quantum Cramer-Rao bound is saturated by the estimation strategy based on the multiparameter method of moments \cite{GessnerNATCOMM2020} based on local photodetection.
Furthermore, our formalism allows to go beyond the homodyne limit of Refs.~\cite{XiaPRL2020,ZhuangPRA2018} and discuss the phase sensitivity of a network of MZIs with respect to the total number of particles $\bar{n}_T$ used: this is crucial to discuss scalings of phase variance that are faster than $O(1/\bar{n}_T)$.

The results of this work are relevant in current experiments realizing squeezed vacuum-light and multimode linear splitting transformations~\cite{GuoNATPHYS2020, XiaPRL2020, NokkalaNJP2018}.  
They pave the way to sensor networks using multiple MZIs -- in both optical and atomic systems -- with a large variety of applications ranging from field and biological sensing, gravitational wave detection, quantum clocks and inertial measurements.

\begin{widetext}

\section{Appendix}

\subsection{Detailed derivation of Eqs. (\ref{MOM}) and (\ref{QFIM})}

In this Appendix, we provide details on the derivation of Eqs.~(\ref{MOM}) and~(\ref{QFIM}).
Our methods are based on a technique to calculate the QFIM that was outlined in Ref.~\cite{GagatsosPRA2016}.
It should be noticed, however, that Ref.~\cite{GagatsosPRA2016} considered a different sensor network configuration: the generalization to a network of MZIs is not straightforward and requires additional algebraic work.
Furthermore, our derivation corrects some flaws which are present in Ref.~\cite{GagatsosPRA2016} and that led to an incorrect final expression (see discussion below). 
Alternative approaches to calculate the QFIM of Gaussian states have been also considered, see Refs.~\cite{JiangPRA2014, BanchiPRL2015, NicholsPRA2018, OhPRR2020}. 

\subsubsection{Preliminary definitions}

We consider the general case where a product of $d'$ single-mode squeezed states is sent to a passive linear network $\hat{A}^{\dagger}$ and transformed according to
\begin{equation} \label{general_initial_state}
    \ket{\Psi}=\hat{A}^{\dagger}\bigotimes_{k=1}^{d'}\ket{\beta_k,\xi_k}.
\end{equation}
Here, $\ket{\beta_k,\xi_k}$ is the single-mode displaced-squeezed state in the mode $k$: 
$\beta_k$ is the coherent amplitude of the state and $\xi_k=r_k e^{i\varphi_k}$ its squeeze parameter. 
In the following, we will assume that $\hat{A}^{\dagger}$ is a Gaussian unitary, that is, a unitary operator which transforms Gaussian states into Gaussian states. 
Because $\hat{A}^{\dagger}$ is also a passive, i.e. particle-number preserving, transformation, if we set  $\hat{\vect{c}}=(\hat{c}_1,\dots,\hat{c}_{d'})^T$, a relation $\hat{A}\hat{\vect{c}}\hat{A}^{\dagger}=\hat{\vect{c}}'=\vect{\mathcal{A}}\hat{\vect{c}}$ must hold, with $\vect{\mathcal{A}}$ a unitary matrix (similarly, $\hat{A} \hat{\vect{c}}^\dag \hat{A}^{\dagger} =\vect{\mathcal{A}}^*\hat{\vect{c}}^\dag$).
Here, $\hat{c}^{\dagger}_k$ and $\hat{c}_k$ are bosonic creation and annihilation operators, respectively,

We recall the definition of the Q-function for the state $\ket{\Phi}$ of a generic $d'$-mode system:
\begin{equation}
    Q(\vect{\alpha})=\frac{\vert \langle \vect{\alpha} \vert \Phi \rangle \vert^2}{\pi},
\end{equation}
where $\ket{\vect{\alpha}}=\bigotimes_{k=1}^{d'}\ket{\alpha_k}$, $\ket{\alpha_k}$ being an arbitrary single-mode coherent state in mode $k$. 
From the point of view of the Q-function, a transformation $\hat{A}^{\dagger}\ket{\Phi}$ is equivalent to a transformation $\hat{A}\ket{\vect{\alpha}}$ of the coherent states; moreover, it is a well know property that a Gaussian passive transformation $\hat{A}$ sends a product of coherent states into another product of coherent states, in particular: $\hat{A}\ket{\vect{\alpha}}=\ket{\vect{\alpha}'}=\ket{\vect{A}\vect{\alpha}}$.
As the notation just used suggests, and according to Ref. \cite{GagatsosPRA2016}, $\vect{A}$ is the matrix that implements the transformation $\vect{\alpha}'=\vect{A}\vect{\alpha}$ of the amplitudes of the coherent states associated with the Q-function. It is possible to show that this matrix is the same as the one which describes the transformation of the annihilation operators implemented by $\hat{A}$, that is $\hat{A}^{\dagger}\hspace{0.1cm}\vect{\hat{c}}\hspace{0.1cm}\hat{A}=\vect{\hat{c}}'=\vect{A}\vect{\hat{c}}$, and, correspondingly, the hermitian conjugate of the matrix which describes the transformation of the annihilation operators implemented by $\hat{A}^{\dagger}$, which we have denoted as $\vect{\mathcal{A}}$ above. The relation $\vect{\mathcal{A}}^{\dagger}=\vect{A}$ will be frequently used in what follows.

Using the Q-function representation of the states, \makeref{general_initial_state}, Ref. \cite{GagatsosPRA2016} showed that
\begin{align}
\langle \hat{n}_i \rangle&=-1+\partial_i\partial_i^*G(\vect{\mu})\Bigr\rvert_{\vect{\mu} = 0},\label{ni_general}\\
\langle \hat{n}_i\hat{n}_j \rangle&=\left[\partial_i\partial_i^*\partial_j\partial_j^*-(1+\delta_{ij})\partial_i\partial_i^*-\partial_j\partial_j^*\right]G(\vect{\mu})\Bigr\rvert_{\vect{\mu} = 0}+1,\label{ninj_general}
\end{align}
where $\vect{\mu}=(\lambda_1,\dots,\lambda_{d'},\lambda_1^*,\dots,\lambda_{d'}^*)^T$ is an arbitrary $2d'$-dimensional complex vector, $\partial_i$, $\partial_i^*$ are shorthand notation for $\partial/\partial \lambda_i$, $\partial/\partial \lambda_i^*$, $\hat{n}_i=\hat{c}^{\dagger}_i\hat{c}_i$ and the expectation values are evaluated in state $\ket{\Psi}$. 
We have $G(\vect{\mu})=e^{\Delta}$, where
\begin{equation} \label{Delta}
    \Delta \equiv \displaystyle\frac{1}{4}\left(\vect{\nu_b}^{\dagger}\vect{M}^{-1}\vect{\mu}+\vect{\mu}^{\dagger}\vect{M}^{-1}\vect{\nu_b}+\vect{\mu}^{\dagger}\vect{M}^{-1}\vect{\mu}\right),
\end{equation}
\be
\vect{\nu_b}=(b_1,\dots,b_{d'},b_1^*,\dots,b_{d'}^*)^T,
\ee
\be
b_j=\sum_k \vect{A}^{\dagger}_{jk}(\beta_k+\beta_k^*e^{i\varphi_k}\tanh{r_k}),
\ee
\be
\vect{M}^{-1}=2\begin{pmatrix} \vect{E}&-\vect{NE}^T \\ -\vect{N}^{\dagger}\vect{E}&\vect{E}^T \end{pmatrix}, \label{M-1}
\ee
\begin{equation}\label{N}
    \vect{N}=\vect{A}^{\dagger}\vect{D}\vect{A}^*,    \end{equation}
\begin{equation}\label{E}
    \vect{E}=\vect{A}^{\dagger}\vect{C}\vect{A},
\end{equation}
\begin{equation}\label{C}
    \vect{C}_{jk}=\delta_{jk}\cosh^2 r_k,
\end{equation}
and
\begin{equation}\label{D}
    \vect{D}_{jk}=\delta_{jk}e^{i\varphi_k}\tanh{r_k}.
\end{equation}
Using Eq. (\ref{M-1}), we find 
\begin{align}
\Delta=\frac{1}{4}\left[\begin{pmatrix}\vect{b}^*&\vect{b}\end{pmatrix}\cdot2\begin{pmatrix} \vect{E}&-\vect{NE}^T \\ -\vect{N}^{\dagger}\vect{E}&\vect{E}^T \end{pmatrix}\begin{pmatrix}\vect{\lambda}\\ \vect{\lambda}^*\end{pmatrix}+\begin{pmatrix}\vect{\lambda}^*&\vect{\lambda}\end{pmatrix}\cdot2\begin{pmatrix} \vect{E}&-\vect{NE}^T \\ -\vect{N}^{\dagger}\vect{E}&\vect{E}^T \end{pmatrix}\begin{pmatrix}\vect{b}\\ \vect{b}^*\end{pmatrix} 
+\begin{pmatrix}\vect{\lambda}^*&\vect{\lambda}\end{pmatrix}\cdot2\begin{pmatrix} \vect{E}&-\vect{NE}^T \\ -\vect{N}^{\dagger}\vect{E}&\vect{E}^T \end{pmatrix}\begin{pmatrix}\vect{\lambda}\\ \vect{\lambda}^*\end{pmatrix}\right], \nonumber
\end{align}
which we can write more explicitly as 
\begin{align}
\Delta &=\frac{1}{2}\left[\vect{b}^*\cdot\left(\vect{E}\vect{\lambda}-\vect{NE}^T\vect{\lambda}^*\right)+\vect{b}\cdot\left(-\vect{N}^{\dagger}\vect{E}\vect{\lambda}+\vect{E}^T\vect{\lambda}^*\right)+\vect{\lambda}^*\cdot\left(\vect{E}\vect{b}-\vect{NE}^T\vect{b}^*\right)+\vect{\lambda}\cdot\left(-\vect{N}^{\dagger}\vect{E}\vect{b}+\vect{E}^T\vect{b}^*\right)\right. \nonumber\\
&\left.\qquad+\vect{\lambda}^*\cdot\left(\vect{E}\vect{\lambda}-\vect{NE}^T\vect{\lambda}^*\right)+\vect{\lambda}\cdot\left(-\vect{N}^{\dagger}\vect{E}\vect{\lambda}+\vect{E}^T\vect{\lambda}^*\right)\right]. \nonumber
\end{align}
We then calculate the first and second partial derivatives of $\Delta$ with respect to $\lambda_i$ and $\lambda_i^*$:
\begin{align}
&\begin{aligned}[b]\partial_i\Delta 
&=\frac{1}{2}\sum_k\left[-\left(\left(\vect{N}^{\dagger}\vect{E}\right)_{ik}+\left(\vect{N}^{\dagger}\vect{E}\right)_{ki}\right)(\vect{\lambda}_k+\vect{b}_k)+2\vect{E}_{ki}(\vect{\lambda}_k^*+\vect{b}_k^*)\right]=\sum_k \vect{E}^*_{ik}(\vect{\lambda}_k^*+\vect{b}_k^*)-(\vect{EN})^*(\vect{\lambda}_k+\vect{b}_k),\end{aligned}\label{Deltaderivative}\\
&\begin{aligned}[b]\partial_i^*\Delta 
&=\frac{1}{2}\sum_k\left[2\vect{E}_{ik}(\vect{\lambda}_k+\vect{b}_k)-\left(\left(\vect{NE}^T\right)_{ik}+\left(\vect{NE}^T\right)_{ki}\right)(\vect{\lambda}_k^*+\vect{b}_k^*)\right]=\sum_k \vect{E}_{ik}(\vect{\lambda}_k+\vect{b}_k)-(\vect{EN})(\vect{\lambda}_k^*+\vect{b}_k^*),\end{aligned}\label{complex_Deltaderivative}\\
&\partial_i^*\partial_j\Delta=\vect{E}_{ij},\\
&\partial_i\partial_j^*\Delta=\vect{E}_{ji}=\vect{E}^*_{ij},\\
&\partial_i^*\partial_j^*\Delta=-\frac{1}{2}\left(\left(\vect{NE}^T\right)_{ij}+\left(\vect{NE}^T\right)_{ji}\right)=-(\vect{EN})_{ij},\\
&\partial_i\partial_j\Delta=-\frac{1}{2}\left(\left(\vect{N}^{\dagger}\vect{E}\right)_{ij}+\left(\vect{N}^{\dagger}\vect{E}\right)_{ji}\right)=-(\vect{EN})^*_{ij}.
\end{align}
To derive the above equations, we have used the following relations:
\begin{align}
    &\vect{E}=\vect{E}^{\dagger} \implies \vect{E}^T=\vect{E}^*, \nonumber \\
    &\vect{N}=\vect{N}^{T} \implies \vect{N}^{\dagger}=\vect{N}^*, \nonumber \\
    &(\vect{EN})^T=\vect{EN}, \nonumber \\
    &\vect{NE}^T=(\vect{EN})^T=\vect{EN}, \nonumber \\
    &\vect{N}^{\dagger}\vect{E}=(\vect{EN})^{\dagger}=\left((\vect{EN})^T\right)^*=(\vect{EN})^*. \nonumber 
\end{align}
We are now ready to work out the partial derivatives of $G(\vect{\mu})$ that appear in Eqs. \makeref{ni_general} and \makeref{ninj_general}:
\begin{align}
\partial_i\partial_i^*G(\vect{\mu})
=(\partial_i\partial_i^*\Delta)e^{\Delta}+(\partial_i\Delta)(\partial_i^*\Delta)e^{\Delta}=\left(\vect{E}_{ii}+(\partial_i\Delta)(\partial_i^*\Delta)\right)e^{\Delta},   \nonumber 
\end{align}
and
\begin{align}
\partial_i\partial_i^*\partial_j\partial_j^*G(\vect{\mu})
=&\partial_i\left[\left(\vect{E}_{jj}+(\partial_j\Delta)(\partial_j^*\Delta)\right)(\partial_i^*\Delta)e^{\Delta}+\left((\partial_i^*\partial_j\Delta)(\partial_j^*\Delta)+(\partial_j\Delta)(\partial_i^*\partial_j^*\Delta)\right)e^{\Delta}\right] \nonumber \\
=&\left[(\vect{EN})_{ij}(\vect{EN})^*_{ij}-(\vect{EN})^*_{ij}(\partial_i^*\Delta)(\partial_j^*\Delta)-(\vect{EN})_{ij}(\partial_i\Delta)(\partial_j\Delta)+\vect{E}_{ij}\vect{E}^*_{ij}+\vect{E}_{ij}(\partial_i\Delta)(\partial_j^*\Delta)+\vect{E}^*_{ij}(\partial_i^*\Delta)(\partial_j\Delta)\right.\nonumber\\
&\left.+\left(\vect{E}_{ii}+(\partial_i\Delta)(\partial_i^*\Delta)\right)\left(\vect{E}_{jj}+(\partial_j\Delta)(\partial_j^*\Delta)\right)\right]e^{\Delta}. \nonumber 
\end{align}
By evaluating the derivatives at $\vect{\mu}=0$, we get
\begin{equation}
\partial_i\partial_i^*G(\vect{\mu})\Bigr\rvert_{\vect{\mu} = 0}=\vect{E}_{ii}+(\partial_i\Delta)_0(\partial_i^*\Delta)_0,  \nonumber 
\end{equation}
where $(\partial_i\Delta)_0$ is shorthand notation for $(\partial_i\Delta)\bigr\rvert_{\vect{\mu} = 0}$, and
\begin{align}
\partial_i\partial_i^*\partial_j\partial_j^*G(\vect{\mu})\Bigr\rvert_{\vect{\mu}=0}=&(\vect{EN})_{ij}(\vect{EN})^*_{ij}-(\vect{EN})^*_{ij}(\partial_i^*\Delta)_0(\partial_j^*\Delta)_0-(\vect{EN})_{ij}(\partial_i\Delta)_0(\partial_j\Delta)_0+\vect{E}_{ij}\vect{E}^*_{ij}+\vect{E}_{ij}(\partial_i\Delta)_0(\partial_j^*\Delta)_0+\vect{E}^*_{ij}(\partial_i^*\Delta)_0(\partial_j\Delta)_0\nonumber\\
&+\left(\vect{E}_{ii}+(\partial_i\Delta)_0(\partial_i^*\Delta)_0\right)\left(\vect{E}_{jj}+(\partial_j\Delta)_0(\partial_j^*\Delta)_0\right). \nonumber 
\end{align}
Finally, going back to Eqs. \makeref{ni_general} and \makeref{ninj_general}, we get
\begin{align}
\langle \hat{n}_i \rangle=-1+\partial_i\partial_i^*G(\vect{\mu})\Bigr\rvert_{\vect{\mu} = 0} =-1+\vect{E}_{ii}+(\partial_i\Delta)_0(\partial_i^*\Delta)_0 \label{ni} 
\end{align}
and 
\begin{align}
\langle \hat{n}_i\hat{n}_j \rangle=&\left[\partial_i\partial_i^*\partial_j\partial_j^*-(1+\delta_{ij})\partial_i\partial_i^*-\partial_j\partial_j^*\right]G(\vect{\mu})\Bigr\rvert_{\vect{\mu}=0}+1 \nonumber\\
=&\partial_i\partial_i^*\partial_j\partial_j^*G(\vect{\mu})\Bigr\rvert_{\vect{\mu} = 0}-(1+\delta_{ij})(\langle \hat{n}_i \rangle+1)-(\langle \hat{n}_j \rangle+1)+1 \nonumber \\
=&(\vect{EN})_{ij}(\vect{EN})^*_{ij}-(\vect{EN})^*_{ij}(\partial_i^*\Delta)_0(\partial_j^*\Delta)_0-(\vect{EN})_{ij}(\partial_i\Delta)_0(\partial_j\Delta)_0+\vect{E}_{ij}\vect{E}^*_{ij}+\vect{E}_{ij}(\partial_i\Delta)_0(\partial_j^*\Delta)_0+\vect{E}^*_{ij}(\partial_i^*\Delta)_0(\partial_j\Delta)_0\nonumber\\
&+\langle \hat{n}_i \rangle \langle \hat{n}_j \rangle-\delta_{ij}\left(\vect{E}_{ii}+(\partial_i\Delta)_0(\partial_i^*\Delta)_0\right).\label{ninj_intermediate}
\end{align}
At this point, we introduce the $d'\times d'$ matrix $\vect{h}$, with elements 
\begin{equation}
    \vect{h}_{ij}=\langle \hat{n}_i\hat{n}_j \rangle-\langle \hat{n}_i \rangle \langle \hat{n}_j \rangle,\label{hij_general}
\end{equation}
whose expression, taking into account the above equations for $\langle \hat{n}_i \rangle$ and $\langle \hat{n}_i\hat{n}_j \rangle$, can be immediately derived:
\begin{align}
\vect{h}_{ij}=&(\vect{EN})_{ij}(\vect{EN})^*_{ij}-(\vect{EN})^*_{ij}(\partial_i^*\Delta)_0(\partial_j^*\Delta)_0-(\vect{EN})_{ij}(\partial_i\Delta)_0(\partial_j\Delta)_0+\vect{E}_{ij}\vect{E}^*_{ij}+\vect{E}_{ij}(\partial_i\Delta)_0(\partial_j^*\Delta)_0+\vect{E}^*_{ij}(\partial_i^*\Delta)_0(\partial_j\Delta)_0\nonumber\\
&-\delta_{ij}\left(\vect{E}_{ii}+(\partial_i\Delta)_0(\partial_i^*\Delta)_0\right). \nonumber
\end{align}
This equation can be rewritten in a compact form by 
introducing the vector
\be
\vect{\gamma}_i\equiv(\partial_i\Delta)_0. \nonumber\\
\ee
Notice that $(\gamma_i)^* = (\partial_i^*\Delta)_0$.
From Eqs. \makeref{Deltaderivative} and \makeref{complex_Deltaderivative}, we get 
$\vect{\gamma}_i=\sum_k \vect{E}^*_{ik}b_k^*-(\vect{EN})^*_{ik}b_k$,
namely 
$\vect{\gamma}=\vect{E}^*\vect{b}^*-(\vect{EN})^*\vect{b}$. 
Finally, making use of vector $\vect{\gamma}$, and of the Hadamard entrywise product $\circ$, we can rewrite $\vect{h}$ in the compact form:
\begin{align}
\vect{h}
=&\vect{EN}\circ (\vect{EN})^*-\vect{EN}\circ\vect{\gamma}\vect{\gamma}^T-(\vect{EN})^*\circ\left(\vect{\gamma}\vect{\gamma}^T\right)^*+\vect{E}\circ \vect{E}^*+\vect{E}\circ\vect{\gamma}\vect{\gamma}^{\dagger}+\vect{E}^*\circ\left(\vect{\gamma}\vect{\gamma}^{\dagger}\right)^*-\left(\vect{E}+\vect{\gamma}\vect{\gamma}^{\dagger}\right)\circ\vect{I}.\label{hij_final}
\end{align}
It is evident that matrix $\vect{h}$ is real and symmetric. An expression similar to Eq. (\ref{hij_final}) was derived in Ref. \cite{GagatsosPRA2016}, see Eq. (14) in that reference. 
There are however important differences with respect to Eq. (\ref{hij_final}) due to flaws in the derivation reported in Ref.~\cite{GagatsosPRA2016}.

\subsubsection{Quantum Fisher information matrix, Eq. (\ref{QFIM})}

Using Eq. \makeref{hij_final}, we now show how to express the QFIM  in terms of the matrices $\vect{E}$, $\vect{EN}$, etc., which were introduced in the previous paragraph and will here be evaluated for the specific case of the Mach-Zehnder sensor network of Fig.~\ref{fig1}. 
In our sensing scheme, the initial state is given by
\begin{equation}
    \ket{\Psi_{\rm in}}=\left(\ket{\alpha_1}\otimes\dots\otimes \ket{\alpha_d}\right)\otimes\left(\ket{0}\otimes\dots\otimes\ket{\xi}\otimes\dots\otimes\ket{0}\right). \nonumber
\end{equation}
It is a product state of coherent states in modes $a_1, ..., a_d$, a squeezed-vacuum state $\ket{\xi}$ in mode $(b_{\rm in})_D$ and the vacuum $\ket{0}$ in modes $(b_{\rm in})_j$ for $j=1, ..., d$ and $j\neq D$.
This initial state should be compared with the product state in Eq. \makeref{general_initial_state}.
In order to facilitate the identification of the two cases, we can set $\hat{a}_j\equiv \hat{c}_j$ and $(\hat{b}_{\rm in})_j\equiv \hat{c}_{j+d}$ $(j=1,\dots,d)$, thus introducing a more homogeneous notation valid for all of the $2d=d'$ input modes of the sensing apparatus. We then identify the $2d \times 2d$ unitary matrix corresponding to the mode transformation performed by the QC as
\begin{equation}
    \vect{U}_{\rm QC}^\dag =\begin{pmatrix}\vect{I}_d&0\\0&\vect{U}^{\dagger} \end{pmatrix}\label{U_NET}. \nonumber
\end{equation}
The $d\times d$ identity matrix $\vect{I}_d$ describes the action of the QC on the coherent states, while $\hat{b}_j=\sum_k (\vect{U}^{\dagger})_{jk}(\hat{b}_{\rm in})_k$ $(j=1,\dots,d)$, $\vect{U}^{\dagger}$ being a unitary $d\times d$ matrix. 
We denote as $\ket{\Psi_0}$ the output state of the QC: $\ket{\Psi_0}=\hat{U}_{\rm QC}^{\dagger}\ket{\Psi_{\rm in}}$.

The phases $\theta_1, \dots, \theta_d$ to be estimated are encoded in $\ket{\Psi_0}$ through the unitary transformation $\otimes_{j=1}^d e^{-i \theta_j(\hat{J}_y)_j}$, where the transformation $e^{-i \theta_j(\hat{J}_y)_j}$ identifies the $j$th MZI in the network, with $\hat{H}_j = (\hat{J}_y)_j = (\hat{a}_j^\dag \hat{b}_j - \hat{b}_j^\dag \hat{a}_j)/2i$. This is equivalent to the phases being encoded in the state $\ket{\Psi}=\otimes_{j=1}^d e^{-i\frac{\pi}{2}(\hat{J}_x)_j}\ket{\Psi_0}$, where $(\hat{J}_x)_j = (\hat{a}_j^\dag \hat{b}_j + \hat{b}_j^\dag \hat{a}_j)/2$, through the unitary transformation $\otimes_{j=1}^d e^{-i \theta_j(\hat{J}_z)_j}$, with $(\hat{J}_z)_j = (\hat{a}_j^\dag \hat{a}_j - \hat{b}_j^\dag \hat{b}_j)/2$. This alternative formulation is more convenient here. Overall, $\ket{\Psi}$ can be expressed as
\begin{equation}
    \ket{\Psi}=\left(\otimes_{j=1}^d e^{-i\frac{\pi}{2}(\hat{J}_x)_j}\right)\hat{U}^{\dagger}_{\rm QC}\ket{\Psi_{\rm in}}. \nonumber
\end{equation}
This equation should be compared with Eq. \makeref{general_initial_state}: the identification $\hat{A}^{\dagger}=(\otimes_{j=1}^d e^{-i\frac{\pi}{2}(\hat{J}_x)_j})\hat{U}^{\dagger}_{\rm QC}$ is straightforward. The action of $\hat{A}^{\dagger}$ on the annihilation operators of the input modes can be represented by the relation $\hat{c}_j'=\sum_k \vect{\mathcal{A}}_{jk}\hat{c}_k$ ($j=1,\dots,2d$) with 
\begin{align}
\vect{\mathcal{A}}=&\frac{1}{\sqrt{2}}\begin{pmatrix}\vect{I}_d&-i\vect{I}_d\\-i\vect{I}_d&\vect{I}_d\end{pmatrix}\begin{pmatrix}\vect{I}_d&0\\0&\vect{U}^{\dagger}\end{pmatrix}=\frac{1}{\sqrt{2}}\begin{pmatrix}\vect{I}_d&-i\vect{U}^{\dagger}\\-i\vect{I}_d&\vect{U}^{\dagger}\end{pmatrix}.\label{network} 
\end{align}
Notice that the matrix $\frac{1}{\sqrt{2}}\begin{pmatrix}\vect{I}_d&-i\vect{I}_d\\-i\vect{I}_d&\vect{I}_d\end{pmatrix}$ in the product above describes an array of balanced beam splitters working in parallel, corresponding to $\otimes_{j=1}^d e^{-i\frac{\pi}{2}(\hat{J}_x)_j}$.
Under the hypotheses of a pure state $\ket{\Psi}$ for the system and of a phase-imprinting transformation of the form $\otimes_{j=1}^d e^{-i \theta_j(\hat{J}_z)_j}$ (see above), the QFIM is given by 
\begin{equation}
\left(\vect{\FQ}\right)_{ij} = 4\left(\langle\Psi\vert(\hat{J}_z)_i
(\hat{J}_z)_j\ket{\Psi}-\langle\Psi\vert(\hat{J}_z)_i \ket{\Psi}\langle\Psi\vert(\hat{J}_z)_j\ket{\Psi}\right) \hspace{0.5cm} (i,j=1,\dots,d).  \nonumber  
\end{equation}
Since $(\hat{J}_z)_j = (\hat{a}_j^{\dagger} \hat{a}_j - \hat{b}_j^{\dagger} \hat{b}_j)/2=(\hat{c}_j^{\dagger} \hat{c}_j-\hat{c}_{j+d}^{\dagger} \hat{c}_{j+d})/2=(\hat{n}_j-\hat{n}_{j+d})/2$, after simple calculations we get to express $\left(\vect{\FQ}\right)_{ij}$ as 
\begin{equation}
    \left(\vect{\FQ}\right)_{ij} = \vect{h}_{i,j} + \vect{h}_{i+d,j+d} - \vect{h}_{i,j+d} - \vect{h}_{i+d, j},  \hspace{0.5cm} (i,j=1,\dots,d)\label{QFIM_hij}
\end{equation}
where the expression of $\vect{h}_{ij}$, $\vect{h}_{ij}=\langle  \Psi \vert \hat{n}_i\hat{n}_j \ket{ \Psi } - \langle  \Psi \vert \hat{n}_i \ket{ \Psi }\langle  \Psi \vert \hat{n}_j \ket{ \Psi }$, is exactly the one already given in Eq. \makeref{hij_general}. Notice that $\vect{\FQ}$ is a $d\times d$ matrix, whose elements, according to Eq. \makeref{QFIM_hij}, can be obtained as combinations of the elements of $\vect{h}$, a $2d\times 2d$ matrix. 

Equation \makeref{hij_final} from the previous paragraph expresses $\vect{h}$ in terms of the two matrices $\vect{E}$ and $\vect{EN}$ and the vector $\vect{\gamma}$. The two matrices are derived by referring to Eqs. from \makeref{N} to \makeref{D}. In particular, from Eqs. \makeref{C} and \makeref{D} we get
{\begin{align}
\vect{C}=\begin{pmatrix}\vect{I}_d&0\\0&\vect{C}_1\end{pmatrix},\label{C_QFIM}
\end{align}}
and
{\begin{align}
\vect{CD}=\begin{pmatrix}0&0\\0&\vect{C}_1\vect{D}_1\end{pmatrix}.\label{CD_QFIM}
\end{align}}
In the above equations, $\vect{C_1}$ and $\vect{C_1D_1}$ are $d \times d$ matrices with elements
\be
(\vect{C_1})_{ij}=\delta_{ij}\left[\left(1-\delta_{Dj}\right)+\delta_{Dj}c^2\right],\label{C1}
\ee
and
\be
(\vect{C_1D_1})_{ij}=\delta_{ij}\delta_{Dj}e^{i\varphi}sc,\label{C1D1}
\ee
respectively,
where $D$ is the index of the input port into which $\ket{\xi}$ is injected, and $s \equiv \sinh{r}$, $c \equiv \cosh{r}$ (these shortcuts will be repeatedly used below). 
On account of Eqs. \makeref{N}, \makeref{E} and \makeref{network} and recalling the fundamental relation $\vect{A}=\vect{\mathcal{A}}^{\dagger}$  discussed above, we have ($\vect{U}\equiv \vect{A_1}$ in the following)
\begin{align}
    \vect{E}=&\vect{A}^{\dagger}\vect{C}\vect{A}=\frac{1}{2}\begin{pmatrix}\vect{I}_d+\vect{E_1}&i(\vect{I}_d-\vect{E_1})\\-i(\vect{I}_d-\vect{E_1})&\vect{I}_d+\vect{E_1}\end{pmatrix}, \nonumber 
\end{align}
with $\vect{E_1}=\vect{A_1}^{\dagger}\vect{C_1}\vect{A_1}$, and
\begin{align}
    \vect{EN}=&\vect{A}^{\dagger}\vect{CD}\vect{A}^*=\frac{1}{2}\begin{pmatrix}-\vect{E_1N_1}&-i\vect{E_1N_1}\\-i\vect{E_1N_1}&\vect{E_1N_1}\end{pmatrix}, \nonumber
\end{align}
with $\vect{E_1N_1}=\vect{A_1}^{\dagger}\vect{C_1D_1}\vect{A_1}^*$.
It is not difficult to see that, when none of the input modes is in a state $\ket{\beta,\xi}$ with both $\beta \neq 0$ and $\xi \neq 0$ -- that is, each mode is either in a coherent or in a squeezed vacuum state -- the expression of $\vect{\gamma}$ can be simplified as $\vect{\gamma}=\vect{b}^*$.
Indeed, if that is the case, then $\vect{E}^*\vect{b}^*=\vect{b}^*$ and $(\vect{EN})^*\vect{b}=0$. In such case one also has $\vect{b}=\vect{A}^{\dagger}\vect{\beta}$ 
and $\vect{\beta}=(\vect{\beta_0}\hspace{0.1cm}0)^T$, with
\begin{align}
\vect{\beta_0}=\begin{pmatrix}\alpha_1\\\vdots\\\alpha_d\end{pmatrix}=\begin{pmatrix}|\alpha_1|e^{i\phi_1}\\\vdots\\|\alpha_d|e^{i\phi_d}\end{pmatrix}. \nonumber
\end{align}
Thus, one finds
\begin{align}
\vect{\gamma}=\frac{1}{\sqrt{2}}\begin{pmatrix}\vect{\beta_0}^*\\i\vect{\beta_0}^*\end{pmatrix}=\frac{1}{\sqrt{2}}\begin{pmatrix}\vect{\gamma_0}\\i\vect{\gamma_0}\end{pmatrix},  \nonumber
\end{align}
where the symbol $\vect{\gamma_0} \equiv \vect{\beta_0}^*$ was introduced. 
We are  interested in the following combinations of $\vect{\gamma}$:
{\begin{equation}
\vect{\gamma}\vect{\gamma}^T=\frac{1}{2}\begin{pmatrix}\vect{\gamma_0}\vect{\gamma_0}^T&i\vect{\gamma_0}\vect{\gamma_0}^T\\i\vect{\gamma_0}\vect{\gamma_0}^T&-\vect{\gamma_0}\vect{\gamma_0}^T\end{pmatrix}, \nonumber
\end{equation}} 
and 
{\begin{equation}
\vect{\gamma}\vect{\gamma}^{\dagger}=\frac{1}{2}\begin{pmatrix}\vect{\gamma_0}\vect{\gamma_0}^{\dagger}&-i\vect{\gamma_0}\vect{\gamma_0}^{\dagger}\\i\vect{\gamma_0}\vect{\gamma_0}^{\dagger}&\vect{\gamma_0}\vect{\gamma_0}^{\dagger}\end{pmatrix}, \nonumber
\end{equation}} 
where
\begin{align}
    \vect{\gamma_0}\vect{\gamma_0}^T=&\begin{pmatrix}|\alpha_1|^2e^{-2i\phi_1}&|\alpha_1||\alpha_2|e^{-i(\phi_1+\phi_2)}&\hdots&|\alpha_1||\alpha_d|e^{-i(\phi_1+\phi_d)}\\
    |\alpha_2||\alpha_1|e^{-i(\phi_2+\phi_1)}&|\alpha_2|^2e^{-2i\phi_2}&\hdots&|\alpha_2||\alpha_d|e^{-i(\phi_2+\phi_d)}\\
    \vdots&\vdots&&\vdots\\
    |\alpha_d||\alpha_1|e^{-i(\phi_d+\phi_1)}&|\alpha_d||\alpha_2|e^{-i(\phi_d+\phi_2)}&\hdots&|\alpha_d|^2e^{-2i\phi_d}\end{pmatrix}\equiv\vect{\phi_+}, \nonumber
\end{align} 
and
\begin{align}
    \vect{\gamma_0}\vect{\gamma_0}^{\dagger}=&\begin{pmatrix}|\alpha_1|^2&|\alpha_1||\alpha_2|e^{-i(\phi_1-\phi_2)}&\hdots&|\alpha_1||\alpha_d|e^{-i(\phi_1-\phi_d)}\\
    |\alpha_2||\alpha_1|e^{-i(\phi_2-\phi_1)}&|\alpha_2|^2&\hdots&|\alpha_2||\alpha_d|e^{-i(\phi_2-\phi_d)}\\
    \vdots&\vdots&&\vdots\\
    |\alpha_d||\alpha_1|e^{-i(\phi_d-\phi_1)}&|\alpha_d||\alpha_2|e^{-i(\phi_d-\phi_2)}&\hdots&|\alpha_d|^2\end{pmatrix}\equiv\vect{\phi_-}. \nonumber
\end{align}

According to Eq. \makeref{hij_final}, which reads
\begin{align}
\vect{h}
=&\vect{EN}\circ (\vect{EN})^*-\vect{EN}\circ\vect{\gamma}\vect{\gamma}^T-(\vect{EN})^*\circ\left(\vect{\gamma}\vect{\gamma}^T\right)^*+\vect{E}\circ \vect{E}^*+\vect{E}\circ\vect{\gamma}\vect{\gamma}^{\dagger}+\vect{E}^*\circ\left(\vect{\gamma}\vect{\gamma}^{\dagger}\right)^*-\left(\vect{E}+\vect{\gamma}\vect{\gamma}^{\dagger}\right)\circ\vect{I}, \nonumber
\end{align}
$\vect{h}$ is obtained by referring to the following matrices:
\begin{align}\label{1_term_QFIM}
    \vect{EN}\circ(\vect{EN})^*=\frac{1}{4}\begin{pmatrix}\vect{E_1N_1}\circ(\vect{E_1N_1})^*&\vect{E_1N_1}\circ(\vect{E_1N_1})^*\\\vect{E_1N_1}\circ(\vect{E_1N_1})^*&\vect{E_1N_1}\circ(\vect{E_1N_1})^*\end{pmatrix},
\end{align}
\begin{align}\label{2_term_QFIM}
    -\vect{EN}\circ\vect{\gamma}\vect{\gamma}^T=\frac{1}{4}\begin{pmatrix}\vect{E_1N_1}\circ\vect{\phi_+}&-\vect{E_1N_1}\circ\vect{\phi_+}\\-\vect{E_1N_1}\circ\vect{\phi_+}&\vect{E_1N_1}\circ\vect{\phi_+}\end{pmatrix},
\end{align}
\begin{align}\label{3_term_QFIM}
    \vect{E}\circ \vect{E}^*=\frac{1}{4}\begin{pmatrix}\vect{I}_d+\left(\vect{E_1}+\vect{E_1}^*\right)\circ\vect{I}_d+\vect{E_1}\circ\vect{E_1}^*&\vect{I}_d-\left(\vect{E_1}+\vect{E_1}^*\right)\circ\vect{I}_d+\vect{E_1}\circ\vect{E_1}^*\\\vect{I}_d-\left(\vect{E_1}+\vect{E_1}^*\right)\circ\vect{I}_d+\vect{E_1}\circ\vect{E_1}^*&\vect{I}_d+\left(\vect{E_1}+\vect{E_1}^*\right)\circ\vect{I}_d+\vect{E_1}\circ\vect{E_1}^*\end{pmatrix},
\end{align}
\begin{align}\label{4_term_QFIM}
    \vect{E}\circ\vect{\gamma}\vect{\gamma}^{\dagger}=\frac{1}{4}\begin{pmatrix}\vect{I}_d\circ\vect{\phi_-}+\vect{E_1}\circ\vect{\phi_-}&\vect{I}_d\circ\vect{\phi_-}-\vect{E_1}\circ\vect{\phi_-}\\\vect{I}_d\circ\vect{\phi_-}-\vect{E_1}\circ\vect{\phi_-}&\vect{I}_d\circ\vect{\phi_-}+\vect{E_1}\circ\vect{\phi_-}\end{pmatrix},
\end{align}
\begin{align}\label{5_term_QFIM}
    -\left(\vect{E}+\vect{\gamma}\vect{\gamma}^{\dagger}\right)\circ\vect{I}_d=\frac{1}{2}\begin{pmatrix}\vect{I}_d+\vect{E_1}\circ\vect{I}_d+\vect{I}_d\circ\vect{\phi_-}&0\\0&\vect{I}_d+\vect{E_1}\circ\vect{I}_d+\vect{I}_d\circ\vect{\phi_-}\end{pmatrix}.
\end{align}
The above set of equations shows that $\vect{h}$ is a sum of $2d \times 2d$ matrices of the form $\begin{pmatrix}\vect{X}^{(k)}&\vect{Y}^{(k)}\\\vect{Y}^{(k)}&\vect{X}^{(k)}\end{pmatrix}$, $\vect{X}^{(k)}$, $\vect{Y}^{(k)}$ being $d \times d$ matrices. Consequently, computing the two sums $\vect{X}=\sum_{k}\vect{X}^{(k)}$ and $\vect{Y}=\sum_{k}\vect{Y}^{(k)}$ is enough to determine $\vect{h}$ completely. We find

\begin{align}
    \vect{X}=\frac{1}{4}\vect{E_1N_1}\circ(\vect{E_1N_1})^*+\frac{1}{4}\vect{E_1}\circ \vect{E_1}^*+\frac{1}{4}\left[\vect{E_1N_1}\circ\vect{\phi_+}+(\vect{E_1N_1})^*\circ\vect{\phi_+}^*\right]+\frac{1}{4}(\vect{E_1}\circ\vect{\phi_-}+\vect{E_1}^*\circ\vect{\phi_-}^*)
    +\frac{1}{4}(\vect{E_1}+\vect{E_1}^*)\circ\vect{I}_d-\frac{1}{2}\vect{E_1}\circ\vect{I}_d-\frac{1}{4}\vect{I}_d \nonumber
\end{align}
and
\begin{align}
    \vect{Y}=\frac{1}{4}\vect{E_1N_1}\circ(\vect{E_1N_1})^*+\frac{1}{4}\vect{E_1}\circ \vect{E_1}^*-\frac{1}{4}\left[\vect{E_1N_1}\circ\vect{\phi_+}+(\vect{E_1N_1})^*\circ\vect{\phi_+}^*\right]-\frac{1}{4}(\vect{E_1}\circ\vect{\phi_-}+\vect{E_1}^*\circ\vect{\phi_-}^*)-\frac{1}{4}(\vect{E_1}+\vect{E_1}^*)\circ\vect{I}_d+\frac{1}{2}\vect{I}_d\circ\vect{\phi_-}+\frac{1}{4}\vect{I}_d. \nonumber
\end{align}
Finally, from Eq. \makeref{QFIM_hij}, and using 
$\vect{h}_{i,j}=\vect{h}_{i+d,j+d}=\vect{X}_{ij}$
and 
$\vect{h}_{i,j+d}=\vect{h}_{i+d,j}=\vect{Y}_{ij}$,
we find
\begin{align}
    \vect{\FQ}=&2\vect{X}-2\vect{Y}
    =\vect{E_1N_1}\circ\vect{\phi_+}^*+(\vect{E_1N_1})^*\circ\vect{\phi_+}^*+\vect{E_1}\circ\vect{\phi_-}+\vect{E_1}^*\circ\vect{\phi_-}^*-\vect{\phi_-}\circ\vect{I}_d+\vect{E_1}^*\circ\vect{I}_d-\vect{I}_d.\label{QFIM_Mach_Zender_2}
\end{align}
Furthermore, we use
\beq
    (\vect{E_1N_1})_{ij}&=&\left(\vect{A_1}^{\dagger}\vect{C_1D_1}\vect{A_1}^*\right)_{ij}
    =\sum_{kl}\left(\vect{U}^{\dagger}\right)_{ik}\left(e^{i\varphi}sc\right)\delta_{kl}\delta_{Dl}\left(\vect{U}^*\right)_{lj}
    =e^{i\varphi}sc\hspace{0.1cm}\left(\vect{U}^{\dagger}\right)_{iD}\left(\vect{U}^*\right)_{Dj}
    \nonumber \\
    &=&e^{i\varphi}sc\hspace{0.1cm}\left(\vect{U}^*\right)_{Di}\left(\vect{U}^*\right)_{Dj} \nonumber
\eeq
and 
\beq 
    (\vect{E_1})_{ij}&=&\left(\vect{A_1}^{\dagger}\vect{C_1}\vect{A_1}\right)_{ij}=\sum_{kl}\left(\vect{U}^{\dagger}\right)_{ik}\delta_{kl}\left[\left(1-\delta_{Dl}\right)+\delta_{Dl}c^2\right]\left(\vect{U}\right)_{lj}=\sum_{kl}\left(\vect{U}^{\dagger}\right)_{ik}\delta_{kl}\left(\vect{U}\right)_{lj}+\sum_{kl}\left(\vect{U}^{\dagger}\right)_{ik}(c^2-1)\delta_{kl}\delta_{Dl}\left(\vect{U}\right)_{lj} \nonumber \\ &=&\delta_{ij}+s^2\left(\vect{U}^*\right)_{Di}\left(\vect{U}\right)_{Dj}, \nonumber
\eeq
derived on the basis of Eq. \makeref{C1D1} and \makeref{C1}, respectively.
Also, since
$(\vect{\phi_{+}})_{ij}=|\alpha_i||\alpha_j|e^{-i\phi_i}e^{-i\phi_j}$,
and
$(\vect{\phi_{-}})_{ij}=|\alpha_i||\alpha_j|e^{-i\phi_i}e^{i\phi_j}$,
one has
\begin{equation}
    (\vect{E_1N_1}\circ\vect{\phi_{+}})_{ij}=e^{i\varphi}sc\hspace{0.1cm}\left(\vect{U}^*\right)_{Di}\left(\vect{U}^*\right)_{Dj}|\alpha_i||\alpha_j|e^{-i\phi_i}e^{-i\phi_j},\label{first_term} 
\end{equation}
and
\begin{align}
    (\vect{E_1}\circ\vect{\phi_{-}})_{ij}=&\left[\delta_{ij}+s^2\left(\vect{U}^*\right)_{Di}\left(\vect{U}\right)_{Dj}\right]|\alpha_i||\alpha_j|e^{-i\phi_i}e^{i\phi_j}=|\alpha_i|^2\delta_{ij}+s^2\left(\vect{U}^*\right)_{Di}\left(\vect{U}\right)_{Dj}|\alpha_i||\alpha_j|e^{-i\phi_i}e^{i\phi_j}.\label{second_term}
\end{align}
For the sake of a lighter notation, we set $(\vect{U})_{Di} \equiv u_i$; the same notation was used in the main text. Plugging Eqs. \makeref{first_term} and \makeref{second_term} into Eq. \makeref{QFIM_Mach_Zender_2} and by some algebraic manipulation, we find
\begin{align}
    (\vect{\FQ})_{ij}=&|\alpha_i||\alpha_j|\left[\left(e^{-i\varphi}u_ie^{i\phi_i} u_je^{i\phi_j}+e^{i\varphi}u^*_ie^{-i\phi_i} u^*_je^{-i\phi_j}\right)sc+2\delta_{ij}+\left(u^*_ie^{-i\phi_i} u_je^{i\phi_j}+u_ie^{i\phi_i} u^*_je^{-i\phi_j}\right)s^2-\delta_{ij}\right]+|u_i|^2\delta_{ij}s^2 \nonumber \\
    =&|\alpha_i||\alpha_j|\left[\Re\left(e^{i\chi_i}u_i\right)\Re\left(e^{i\chi_j}u_j\right)\left(e^{2r}-1\right)+\Im\left(e^{i\chi_i}u_i\right)\Im\left(e^{i\chi_j}u_j\right)\left(e^{-2r}-1\right)\right]+\left(|\alpha_i|^2+|u_i|^2\bar{n}_s\right)\delta_{ij}\label{QFIM_final}, 
\end{align}
where $\chi_i \equiv \phi_i-\varphi/2$ and $s^2 \equiv \sinh^2 r=\bar{n}_s$. By imposing the condition $\Im(e^{i\chi_j}u_j)= 0$, for $j=1, ..., d$, one finally recovers Eq. \makeref{QFIM} of the main text. 

\subsubsection{Inverse moment matrix, Eq. (\ref{MOM})}

The so called \textit{moment matrix} corresponds to the covariance matrix of a particular set of estimators of the unknown parameters $\theta_i$, $i=1,\dots,d$. It is defined as follows \cite{GessnerNATCOMM2020}:
\begin{equation}
    \vect{\mathcal{M}}=\vect{G}^T\vect{\Gamma}^{-1}\vect{G},
\end{equation}
where
\begin{equation}
    \vect{G}_{ij}=\frac{\partial \langle\hat{X}_i\rangle_{\rho(\vect{\theta})}}{\partial \theta_j} \hspace{0.5cm} (i=1,\dots,K; \hspace{0.1cm} j=1,\dots,d), \label{G}
\end{equation}
and
\begin{equation}
    \vect{\Gamma}_{ij}=\langle \hat{X}_i\hat{X}_j\rangle_{\rho(\vect{\theta})}-\langle\hat{X}_i\rangle_{\rho(\vect{\theta})}\langle\hat{X}_j\rangle_{\rho(\vect{\theta})} \hspace{0.5cm} (i,j=1,\dots,K), \label{Gamma}
\end{equation}
$\rho(\vect{\theta})$ being the output state of the whole sensor network. The $\hat{X}_i$, $i=1,\dots,K$, are $K$ hermitian operators which correspond to measurements performed on the output state $\rho(\vect{\theta})$. We choose $\hat{X}_i \equiv (\hat{J}_z)_i=\left(\hat{a}_i^{\dagger}\hat{a}_i-\hat{b}_i^{\dagger}\hat{b}_i\right)/2$, with $i=1,\dots,d$. In such case, the two matrices $\vect{G}$ and $\vect{\Gamma}$ can be evaluated through the \textit{Q function}-based technique already illustrated in the previous section. This time, we need to take into account the complete transformation performed by the network on the input state, including the phase--encoding stage. Such transformation is described by the matrix 
\begin{align} \label{transformation_moment_matrix}
    \vect{A}^{\dagger}=&\begin{pmatrix}\vect{\tilde{C}}&\vect{\tilde{S}}\\-\vect{\tilde{S}}&\vect{\tilde{C}}\end{pmatrix}\begin{pmatrix}\vect{I}_d&0\\0&\vect{A_1}^{\dagger}\end{pmatrix}
    = \begin{pmatrix}\vect{\tilde{C}}&\vect{\tilde{S}}\vect{A_1}^{\dagger}\\-\vect{\tilde{S}}&\vect{\tilde{C}}\vect{A_1}^{\dagger}\end{pmatrix},
\end{align}
where $\vect{\tilde{C}}_{ij}=\cos{\left(\theta_i/2\right)}\delta_{ij}$ and $\vect{\tilde{S}}_{ij}=\sin{\left(\theta_i/2\right)}\delta_{ij}$.
Let $\ket{\Psi(\vect{\theta})}$ be the output state of our MZI sensor network. On the basis of Eq. \makeref{Gamma} and having set $\hat{X}_i \equiv (\hat{J}_z)_i$, we have
\begin{equation}
\vect{\Gamma}_{ij} =\langle\Psi(\vect{\theta})\vert(\hat{J}_z)_i
(\hat{J}_z)_j\ket{\Psi(\vect{\theta})}-\langle\Psi(\vect{\theta})\vert(\hat{J}_z)_i\ket{\Psi(\vect{\theta}})\langle\Psi(\vect{\theta})\vert(\hat{J}_z)_j\ket{\Psi(\vect{\theta})} \hspace{0.5cm} (i,j=1,\dots,d),
\end{equation}
which can be rewritten as 
\begin{equation}
    \vect{\Gamma}_{ij} =\frac{1}{4} \left(\vect{h}_{i,j} + \vect{h}_{i+d,j+d} - \vect{h}_{i,j+d} - \vect{h}_{i+d, j}\right)  \hspace{0.5cm} (i,j=1,\dots,d),
\end{equation}
where $\vect{h}_{ij}=\langle\Psi(\vect{\theta})\vert\hat{n}_i\hat{n}_j \ket{\Psi(\vect{\theta})}-\langle\Psi(\vect{\theta})\vert\hat{n}_i\ket{\Psi(\vect{\theta})}\langle\Psi(\vect{\theta})\vert\hat{n}_j\ket{\Psi(\vect{\theta})}$. Notice that $\vect{h}_{ij}$ can still be computed by means of Eq. \makeref{hij_final}, provided that the expression of matrix $\vect{A}^{\dagger}$ is that given in Eq. \makeref{transformation_moment_matrix}. Matrices $\vect{C}$ and $\vect{CD}$, on the other hand, being only related to the form of the input state, have the same expressions as in Eqs. from \makeref{C_QFIM} to \makeref{C1D1}. So we get
\begin{align}
    \vect{E}=\vect{A}^{\dagger}\vect{C}\vect{A}=\begin{pmatrix}\vect{\tilde{C}}^2+\vect{\tilde{S}}\vect{E_1}\vect{\tilde{S}}&-\left(\vect{\tilde{S}}\vect{\tilde{C}}-\vect{\tilde{S}}\vect{E_1}\vect{\tilde{C}}\right)\\-\left(\vect{\tilde{S}}\vect{\tilde{C}}-\vect{\tilde{C}}\vect{E_1}\vect{\tilde{S}}\right)&\vect{\tilde{S}}^2+\vect{\tilde{C}}\vect{E_1}\vect{\tilde{C}}\end{pmatrix}\label{E_moment_matrix}
\end{align}
and
\begin{align}
    \vect{EN}=&\vect{A}^{\dagger}\vect{CD}\vect{A}^*=\begin{pmatrix}\vect{\tilde{S}}\vect{E_1N_1}\vect{\tilde{S}}&\vect{\tilde{S}}\vect{E_1N_1}\vect{\tilde{C}}\\\vect{\tilde{C}}\vect{E_1N_1}\vect{\tilde{S}}&\vect{\tilde{C}}\vect{E_1N_1}\vect{\tilde{C}}\end{pmatrix}. \nonumber
\end{align}
We also have 
\begin{align}
\vect{\gamma}=&\vect{b}^*=\begin{pmatrix}\vect{\tilde{C}}\vect{\beta_0}^*\\-\vect{\tilde{S}}\vect{\beta_0}^*\end{pmatrix}=\begin{pmatrix}\vect{\tilde{C}}\vect{\gamma_0}\\-\vect{\tilde{S}}\vect{\gamma_0}\end{pmatrix}, \nonumber
\end{align}
so that
\begin{align}
\vect{\gamma}\vect{\gamma}^{\dagger}=\begin{pmatrix}\vect{\tilde{C}}\vect{\phi_-}\vect{\tilde{C}}&-\vect{\tilde{C}}\vect{\phi_-}\vect{\tilde{S}}\\-\vect{\tilde{S}}\vect{\phi_-}\vect{\tilde{C}}&\vect{\tilde{S}}\vect{\phi_-}\vect{\tilde{S}}\end{pmatrix},\label{gammad_moment_matrix}
\end{align}
and
\begin{align}
\vect{\gamma}\vect{\gamma}^T=\begin{pmatrix}\vect{\tilde{C}}\vect{\phi_+}\vect{\tilde{C}}&-\vect{\tilde{C}}\vect{\phi_+}\vect{\tilde{S}}\\-\vect{\tilde{S}}\vect{\phi_+}\vect{\tilde{C}}&\vect{\tilde{S}}\vect{\phi_+}\vect{\tilde{S}}\end{pmatrix}. \nonumber
\end{align}
At this point, we set $\theta_i=\pi/2$, a choice which is expected to lead to minimum estimation uncertainty [maximum slope condition, see Eq. (\ref{Gij}) below]. 
We thus obtain the same matrices as in Eqs. \makeref{1_term_QFIM}-\makeref{5_term_QFIM}, while, for what concerns Eq. \makeref{2_term_QFIM}, we find
\begin{align}
    -\vect{EN}\circ\vect{\gamma}\vect{\gamma}^T=\frac{1}{4}\begin{pmatrix}-\vect{E_1N_1}\circ\vect{\phi_+}&\vect{E_1N_1}\circ\vect{\phi_+}\\\vect{E_1N_1}\circ\vect{\phi_+}&-\vect{E_1N_1}\circ\vect{\phi_+}\end{pmatrix}, \nonumber
\end{align}
which has opposite sign with respect to the corresponding matrix used in the QFIM case. As a consequence, we eventually obtain an expression for $\vect{\Gamma}$ which is almost identical to the QFIM, the most notable difference being the swap of the real and imaginary part of the term $e^{i\chi_i}u_i$ (compare with Eq. \makeref{QFIM_final}):
\begin{align} \label{Gamma_final}
    \vect{\Gamma}_{ij}=\frac{1}{4}\left\{|\alpha_i||\alpha_j|\left[\Im\left(e^{i\chi_i}u_i\right)\Im\left(e^{i\chi_j}u_j\right)\left(e^{2r}-1\right)+\Re\left(e^{i\chi_i}u_i\right)\Re\left(e^{i\chi_j}u_j\right)\left(e^{-2r}-1\right)\right]+\left(|\alpha_i|^2+|u_i|^2\bar{n}_s\right)\delta_{ij}\right\}.
\end{align}
Consider now Eq. \makeref{G}, rewritten for the specific choice $\hat{X}_i\equiv(\hat{J}_z)_i$:
\begin{align} \label{Gij}
    \vect{G}_{ij}=&\frac{\partial \langle\Psi(\vect{\theta})\vert(\hat{J}_z)_i\ket{\Psi(\vect{\theta})}}{\partial \theta_j}=\frac{1}{2}\frac{\partial \langle\Psi(\vect{\theta})\vert(\hat{n}_i-\hat{n}_{i+d})\ket{\Psi(\vect{\theta})}}{\partial \theta_i}\delta_{ij}, \hspace{0.5cm} (i=1,\dots,d). 
\end{align}
A formula for the evaluation of $\langle\Psi(\vect{\theta})\vert\hat{n}_i\ket{\Psi(\vect{\theta})}$ was derived in Eq. \makeref{ni}:
\begin{align}
    \langle\Psi(\vect{\theta})\vert\hat{n}_i\ket{\Psi(\vect{\theta})}=-1+\vect{E}_{ii}+\vect{\gamma}_i\vect{\gamma}_i^* \hspace{0.5cm} (i=1,\dots,2d). \nonumber
\end{align}
The same information is conveniently condensed in the (diagonal) matrix 
\begin{align}
    -\vect{I}_d+\left(\vect{E}+\vect{\gamma}\vect{\gamma}^{\dagger}\right)\circ\vect{I}_d, \nonumber
\end{align}
whose diagonal elements are the $\langle\Psi(\vect{\theta})\vert\hat{n}_i\ket{\Psi(\vect{\theta})}$, $i=1, \dots, 2d$. 
Making use of Eqs. \makeref{E_moment_matrix} and \makeref{gammad_moment_matrix}, one finds
\begin{align}
    -\vect{I}_d+\left(\vect{E}+\vect{\gamma}\vect{\gamma}^{\dagger}\right)\circ\vect{I}_d=&\begin{pmatrix}-\vect{I}_d+\vect{\tilde{C}}^2+\left(\vect{\tilde{S}}\vect{E_1}\vect{\tilde{S}}\right)\circ\vect{I}_d+\vect{\tilde{C}}\vect{\phi_-}\vect{\tilde{C}}&0\\0&-\vect{I}_d+\vect{\tilde{S}}^2+\left(\vect{\tilde{C}}\vect{E_1}\vect{\tilde{C}}\right)\circ\vect{I}_d+\vect{\tilde{S}}\vect{\phi_-}\vect{\tilde{S}}\end{pmatrix} \nonumber \\ &\begin{pmatrix}-\vect{I}_d+\vect{\tilde{C}}^2+\vect{\tilde{S}}^2\left(\vect{E_1}\circ\vect{I}_d\right)+\vect{\tilde{C}}\vect{\phi_-}\vect{\tilde{C}}&0\\0&-\vect{I}_d+\vect{\tilde{S}}^2+\vect{\tilde{C}}^2\left(\vect{E_1}\circ\vect{I}_d\right)+\vect{\tilde{S}}\vect{\phi_-}\vect{\tilde{S}}\end{pmatrix}, \nonumber
\end{align}
where the general result $(\vect{Z}\vect{Y}\vect{Z})\circ\vect{I}_d=\vect{Z}^2(\vect{Y}\circ\vect{I}_d)$ has been used, $\vect{Z}$ being a diagonal matrix. Note that, for $i=1,\dots,d$, the expectation value  $\langle\Psi(\vect{\theta})\vert(\hat{n}_i-\hat{n}_{i+d})\ket{\Psi(\vect{\theta})}$ corresponds to the difference between the two diagonal blocks of this matrix, so that, after simple algebraic manipulation, we get
\begin{align}
    \vect{G}_{ij}=&\frac{1}{2}\frac{\partial \langle\Psi(\vect{\theta})\vert(\hat{n}_i-\hat{n}_{i+d})\ket{\Psi(\vect{\theta})}}{\partial \theta_i}\delta_{ij}
    =\frac{1}{2}\sin\theta_i\left[\vect{E_1}\circ\vect{I}_d-\left(1+|\alpha_i|^2\right)\right]_{ii}\delta_{ij}.\nonumber
\end{align}
We set again $\theta_i=\pi/2$, ending up with: 
\begin{align}\label{G_final}
    \vect{G}_{ij}=\frac{1}{2}\left(|u_i|^2s^2-|\alpha_i|^2\right)\delta_{ij}.
\end{align}
At this point, if we aimed at obtaining the explicit expression of the moment matrix, we would still have to compute the inverse of matrix $\vect{\Gamma}$. On the other hand, we only need $\vect{G}^{-1}$ to derive $\vect{M}^{-1}=(\vect{G}^{-1})^T\vect{\Gamma}\vect{G}^{-1}$, the advantage being that $\vect{G}$ is diagonal and thus easily inverted. From Eqs. \makeref{Gamma_final} and \makeref{G_final}, one readily gets
\begin{align}
    \left(\vect{M}^{-1}\right)_{ij}=&|\alpha_i||\alpha_j|\frac{\Im{(e^{i\chi_i}u_i)}} {|\alpha_i|^2-|u_i|^2\bar{n}_s}\frac{\Im{(e^{i\chi_j} u_j)}}{|\alpha|_j^2-|u_j|^2\bar{n}_s}(e^{2r}-1)+|\alpha_i||\alpha_j|\frac{\Re{(e^{i\chi_i}u_i)}} {|\alpha_i|^2-|u_i|^2\bar{n}_s}\frac{\Re{(e^{i\chi_j} u_j)}}{|\alpha|_j^2-|u_j|^2\bar{n}_s}(e^{-2r}-1)+\frac{|\alpha_i|^2+|u_i|^2\bar{n}_s}{(|\alpha_i|^2-|u_i|^2\bar{n}_s)^2}\delta_{ij},
\end{align}
where $\chi_i \equiv \phi_i-\varphi/2$ and $s^2 \equiv \sinh^2 r=\bar{n}_s$. 
Under the condition $\Im(e^{i\chi_j}u_j)= 0$, for $j=1, ..., d$, one finally recovers Eq. \makeref{MOM}. 

\end{widetext}

\subsection{Demonstration of Eqs. (\ref{ubound3LB}) and (\ref{ubound3UB})}
\label{A}

First, let us recall the general expression of $\Delta^2(\vect{v} \cdot \vect{\theta})_{\rm emom}$, obtained using Eq.~(\ref{MOM}):
\beq \label{Eq1APPA}
&& \Delta^2(\vect{v} \cdot \vect{\theta})_{\rm emom} = \vect{v}^T\vect{\mathcal{M}}(\tilde{\vect{u}}, \vert \alpha_1 \vert^2, ..., \vert \alpha_d \vert^2, \bar{n}_s )^{-1}\vect{v} \nonumber \\
&& \quad = (e^{-2r}-1)\left(\sum_{j=1}^d \frac{|\alpha_j|\tilde{u}_j v_j}{|\alpha_j|^2-\tilde{u}_j^2\bar{n}_s}\right)^2+\sum_{j=1}^d \frac{|\alpha_j|^2+\tilde{u}_j^2\bar{n}_s}{\left(|\alpha_j|^2-\tilde{u}_j^2\bar{n}_s\right)^2}v_j^2. \nonumber \\
\eeq
Also, as a general result, we notice that the minimization problem $\min_{\vect{U}}\Delta^2(\vect{v} \cdot \vect{\theta})_{\rm emom}$ is characterized by a symmetry of a simple type: the value of the minimum is invariant under the transformation $v_i\to-v_i$ performed on an arbitrary number of components of $\vect{v}$. 
This can be proved as follows. Let $\vect{\tilde{u}}_{\vect{v}}$ be the specific vector $\vect{\tilde{u}}$ that minimizes $\Delta^2(\vect{v} \cdot \vect{\theta})_{\rm emom}$. Inverting the sign of $v_i$ for a certain $i\in\{1,\dots,d\}$ [$v_i'=-v_i$] will leave the second term in the second line of Eq. \makeref{Eq1APPA} unaltered, only the first term will change. However, it is easy to realize that the only effect of this transformation is to modify the minimum point of $\Delta^2(\vect{v}' \cdot \vect{\theta})_{\rm emom}$, which is now $\vect{\tilde{u}}_{\vect{v'}}$ with $(\vect{\tilde{u}}_{\vect{v'}})_i=-(\vect{\tilde{u}}_{\vect{v}})_i$, while the minimum itself will remain the same as for $\Delta^2(\vect{v} \cdot \vect{\theta})_{\rm emom}$. Also, this argument can be immediately extended to a transformation that inverts the sign of an arbitrary number of components of $\vect{v}$. 
The same symmetry is easily seen to apply also to $\min_{ r_1', \dots, r_d'}\Delta^2(\vect{v} \cdot \vect{\theta})_{\rm smom}$, the sequential strategy case.

The lower bound (\ref{ubound3LB}) is obtained by noticing that
\be
\min_{\vect{U}, \vert \alpha_1 \vert^2, ... \vert \alpha_d \vert^2}\Delta^2(\vect{v} \cdot \vect{\theta})_{\rm emom} \geq \min_{\vect{U}, \vert \alpha_1 \vert^2, ... \vert \alpha_d \vert^2}\Delta^2(\vect{v}_j \cdot \vect{\theta})_{\rm emom} \nonumber 
\ee
where we have indicated $\vect{v}_j$ as the vector with elements $v_j = 1/\sqrt{d}$ and $v_{i\neq j}=0$ and corresponds to the estimation of a single phase shift $\vect{v}_j \cdot \vect{\theta} = \theta_j/\sqrt{d}$.
This inequality is supported by numerical evidence, see Fig.~\ref{fig3}.
Specifically, from Eq.~(\ref{Eq1APPA}), we have
\beq \label{AppBeq}
\Delta^2(\vect{v}_j \cdot \vect{\theta})_{\rm emom} &=& \frac{ (e^{-2r}-1)|\alpha_j|^2\tilde{u}_j^2}{d(|\alpha_j|^2-\tilde{u}_j^2\bar{n}_s)^2}+ \frac{|\alpha_j|^2+\tilde{u}_j^2\bar{n}_s}{d(|\alpha_j|^2-\tilde{u}_j^2\bar{n}_s)^2} \nonumber \\
&& \approx 
\frac{(e^{-2r}-1) \tilde{u}_j^2 +1}{d|\alpha_j|^2}
+ \frac{\tilde{u}_j^2\bar{n}_s}{d|\alpha_j|^4}, 
\eeq
where the approximate expression in the second line is obtained for $|\alpha_j|^2 \gg \bar{n}_s$.
Rewriting the numerator of Eq.~(\ref{AppBeq}) as $|\alpha_j|^2+[\bar{n}_s - (1 - e^{-2r})|\alpha_j|^2]\tilde{u}_j^2$, we see that the minimum in the interval $0\leqslant\tilde{u}^2_j\leqslant1$ is clearly achieved for $\tilde{u}^2_j=1$ if the term between square brackets is negative, namely $|\alpha_j|^2>\bar{n}_s/(1-e^{-2r})$, or simply $|\alpha_j|^2>\bar{n}_s$ for $r\gg 1$.
Moreover, the minimum of Eq.~(\ref{AppBeq}) with respect to $|\alpha_j|^2$ is clearly obtained by setting $|\alpha_j|^2$ to its maximum value, which corresponds to $|\alpha_j|^2=d\bar{n}_c$. Thus, we have
\be
\min_{\vect{U}, \vert \alpha_1 \vert^2, ... \vert \alpha_d \vert^2}\Delta^2(\vect{v}_j \cdot \vect{\theta})_{\rm emom} = \frac{e^{-2r}}{d^2\bar{n}_c}
+ \frac{\bar{n}_s}{d^3\bar{n}_c^2}. \nonumber
\ee
We recall that the above equation has been derived under the condition $|\alpha_j|^2 = d\bar{n}_c \gg \bar{n}_s$ that implies $\bar{n}_T = d \bar{n}_c + \bar{n}_s \approx d\bar{n}_c$ and this $\bar{n}_T \gg \bar{n}_s$.
We thus recover Eq.~(\ref{ubound3LB}). 

The upper bound~(\ref{ubound3UB}) is obtained for the sub-optimal conditions i) $\vert \alpha_j \vert^2 = \bar{n}_c$ for all $j$, where $\bar{n}_c = (\bar{n}_T - \bar{n}_s)/d$, and ii) $\vect{\tilde{u}} = \vect{v} \sqrt{d}$.
The bound reads 
\beq
&& \min_{\vect{U}, \vert \alpha_1 \vert^2, ... \vert \alpha_d \vert^2}\Delta^2(\vect{v} \cdot \vect{\theta})_{\rm emom} \leq \vect{v}^T\vect{\mathcal{M}}(\vect{v}\sqrt{d}, \bar{n}_c, \bar{n}_s )^{-1}\vect{v}\nonumber \\
&& = d\bar{n}_c (e^{-2r}-1)\left(\sum_{j=1}^d \frac{v_j^2}{\bar{n}_c-dv_j^2\bar{n}_s}\right)^2+\sum_{j=1}^d \frac{\bar{n}_c+dv_j^2\bar{n}_s}{(\bar{n}_c-dv_j^2\bar{n}_s)^2}v_j^2. \nonumber
\eeq
Taking into account the normalization $\sum_{j=1}^d v_j^2 = 1/d$, we have $d v_j^2 \leq 1$ for all $j$. 
Therefore, 
\be
\left(\sum_{j=1}^d \frac{v_j^2}{\bar{n}_c-dv_j^2\bar{n}_s}\right)^2 \leq \frac{1}{d^2 (\bar{n}_c - \bar{n}_s)^2} \nonumber 
\ee
and 
\be
\sum_{j=1}^d \frac{\bar{n}_c+dv_j^2\bar{n}_s}{(\bar{n}_c-dv_j^2\bar{n}_s)^2}v_j^2 \leq \frac{d \bar{n}_c + \mathcal{W} \bar{n}_s}{d^2 (\bar{n}_c - \bar{n}_s)^2}, \nonumber
\ee
where $\mathcal{W} = d^3 \sum_{j=1}^d v_j^4$.
Notice that $\mathcal{W} \geq 1$, with $\mathcal{W} = 1$ for $v_j= \pm 1/d$ for all $j$.
Combining the above equations and using $d \bar{n}_c = \bar{n}_T - \bar{n}_s$, 
we find
\be
\min_{\vect{U}, \vert \alpha_1 \vert^2, ... \vert \alpha_d \vert^2}\Delta^2(\vect{v} \cdot \vect{\theta})_{\rm emom} 
\leq 
\frac{(\bar{n}_T-\bar{n}_s) e^{-2r} + \bar{n}_s \mathcal{W}}{[\bar{n}_T - (d+1) \bar{n}_s]^2}.
\nonumber
\ee
Finally, we recover  Eq.~(\ref{ubound3UB}) under the condition $\bar{n}_T \gg (d+1) \bar{n}_s$.

\subsection{Demonstration of Eqs.~(\ref{ubound7LB}) and~(\ref{ubound7UB})}

The lower bound to $\min_{\vect{U}, \vert \alpha_1 \vert^2, ..., \vert \alpha_d \vert^2} \Delta^2(\vect{v} \cdot \vect{\theta})_{\rm eQCR}$ is obtained by first using the Cauchy-Schwarz inequality 
\be \label{CZ}
\Delta^2(\vect{v} \cdot \vect{\theta})_{\rm eQCR} = \vect{v}^T \vect{\FQ}^{-1}\vect{v} \geq \frac{\vert \vect{v} \vert^4}{(\vect{v}^T \vect{\FQ}\vect{v})}.
\ee
From Eq.~(\ref{QFIM}) we then have 
\be \label{AppB1}
\vect{v}^T \vect{\FQ}\vect{v} = (e^{2r} -1) \bigg(\sum_{j=1}^d \vert \alpha_j \vert \tilde{u}_j v_j \bigg)^2 + \sum_{j=1}^d \vert \alpha_j \vert^2 v_j^2 + \bar{n}_s \sum_{j=1}^d \tilde{u}_j^2 v_j^2.
\ee
We have
\be \label{AppB2}
\bigg(\sum_{j=1}^d \vert \alpha_j \vert \tilde{u}_j v_j \bigg)^2 \leq \bigg(\sum_{j=1}^d v_j^2 \bigg) \bigg(\sum_{j=1}^d  \vert \alpha_j \vert^2 \tilde{u}_j^2 \bigg) \leq \frac{\sum_{j=1}^d  \vert \alpha_j \vert^2}{d}, \nonumber
\ee
where the first inequality is due to Cauchy-Schwarz and the second is a consequence of $\tilde{u}_j^2 \leq 1$ and $\sum_{j=1}^d v_j^2 = 1/d$.
Using $v_j^2 \leq 1/d$, we also have 
\be \label{AppB3}
\sum_{j=1}^d \vert \alpha_j \vert^2 v_j^2 \leq \frac{\sum_{j=1}^d \vert \alpha_j \vert^2}{d}, \nonumber
\ee
and 
\be \label{AppB4}
\sum_{j=1}^d \tilde{u}_j^2 v_j^2 \leq \frac{1}{d}. \nonumber
\ee
Combining Eq.~(\ref{AppB1}) with the above inequalities gives
\be
\vect{v}^T \vect{\FQ}\vect{v} \leq  \frac{e^{2r} \sum_{j=1}^d \vert \alpha_j \vert^2+\bar{n}_s}{d}. \nonumber
\ee
Taking into account that $\sum_{j=1}^d \vert \alpha_j \vert^2 = \bar{n}_T - \bar{n}_s$ and $\vert \vect{v} \vert^4 = 1/d^2$, from Eq. (\ref{CZ}), we obtain
\be 
\Delta^2(\vect{v}\cdot \vect{\theta})_{\rm eQCR} \geq 
\frac{1}{d[\bar{n}_T e^{2r} - \bar{n}_s (e^{2r}-1)]}. \nonumber
\ee
The lower bound is valid for all $\vect{U}$ and all $\vert \alpha_1 \vert^2, ..., \vert \alpha_d \vert^2$ and thus also for the optimal configuration, 
\be
\min_{\vect{U}, \, \vert \alpha_1 \vert^2, ..., \vert \alpha_d \vert^2} \Delta^2(\vect{v} \cdot \vect{\theta})_{\rm eQCR} \geq \frac{1}{d[\bar{n}_T e^{2r} - \bar{n}_s (e^{2r}-1)]},
\ee
which corresponds to the lower bound (\ref{ubound7LB}).

To derive Eq. (\ref{ubound7UB}), we first use the Sherman-Morrison formula to invert Eq.~(\ref{QFIM}):
\begin{align} \label{FQinv}
    \vect{\FQ}^{-1}=&\frac{1}{|\alpha_i|^2+\tilde{u}_i^2\bar{n}_s}\delta_{ij}\nonumber\\&-\frac{\left(e^{2r}-1\right)}{1+\mathcal{K}\left(e^{2r}-1\right)}\frac{|\alpha_i||\alpha_j|\tilde{u}_i\tilde{u}_j}{\left(|\alpha_i|^2+\tilde{u}_i^2\bar{n}_s\right)\left(|\alpha_j|^2+\tilde{u}_j^2\bar{n}_s\right)},
\end{align}
where $\mathcal{K}=\sum_{j=1}^d \frac{|\alpha_j|^2\tilde{u}_j^2}{|\alpha_j|^2+\tilde{u}_j^2\bar{n}_s}$.
From Eq. (\ref{FQinv}), it is possible to derive the general expression for $\Delta^2(\vect{v} \cdot \vect{\theta})_{\rm eQCR}$: 
\beq \label{EqAppB}
&& \Delta^2(\vect{v} \cdot \vect{\theta})_{\rm eQCR} = \vect{v}^T\vect{\FQ}(\vect{\tilde{u}}, \vert \alpha_1 \vert^2, ..., \vert \alpha_d \vert^2, \bar{n}_s )^{-1}\vect{v} \nonumber \\
&& \quad = \sum_{j=1}^d \frac{v_j^2}{|\alpha_j|^2+\tilde{u}_j^2\bar{n}_s}-\frac{\left(e^{2r}-1\right)}{1+\mathcal{K}\left(e^{2r}-1\right)}\left(\sum_{j=1}^d \frac{|\alpha_j|\tilde{u}_j v_j}{|\alpha_j|^2+\tilde{u}_j^2\bar{n}_s}\right)^2. \nonumber \\
\eeq
Similar to the derivation of  Eq.~(\ref{ubound3UB}), the upper bound (\ref{ubound7UB}) is obtained by a specific configuration of the sensor network that simplifies the above equation: i) $\vert \alpha_j \vert^2 = \bar{n}_c$ for all $j$, where $\bar{n}_c = (\bar{n}_T - \bar{n}_s)/d$, and ii) $\vect{\tilde{u}} = \vect{v}\sqrt{d}$.
In this case, we obtain
\be \label{first_form_ubound}
\min_{\vect{U}, \,  \vert \alpha_1 \vert^2, ... \vert \alpha_d \vert^2}\Delta^2(\vect{v} \cdot \vect{\theta})_{\rm eQCR} \leq 
\frac{\mathcal{K}'}{1 + d\bar{n}_c(e^{2r}-1)   \mathcal{K}'}.
\ee
with $\mathcal{K}' = \sum_{j=1}^d \frac{v_j^2}{\bar{n}_c + d v_j^2\bar{n}_s}$.
To derive an upper bound that is independent from $\vect{v}$, we notice that $\mathcal{K}'/(1+x \mathcal{K}')$ is a monotonic growing function of $\mathcal{K}'$, implying that $f(\mathcal{K}_1') < f(\mathcal{K}_2')$ if $\mathcal{K}_1' < \mathcal{K}_2'$.
We have $\bar{n}_c + d v_j^2\bar{n}_s \geq \bar{n}_c$ and thus $\mathcal{K}' \leq 1/(d \bar{n}_c)$.
Therefore, 
\be
\min_{\vect{U}, \,  \vert \alpha_1 \vert^2, ... \vert \alpha_d \vert^2}\Delta^2(\vect{v} \cdot \vect{\theta})_{\rm eQCR} \leq 
\frac{1}{d \bar{n}_c e^{2r}}.
\ee
We recover the upper bound (\ref{ubound7UB}) when noticing that $d\bar{n}_c = \bar{n}_T - \bar{n}_s$.

\subsection{Demonstration of Eqs. \makeref{Dthetaave} and \makeref{DthetaaveQCR}}

Let us consider Eq. (\ref{Eq1APPA}) for $|\alpha_j|^2=\bar{n}_c$, for all $j$.
In the regime $\bar{n}_c\gg\bar{n}_s$ [or, equivalently, $\bar{n}_T = d \bar{n}_c + \bar{n}_s \gg (d+1) \bar{n}_s$], we find 
\begin{align}\label{to_be_minimized}
    \Delta^2(\vect{v} \cdot \vect{\theta})_{\rm emom}
    =\frac{e^{-2r}-1}{\bar{n}_c}\left(\sum_{i=1}^d \tilde{u}_iv_i\right)^2+\frac{1}{d\bar{n}_c}+\frac{\bar{n}_s}{\bar{n}_c^2}\sum_{i=1}^d\tilde{u}_i^2v_i^2.
\end{align}
We now show that the choice $\vect{\tilde{u}}= \sqrt{d} \vect{v}$ is optimal for the estimation of the generalized average phase $\vect{v}_{\rm ave} \cdot \vect{\theta}$, thus proving Eq. \makeref{Dthetaave}.

Let us denote Eq. \makeref{to_be_minimized} with $f(\vect{\tilde{u}})$, as a function of $\vect{\tilde{u}}$. 
The goal here is to minimize $f(\vect{\tilde{u}})$ with respect to $\vect{\tilde{u}}$ under the normalization condition $g(\vect{\tilde{u}})=\sum_{i=1}^d \tilde{u}_i^2-1=0$. 
This problem can be solved through the method of Lagrange multipliers. 
Let $\lambda$ denote the Lagrange multiplier, the computation of the partial derivative with respect to $\tilde{u}_i$ of the Lagrangian function $\mathcal{L}(\vect{\tilde{u}},\lambda)=f(\vect{\tilde{u}})-\lambda g(\vect{\tilde{u}})$ gives 
\begin{equation}
    \frac{\partial}{\partial\tilde{u}_i}\mathcal{L}(\vect{\tilde{u}},\lambda)=\frac{e^{-2r}-1}{\bar{n}_c}2\left(\sum_{j=1}^d \tilde{u}_jv_j\right)v_i+2\frac{\bar{n}_s}{\bar{n}_c^2}\tilde{u}_iv_i^2-2\lambda\tilde{u}_i.
\end{equation} 
The constrained minimum problem is solved by the pair $(\vect{\tilde{u}},\lambda)$ which satisfies $(\partial/\partial\tilde{u}_i)\mathcal{L}(\vect{\tilde{u}},\lambda)=0$ and $g(\vect{\tilde{u}})=0$ at the same time. Note that $\vect{\tilde{u}}=\sqrt{d}\vect{v}$ satisfies the constraint $g(\vect{\tilde{u}})=0$. Therefore, the choice $\vect{\tilde{u}}=\sqrt{d}\vect{v}$ is optimal if and only if a value of $\lambda$ can be found such that $(\partial/\partial\tilde{u}_i)\mathcal{L}\left(\vect{\tilde{u}},\lambda\right)\vert_{\vect{\tilde{u}}=\sqrt{d}\vect{v}}=0$.
We thus search for the unique value of $\lambda$ that simultaneously solves all the equations in the following set:
\begin{equation}\label{condition_minimum}
    \frac{e^{-2r}-1}{\bar{n}_c}+2d\frac{\bar{n}_s}{\bar{n}_c^2}v_i^2-2d\lambda=0, 
\end{equation}
with $i$ such that $v_i \neq 0$, since we have divided by $v_i\neq 0$ -- if $v_i=0$, the $i$-th equation is identically satisfied.  If such $\lambda$ actually exists, by subtracting consecutive pairs of equations contained in \makeref{condition_minimum}, we will end up with 
\begin{equation} \label{conditions}
    2d\frac{\bar{n}_s}{\bar{n}_c^2}\left(v_i^2-v_{i+1}^2\right)=0,
\end{equation}
where $v_i, v_{i+1} \neq 0$.
On the other hand, if the above conditions are met, we can immediately determine the sought value of $\lambda$ by solving any of the [linear] equations in \makeref{condition_minimum}. Therefore, Eq. \makeref{conditions} contains necessary and sufficient conditions for $\vect{\tilde{u}}=\sqrt{d}\vect{v}$ to give the optimal QC, which can be fulfilled in two different ways:
\be
\frac{{\bar{n}}_s}{\bar{n}_c^2}=0, \qquad {\rm or} \qquad v_i^2-v_{i+1}^2=0,\nonumber
\ee
with $v_i, v_{i+1} \neq 0$. 
The first alternative may be interpreted as referring to the situation when the last term in Eq. \makeref{to_be_minimized} is so small with respect to the rest that it can be neglected. 
If that is the case, the system in \makeref{condition_minimum} admits one solution $\lambda$ for every value of $\vect{v}$, implying that $\vect{\tilde{u}}=\sqrt{d}\vect{v}$ must be optimal in the estimation of any linear combination of phases. 
On the other hand, if such term cannot be neglected, we need that  $v_i^2-v_{i+1}^2=0$ for all $v_i, v_{i+1}\neq 0$, namely, that all the non-vanishing components of $\vect{v}$ are equal in modulus. As anticipated, this condition identifies the generalized average phase estimation problem, with an arbitrary number of modes between $1$ and $d$. Evaluated at its minimum point and for $\vect{v}=\vect{v}_{\rm ave}$, Eq. \makeref{to_be_minimized} becomes 
\be 
\min_{\vect{U},\, |\alpha_1|^2,\dots,|\alpha_d|^2}\Delta^2(\vect{v}_{\rm ave} \cdot \vect{\theta})_{\rm emom} 
 =  \frac{e^{-2r}\bar{n}_c +\bar{n}_s/d}{d\bar{n}_c^2}.
\ee

Finally, we prove Eq. \makeref{DthetaaveQCR} for $\bar{n}_c\gg\bar{n}_s$ [equivalent to $\bar{n}_T \gg (d+1) \bar{n}_s$].
Setting $|\alpha_j|^2=\bar{n}_c$ for all $j$ in Eq. \makeref{EqAppB} and taking the limit $\bar{n}_c\gg\bar{n}_s$, one gets
\begin{equation}\label{to_be_minimized_QFIM}
    \Delta^2(\vect{v} \cdot \vect{\theta})_{\rm eQCR}=\frac{e^{-2r}-1}{\bar{n}_c}\left(\sum_{i=1}^d\tilde{u}_iv_i\right)^2+\frac{1}{d\bar{n}_c}.
\end{equation}
This is just Eq. \makeref{to_be_minimized} lacking the last term. As already discussed above, the function in Eq. \makeref{to_be_minimized_QFIM} is minimized by choosing $\vect{\tilde{u}}=\sqrt{d}\vect{v}$, which corresponds to the condition used to get the upper bound. The saturation of the bound is only realized for $\vect{v}=\vect{v}_{\rm ave}$ since only in that case an even distribution of photons between the coherent modes, $|\alpha_j|^2=\bar{n}_c$ for all $j$, can be used to attain the minimum of $\Delta^2(\vect{v} \cdot \vect{\theta})_{\rm eQCR}$.
The right hand side of Eq.~(\ref{DthetaaveQCR}) is obtained, for any value of $\bar{n}_T$, assuming that  conditions $\vect{\tilde{u}}=\sqrt{d}\vect{v}_{\rm ave}$ and $\vert \alpha_j \vert^2$ for all $j$ identify the optimal configuration of the sensor network, for any $\bar{n}_T$. 
These assumptions are confirmed by the results of numerical simulations shown in Fig. \ref{fig2}.

\subsection{Demonstration of Eq. \makeref{gain2an}}

To derive Eq. \makeref{gain2an}, we refer to Eq. \makeref{gain2} which gives the definition of $\mathcal{G}_2(\vect{v})$. We work in the regime $\bar{n}_c\gg e^{2r}\bar{n}_s$, which implies also $\bar{n}_c\gg e^{2r'_i}(\bar{n}'_s)_i$ for $i=1,\dots,d$ since, under the constraint $\mathcal{C}_2$, we have $\sum_{i=1}^d(\bar{n}'_s)_i=\bar{n}_s$. The sensitivity of the entangled strategy, $\Delta^2(\vect{v} \cdot \vect{\theta})_{\rm emom}$, can be obtained from Eq. \makeref{to_be_minimized} by neglecting the last term, which needs not to be considered in such regime. The expression of $\Delta^2(\vect{v} \cdot \vect{\theta})_{\rm smom}$, the sensitivity of the separable strategy, is derived from Eq. \makeref{sMOM}. In the limit considered, we only keep the first term both in the numerator and in the denominator of the inverse moment matrix, getting
\begin{equation}\label{smomc2}
    \Delta^2(\vect{v} \cdot \vect{\theta})_{\rm smom}=\vect{v}^{T}\vect{\mathcal{M}^{-1}}\vect{v}=\sum_{i=1}^d\frac{e^{-2r'_i}}{\bar{n}_c}v_i^2.
\end{equation}
We then optimize the sensitivities of both strategies. In the previous section we proved that, when the rightmost term in Eq. \makeref{to_be_minimized} is negligible, the entangled strategy is optimized by the choice $\vect{\tilde{u}}=\sqrt{d}\vect{v}$, with
\begin{equation}
\min_{\vect{U}} \Delta^2(\vect{v} \cdot \vect{\theta})_{\rm emom}=\frac{e^{-2r}}{d\bar{n}_c}.\end{equation}
In order to make the minimization of Eq. \makeref{smomc2} as simple to perform, we need an extra condition that allows us to express $e^{2r'_i}$ as a simple function of $(\bar{n}'_s)_i=\sinh^2r'_i$. As a first case, we assume $r'_i\gg 1$, so that $e^{2r'_i}\approx 4(\bar{n}'_s)_i$. The method of Lagrange multipliers is very easily applied to this case, predicting the minimum 
\begin{equation}
    \min_{r'_1, ..., r'_d} \Delta^2(\vect{v} \cdot \vect{\theta})_{\rm smom}=\frac{e^{-2r}}{\bar{n}_c}\left(\sum_{i=1}^d|v_i|\right)^2,
\end{equation}
which is achieved for $(\bar{n}'_s)_i=\bar{n}_s|v_i|/\sum_{i=1}^d|v_i|$. We also find $\lambda=-(\sum_{i=1}^d|v_i|)^2/4\bar{n}_s^2$, $\lambda$ denoting the Lagrange multiplier associated with this optimization problem. It is easy to see that the ratio between the two sensitivities corresponds to Eq. \makeref{gain2an}. Notice that, for $v_i\to 0$, one has $(\bar{n}'_s)_i\to 0$ too, so that the approximation $r'_i\gg 1$ cannot be valid in such limit. We thus consider a second case, the limit $r'_i\ll 1$, which justifies the truncated series expansion $e^{-2r'_i}\approx 1-2r'_i\approx1-2\sqrt{(\bar{n}'_s)_i}$. Applying again the method of Lagrange multipliers, this time we find
\begin{equation}
    \min_{r'_1, ..., r'_d} \Delta^2(\vect{v} \cdot \vect{\theta})_{\rm smom}=\frac{1}{d\bar{n}_c}-2\frac{\sqrt{\bar{n}_s}}{\bar{n}_c}\sqrt{\sum_{i=1}^d v_i^4}
\end{equation}
and $(\bar{n}'_s)_i=\bar{n}_s v_i^4/\sum_{i=1}^d v_i^4$, with $\lambda=\sqrt{\sum_{i=1}^d v_i^4}/\sqrt{\bar{n}_s}$. 
The expression of the gain is easily verified to be 
\begin{equation}\label{gain2new2}
    \mathcal{G}_2(\vect{v})= \left(1-2d\sqrt{\bar{n}_s}\sqrt{\sum_{i=1}^d v_i^4}\right)e^{2r}.
\end{equation}
In principle, both of the formulas for $\mathcal{G}_2(\vect{v})$, Eqs. (\ref{gain2an}) and (\ref{gain2new2}), are not expected to provide accurate predictions for non-uniform vectors $\vect{v}$, when $r'_i\gg 1$ and $r'_i\ll 1$ occur simultaneously, for different components $v_i$ of the same $\vect{v}$. 
So, it is interesting to note that Eqs. \makeref{gain2an} and \makeref{gain2new2} still give the correct value of the gain, $\mathcal{G}_2(\vect{v})=1$, when applied to the estimation of a single phase. 
Equation \makeref{gain2an} is found in good agreement with numerical results for all values of $\vect{v}$.

\subsection{Demonstration of Eqs. \makeref{optsep2} and \makeref{gaindinf}}

We demonstrate here Eq. \makeref{gaindinf}, which refers to the gain $\mathcal{G}_2(\vect{v}_{\rm ave})$ achieved in the estimation of the generalized average phase. Starting again from Eq. \makeref{sMOM}, with $|\alpha_j|^2=\bar{n}_c$, this time we can impose the less strict condition $\bar{n}_c\gg\bar{n}_s, (\bar{n}_s')_i$. As a consequence, we are only allowed to neglect the second term in the denominator of the inverse moment matrix, ending up with 
\begin{equation}\label{smomc3}
    \Delta^2(\vect{v} \cdot \vect{\theta})_{\rm smom}=\vect{v}^{T}\vect{\mathcal{M}^{-1}}\vect{v}=\sum_{i=1}^d\frac{\bar{n}_ce^{-2r'_i}+(\bar{n}_s')_i}{\bar{n}_c^2}v_i^2.
\end{equation}
Making use of the general relation $e^{2r_i'}=1+2(\bar{n}_s')_i+\sqrt{[1+2(\bar{n}_s')_i]^2-1}$, with $(\bar{n}_s')_i=\sinh^2r_i'$, which holds for any value of $r'_i$, and setting $\vect{v}=\vect{v}_{\rm ave}$, or $v_i^2=1/d^2$, we get
\begin{equation}
    \Delta^2(\vect{v}_{\rm ave} \cdot \vect{\theta})_{\rm smom}=\frac{1}{d^2\bar{n}_c^2}\sum_{i=1}^d   \frac{\bar{n}_c}{X_i+\sqrt{X^2_i-1}}+\frac{X_i-1}{2}.
\end{equation}
Here, we have set $X_i\equiv1+2(\bar{n}_s')_i$ for convenience. The function above has to be minimized with respect to the set of variables $\{X_i\}_{i=1,\dots,d}$ under the constraint $\sum_{i=1}^d X_i=X$, which comes from the original constraint $\sum_{i=1}^d(\bar{n}'_s)_i=\bar{n}_s$. Notice that $X=d+2\bar{n}_s$. Relying again on the method of Lagrange multipliers, we are able to show that the minimum point is $X_i=X/d$, which corresponds, as expected, to splitting $\bar{n}_s$ evenly between the $d$ MZIs used in the separable strategy: $(\bar{n}_s')_i=\bar{n}_s/d$. In this case, we find the following value for the Lagrange multiplier: $\lambda=1/(d^2\bar{n}_c^2)[\bar{n}_c(1-X/\sqrt{X^2-d^2})+1/2]$. Evaluated at its minimum point and for $\vect{v}=\vect{v}_{\rm ave}$, Eq. \makeref{smomc3} becomes
\be 
\min_{r'_1, ..., r'_d} \Delta^2(\vect{v}_{\rm ave} \cdot \vect{\theta})_{\rm smom} = \frac{\bar{n}_c e^{-2 r'} + \bar{n}_s/d}{d\bar{n}_c^2},
\ee
with $r' = {\rm arcsinh} \sqrt{\bar{n}_s/d}$. Taking the ratio with Eq. \makeref{optent2}, which expresses the optimal sensitivity of the entangled strategy, we find Eq. \makeref{gaindinf}.

\subsection{Further discussion on Eq. (\ref{gain4_1})}

\begin{figure}[t!]
  \includegraphics[width=\columnwidth]{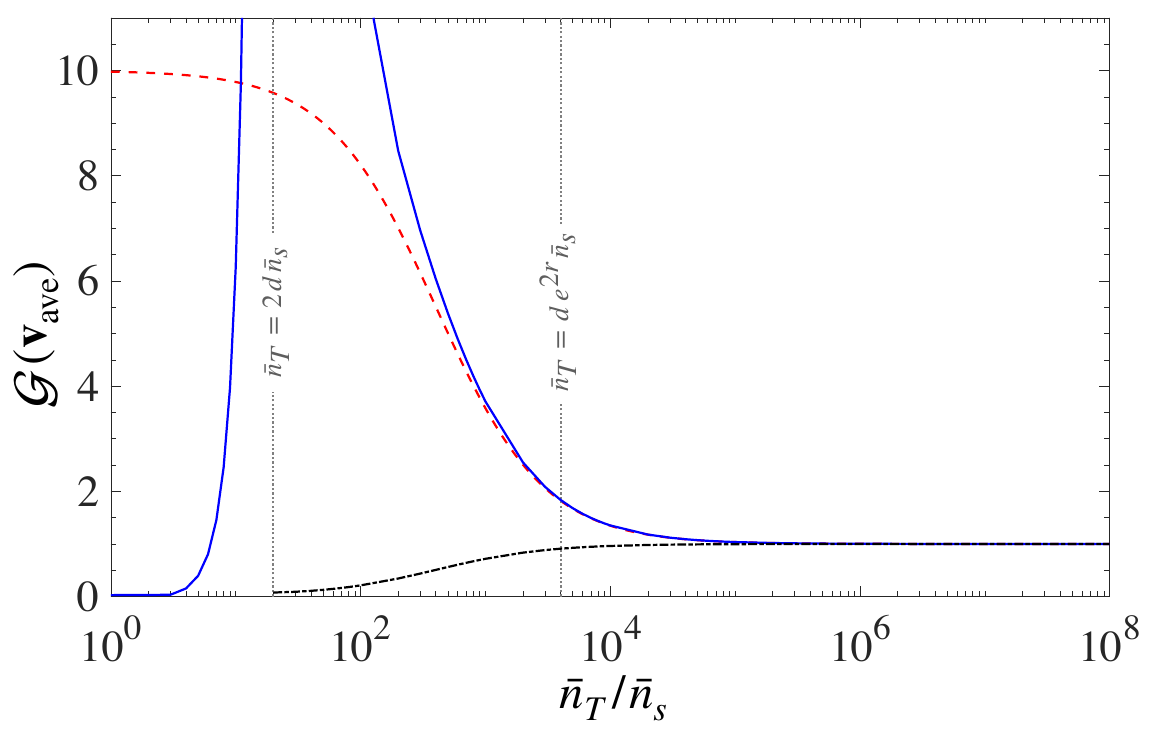} 
\caption{$\mathcal{G}_4(\vect{v}_{\rm ave})$ as a function of $\bar{n}_T$. 
The solid blue line is the result of a  numerical optimization.
The red dashed line is the analytical Eq.~(\ref{gain4_1}).
The dot-dashed black line is Eq.~(\ref{gain4CRB}). 
Vertical dotted lines mark interesting values of $\bar{n}_T$~(see text).
Here $d=10$ and $\bar{n}_s=100$.}
\label{fig8}
\end{figure}

In Fig.~\ref{fig8} we further clarify the behaviour of $\mathcal{G}_4(\vect{v}_{\rm ave})$ in a broad parameter regime.
The dashed red line is the analytical Eq.~(\ref{gain4_1}), which is expected to be accurate for $\bar{n}_T \gg (d+1) \bar{n}_s$.
The analytical formula predicts $\mathcal{G}_4(\vect{v}_{\rm ave})=1$ for $\bar{n}_T \gg d e^{2r} \bar{n}_s$ and  a gain up to $\mathcal{G}_4(\vect{v}_{\rm ave})=d$ when $\bar{n}_T \ll (d+1) \bar{n}_s$.
The solid line is $\mathcal{G}_4(\vect{v}_{\rm ave})$ where minimization over $\vect{U}$ is performed numerically. 
The quantity  diverges at $\bar{n}_T = 2 d \bar{n}_s$ due to the divergence of Eq.~(\ref{Dthetasep4}), while $\min_{\vect{U}} \Delta^2(\vect{v} \cdot \vect{\theta})_{\rm emom}$ remains finite. 
The dot dashed line shows the gain of the entangled strategy (with sensitivity calculate within the multimode moment-matrix approach) with respect to the quantum Cramer-Rao bound of the separable strategy:
\be \label{gain4CRB}
\mathcal{\tilde{G}}_4(\vect{v}) = 
\frac{ \Delta^2(\vect{v} \cdot \vect{\theta})_{\rm sQCR}}{\min_{\vect{U}} \Delta^2(\vect{v} \cdot \vect{\theta})_{\rm emom}},
\ee
where 
\be
\Delta^2(\vect{v} \cdot \vect{\theta})_{\rm sQCR} = \frac{1}{d[\bar{n}_c' e^{2r} + \bar{n}_s]},
\ee
under the constraints considered here.
As we see from the figure, we obtain $\mathcal{\tilde{G}}_4(\vect{v}_{\rm ave})\leq 1$, even where $\mathcal{G}_4(\vect{v}_{\rm ave})$ diverges.
In particular, $\mathcal{G}_4(\vect{v}_{\rm ave})=\mathcal{\tilde{G}}_4(\vect{v}_{\rm ave})=1$ for $\bar{n}_T \gg d e^{2r} \bar{n}_s$.

\subsection{Demonstration of Eqs. (\ref{ineq0}) and (\ref{ineq1})} 

Let us demonstrate Eq. (\ref{ineq0}): the demonstration of Eq. (\ref{ineq1}) is analogous. 
Equation (\ref{ineq0}) is a direct consequence of 
\be \label{ineq2}
\Delta^2(\vect{v} \cdot \vect{\theta})_{\rm emom} = \vect{v}^T \vect{\mathcal{M}}^{-1} \vect{v} \geq 
\vert \vect{v} \vert^4 / \vect{v}^T \vect{\mathcal{M}} \vect{v},
\ee
$\vect{v}^T\vect{\mathcal{M}}\vect{v} \leq 
\vect{v}_{\mu_{\rm max}}^T\vect{\mathcal{M}}\vect{v}_{\mu_{\rm max}} = \mu_{\rm max}$ and $\vert \vect{v} \vert^2 = 1/d$. 
These imply the bound
$\Delta^2(\vect{v} \cdot \vect{\theta})_{\rm emom} \geq 1/(\mu_{\rm max}d)$, which is saturable for $\vect{v}/\vert \vect{v} \vert = \vect{v}_{\mu_{\rm max}}$.
The inequality (\ref{ineq2}) follows from the Cauchy-Schwarz inequality $(\vect{f}^T\vect{f})(\vect{g}^T\vect{g})\geq (\vect{f}^T\vect{g})^2$ with $\vect{f}=\vect{\mathcal{M}}^{1/2} \vect{v}$ and 
$\vect{g}=\vect{\mathcal{M}}^{-1/2}\vect{v}$ and is saturated if and only if $\vect{f} = \lambda \vect{g}$ for some real number $\lambda$, namely
if and only if $\vect{v}$ is an eigenvector of $\vect{\mathcal{M}}$.
Note that $\vect{\mathcal{M}}^{1/2}$ is well definite since $\vect{\mathcal{M}}\geq 0$: 
this follows from $\vect{\mathcal{M}}= \vect{C}^T \vect{\Gamma}^{-1} \vect{C}$, 
$\vect{\Gamma} \geq 0$ being a covariance matrix, and $\vect{\Gamma}^{-1} \geq 0$ being the inverse of a positive semidefinite matrix.


\begin{thebibliography}{400}
\bibitem{notaQFIM}
The matrix $\vect{\FQ}$ has elements $(\vect{\FQ})_{ij} = {\rm Tr}[\hat{\rho} (\hat{L}_i \hat{L}_j + \hat{L}_j \hat{L}_i)]$,
and the Hermitian operators $\hat{L}_i$ ($i=1, ..., d$) are defined by the relation 
$\tfrac{d \hat{\rho}_{\vect{\theta}}}{d\theta_j} = \tfrac{ \hat{L}_j \hat{\rho}_{\vect{\theta}} + \hat{\rho}_{\vect{\theta}} \hat{L}_j}{2}$, and
$\hat{\rho}_{\vect{\theta}}$ is the sensor's output state \cite{HelstromBOOK}.
\bibitem{notaEq13} 
Taking also into account the conditions for the derivation of Eq.~(\ref{ubound3UB}), see Appendix B, the upper bound in Eq.~(\ref{ubound4}) holds under the additional condition $\bar{n}_T \gg (d+1) \bar{n}_s$, and thus for $e^{2r} \mathcal{W} \gtrsim d+1$.
\bibitem{notaEq15}
To be more precise, the lower bound in Eq.~(\ref{ubound6}) holds for $\bar{n}_s \ll \bar{n}_T \ll \bar{n}_s e^{2r}$, where the first term in Eq.~(\ref{ubound7LB}) can be neglected with respect to first one. 
The upper bound in Eq.~(\ref{ubound6}) instead requires $(d+1)\bar{n}_s \ll \bar{n}_T \ll \bar{n}_s e^{2r} \mathcal{W}$, under which condition the first term in Eq.~(\ref{ubound7UB}) can be neglected. 
We conclude by noticing that $\mathcal{W} \geq 1$.
\bibitem{nota3}
The full expression for 
$\tilde{u}^T \mathcal{M}(\tilde{u})^{-1} \tilde{u}$ when taking 
$\vert \alpha_j \vert^2 = \bar{n}_c$ is 
\be
\quad \tilde{u}^T \mathcal{M}(\tilde{u})^{-1} \tilde{u} = \bar{n}_c(e^{-2r}-1)
\bigg( \sum_{j=1}^d \frac{\tilde{u}_j^2}{\bar{n}_c - \tilde{u}_j^2 \bar{n}_s}\bigg)^2
+
\sum_{j=1}^d \frac{(\bar{n}_c  + \tilde{u}_j^2 \bar{n}_s)\tilde{u}_j^2 }{(\bar{n}_c - \tilde{u}_j^2 \bar{n}_s)^2}, \nonumber 
\ee
that reduces to Eq. (\ref{ubound}) for $\bar{n}_c - \tilde{u}_j^2 \bar{n}_s \approx \bar{n}_c$ and taking into account that $\sum_{j=1}^d \tilde{u}_j^2=1$.
For a random choice of splitting network, the condition $\bar{n}_c - \tilde{u}_j^2 \bar{n}_s \approx \bar{n}_c$ can be understood as $d\bar{n}_c \gg \bar{n}_s$ since, in average, $\mathcal{E}_{\rm QC}[\tilde{u}_j^2] = 1/d$. 
Furthermore, for $d\bar{n}_c \gg \bar{n}_s$, we also have $d\bar{n}_c \approx \bar{n}_T$, therefore the condition to recover Eq. (\ref{ubound}) is $\bar{n}_T \gg \bar{n}_s$.
\bibitem{nota2}
Let us consider Eq.~(\ref{MOM}) with $\vert \alpha_j \vert^2 = \bar{n}_c$ for all $j$.
Taking into account that $\mathcal{E}_{\rm QC}[\tilde{u}_j^2] \sim 1/d$, when $d \bar{n}_c \gg \bar{n}_s$, we can approximate $\bar{n}_c - \tilde{u}_j^2 \bar{n}_s \approx \bar{n}_c$ in the denominator of both terms of Eq.~(\ref{MOM}).
Notice that in this case $\bar{n}_T = d \bar{n}_c - \bar{n}_s \approx d \bar{n}_c$.
Furthermore, for $\bar{n}_T \gg \bar{n}_s e^{2r}$, we can neglect $\tilde{u}_j^2\bar{n}_s $ (that is $\leq \bar{n}_s$) with respect to both $\bar{n}_c$ and $\bar{n}_c e^{-2r}$.
\end{thebibliography}
\end{document}